\documentclass[aps,prc,superscriptaddress,showpacs,floatfix,nofootinbib]{revtex4}
\usepackage{amsmath,amssymb,graphicx}
\usepackage{dcolumn} 

\newcommand{\mean}[1]{\left\langle #1 \right\rangle} 

\begin{document}
\voffset=0.5 in  

\renewcommand{\topfraction}{0.99}
\renewcommand{\bottomfraction}{0.99}

\renewcommand{\textfraction}{0.01}

\title{
Directed and elliptic flow of charged pions and protons in Pb + Pb
collisions at 40 and 158$A$ GeV}
\affiliation{NIKHEF, Amsterdam, Netherlands.}
\affiliation{Department of Physics, University of Athens, Athens, Greece.}
\affiliation{Lawrence Berkeley National Laboratory, Berkeley, CA, USA.}
\affiliation{Universit{\'e} Libre de Bruxelles, Brussels, Belgium.}
\affiliation{Comenius University, Bratislava, Slovakia.}
\affiliation{KFKI Research Institute for Particle and Nuclear Physics, Budapest, Hungary.}
\affiliation{MIT, Cambridge, MA, USA.}
\affiliation{Institute of Nuclear Physics, Cracow, Poland.}
\affiliation{Gesellschaft f\"{u}r Schwerionenforschung (GSI), Darmstadt, Germany.}
\affiliation{Wayne State University, Detroit, MI, USA.}
\affiliation{Joint Institute for Nuclear Research, Dubna, Russia.}
\affiliation{Fachbereich Physik der Universit\"{a}t, Frankfurt, Germany.}
\affiliation{CERN, Geneva, Switzerland.}
\affiliation{University of Houston, Houston, TX, USA.}
\affiliation{\'{S}wietokrzyska Academy, Kielce, Poland.}
\affiliation{Fachbereich Physik der Universit\"{a}t, Marburg, Germany.}
\affiliation{Max-Planck-Institut f\"{u}r Physik, Munich, Germany.}
\affiliation{L.P.N.H.E., Universit{\'e} Pierre et Marie Curie, Paris, France.}
\affiliation{Institute of Particle and Nuclear Physics, Charles University, Prague, Czech Republic.}
\affiliation{SPhT, CEA-Saclay, Gif-sur-Yvette, France.}
\affiliation{Nuclear Physics Laboratory, University of Washington, Seattle, WA, USA.}
\affiliation{Institute for Nuclear Studies, Warsaw, Poland.}
\affiliation{Institute for Experimental Physics, University of Warsaw, Warsaw, Poland.}
\affiliation{Rudjer Boskovic Institute, Zagreb, Croatia.}

\author{C.~Alt}
	\affiliation{Fachbereich Physik der Universit\"{a}t, Frankfurt, Germany.}
\author{T.~Anticic} 
        \affiliation{Rudjer Boskovic Institute, Zagreb, Croatia.} 
\author{B.~Baatar}
	\affiliation{Joint Institute for Nuclear Research, Dubna, Russia.}
\author{D.~Barna} 
        \affiliation{KFKI Research Institute for Particle and Nuclear Physics, Budapest, Hungary.}
\author{J.~Bartke} 
        \affiliation{Institute of Nuclear Physics, Cracow, Poland.} 
\author{M.~Behler} 
	\affiliation{Fachbereich Physik der Universit\"{a}t, Marburg, Germany.}
\author{L.~Betev} 
        \affiliation{Fachbereich Physik der Universit\"{a}t, Frankfurt, Germany.} 
\author{H.~Bia{\l}\-kowska} 
        \affiliation{Institute for Nuclear Studies, Warsaw, Poland.} 
\author{A.~Billmeier} 
        \affiliation{Fachbereich Physik der Universit\"{a}t, Frankfurt, Germany.}
\author{C.~Blume} 
        \affiliation{Fachbereich Physik der Universit\"{a}t, Frankfurt, Germany.} 
% ...

\collaboration{NA49 Collaboration} \noaffiliation

\date{\today}

\begin{abstract}
Directed and elliptic flow measurements for charged pions and protons
are reported as a function of transverse momentum, rapidity, and
centrality for 40 and 158$A$ GeV Pb + Pb collisions as recorded
by the NA49 detector. Both the standard method of correlating
particles with an event plane, and the cumulant method of studying
multiparticle correlations are used. In the standard method the
directed flow is corrected for conservation of momentum. In the
cumulant method elliptic flow is reconstructed from genuine 4, 6, and
8-particle correlations, showing the first unequivocal evidence for
collective motion in A+A collisions at SPS energies.
\end{abstract}


\maketitle

\section{Introduction}
In non-central collisions, collective flow leads to characteristic
azimuthal correlations between particle momenta and the reaction
plane.  The geometry of a non-central collision between spherical
nuclei is uniquely specified by the collision axis and the impact
parameter vector ${\bf b}$.  In particular, the latter defines a
unique reference direction in the transverse plane.  Directed flow
($v_1$) and elliptic flow ($v_2$) cause correlations between the
momenta of outgoing particles with this reference
direction~\cite{VoZh96}. They are defined by
\begin{equation}
\label{defv1v2}
v_1\equiv\mean{\cos(\phi - \Phi_{RP})},\ \ 
v_2\equiv\mean{\cos(2(\phi - \Phi_{RP}))},
\end{equation}
where $\phi$ denotes the azimuthal angle of an outgoing particle and
$\Phi_{RP}$ is the azimuthal angle of ${\bf b}$. Angular brackets
denote an average over particles and events.

Since primary collisions between the incoming nucleons are expected to
be insensitive to the direction of impact parameter at high energy,
azimuthal correlations are believed to result from secondary interactions,
or final state interactions.  As such, they are sensitive probes of
``thermalization,'' which should be achieved if final state
interactions are strong enough.  In the energy regime in which
relatively few new particles are created, the flow effects are due to
the nucleons which participate in the collision. Thus at low energy
flow is used to study the properties of compressed nuclear matter and
more specifically the nuclear equation of state~\cite{WRHR}.  In
nuclear collisions at ultrarelativistic energies the number of newly
created particles is so large that their behavior will dominate the
observable flow effects.

For the interpretation of experimental results on flow, theoretical
tools are needed. There are two types of models to describe final
state interactions, based either on hydrodynamics or on a microscopic
transport (or cascade) approach. Hydrodynamics is adequate when the
mean free path of particles is much smaller than the system
size. Then, the interactions between the various particles in the
system can be expressed in terms of global thermodynamic quantities,
{\it i.e.}, an equation of state. In this essentially macroscopic
description, the collective motion results from a pressure gradient in
the reaction volume, the magnitude of which depends upon the
compressibility of the underlying equation of state~\cite{OL92}.
Since partonic and hadronic matter are expected to have different
compressibilities, it may be possible to deduce from a flow
measurement whether it originates from partonic or hadronic matter, or
from the hadronization process taking place during the transition
between the two~\cite{WRHR,DR96,OL98,PD99}. Microscopic cascade models
are more appropriate when the mean free path of the particles is of
the same order, or larger, than the size of the system, which is often
the case in heavy ion collisions. They require a more detailed
knowledge of the interactions (cross sections, etc.)  of the various
particles in the medium. Detailed flow analyses may help to falsify or
confirm the corresponding model assumptions.

In particular, the study of the energy dependence of flow is
considered~\cite{Kolb:99,Kolb:00,Teaney:01} as a promising strategy in
the search for evidence for the hypothesis that the onset of
deconfinement occurs at low SPS energies.  More generally, the energy
scan project~\cite{NA49-ADD1} at the CERN Super Proton Synchrotron
(SPS) was dedicated to the search for the onset of deconfinement in
heavy ion collisions.  In fact anomalies observed in the energy
dependence of total kaon and pion yields \cite{NA49-energy} can
be understood as due to the creation of a transient state of
deconfined matter at energies larger than about 40$A$ GeV
\cite{Gazdzicki:99}.

In that context, this paper presents the most detailed analysis so far
of directed and elliptic flow of pions and protons at various SPS
energies.  The first publication from NA49 on anisotropic
flow~\cite{NA49PRL} was based on a small set of 158$A$ GeV data
with a medium impact parameter trigger with only the tracks in the
Main TPCs used in the analysis.  Subsequently a method was found for
improving the second harmonic event plane resolution and revised
results were posted on the web~\cite{NA49Web}.  The present analysis
of 158$A$ GeV data is both more detailed and more accurate than
the previous one: it uses much larger event statistics, a minimum bias
trigger, integration over transverse momentum $p_t$ or rapidity $y$ 
using cross sections as weights, and improved methods of analysis.
Moreover, in this paper the NA49 results on flow at 40$A$ GeV 
are published for the first time.  Preliminary results from this 
analysis have been presented in 
Refs.~\cite{NA49QM99,Wetzler:QM02,Borghini:QM02}.

Two types of methods are used in the flow analysis: the so-called
``standard'' method~\cite{Danielewicz:hn,Ollitrault:1997di,Poskanzer:1998yz}
requires for each individual collision an ``event plane'' which is an
estimator of its reaction plane. Outgoing particles are then
correlated with this event plane.  This method, however, neglects
other sources of correlations: Bose-Einstein (Fermi-Dirac) statistics,
global momentum conservation, resonance decays, jets, etc.  The
effects of these ``nonflow'' correlations may be large at the SPS, as
shown in Ref.~\cite{Dinh:1999mn,Borghini:2000cm}.  The standard method
has been improved to take into account part of these effects. In
particular, correlations from momentum conservation are now subtracted
following the procedure described in Ref.~\cite{Borghini:2002mv}.
Recently, a new method has been proposed which allows to get rid of
nonflow correlations systematically, independent of their physical
origin~\cite{Borghini:2000sa,Borghini:2001vi,Borghini:2002vp}.  This
method extracts directed and elliptic flow from genuine multiparticle
azimuthal correlations, which are obtained through a cumulant
expansion of measured multiparticle correlations.  The results
obtained with both methods will be presented and compared.

The paper is organized as follows: Sec. \ref{s:experiment} covers
the experiment, Sec. \ref{s:data} describes the data sets, the
selection criteria for events and particles, and the acceptance of the
detector, in Sec.~\ref{s:methods} the two methods of flow
determination are explained, and Sec. \ref{s:results} contains the
results on elliptic and directed flow as function of centrality and
beam energy.  Sec. \ref{s:models} focuses the discussion on model
calculations and Sec. \ref{s:summary} summarizes the paper.

\section{Experiment}
\label{s:experiment}

\begin{figure}[hbt!]
\includegraphics*[width=0.89\textwidth]{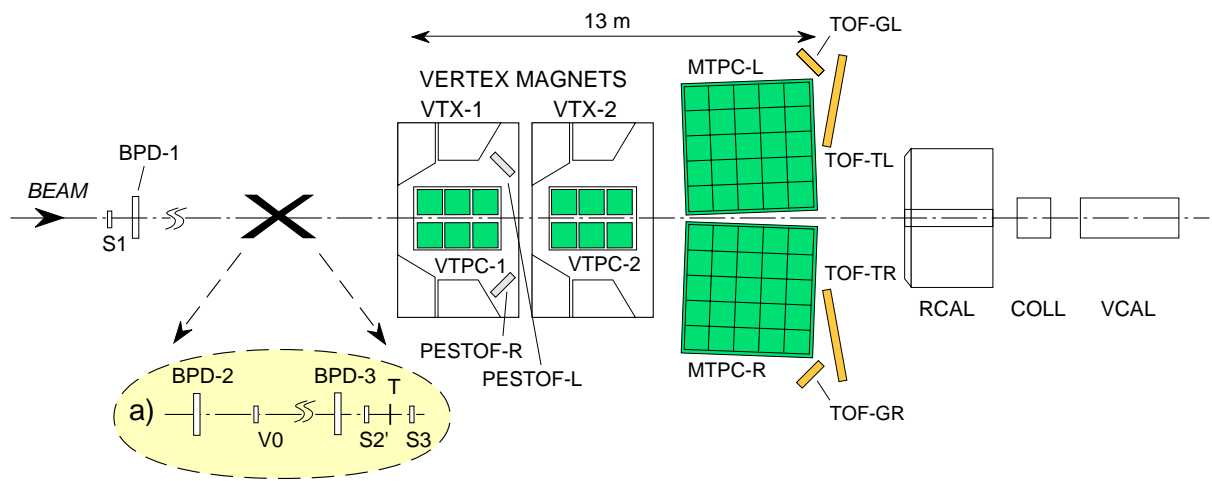}
\caption{\label{fig:setup}  
(Color online) 
 The experimental set--up of the NA49 experiment \protect\cite{na49_nim}.}
\end{figure}

The NA49 experimental set--up \cite{na49_nim} is shown in
Fig. \ref{fig:setup}.  It consists of four large-volume Time
Projection Chambers (TPCs).  Two of these, the Vertex TPCs (VTPC-1 and
VTPC-2), are placed in the magnetic field of two super-conducting
dipole magnets (VTX--1 and VTX--2).  This allows separation of
positively and negatively charged tracks and a precise measurement of
the particle momenta ($p$) with a resolution of $\sigma(p)/p^2 \cong
(0.3-7)\cdot10^{-4}$ (GeV/$c$)$^{-1}$.  The other two TPCs (MTPC-L and
MTPC-R), positioned downstream of the magnets, were optimized for high
precision detection of the ionization energy loss $dE/dx$ (relative
resolution of about 4\%) which provides a means to measure the
particle mass. The TPC data yield spectra of identified hadrons above
midrapidity.  The magnet settings at 158$A$ GeV were: $B$(VTX--1)
$\cong$ 1.5 T and $B$(VTX--2) $\cong$ 1.1 T. In order to optimize the
NA49 acceptance at 40$A$ GeV the magnetic fields of VTX--1 and
VTX--2 were lowered in proportion to the beam momentum. Data were
taken for both field polarities.

The target (T), a thin lead foil (224 mg/cm$^2$, approx. 0.47\% of
Pb-interaction length), was positioned about 80 cm upstream from
VTPC--1. Beam particles were identified by means of their charge as
seen by a gas Cherenkov counter (S2') in front of the target.  An
identical veto-counter directly behind the target (S3) is used to
select minimum bias collisions by requiring a reduction of the
Cherenkov signal by a factor of about 6. Since the Cherenkov signal is
proportional to $Z^2$, this requirement ensures that the projectile
has interacted with a minimal constraint on the type of interaction.
This limits the triggers on non-target interactions to rare beam-gas
collisions, the fraction of which proved to be small after cuts, even
in peripheral Pb+Pb collisions. The counter gas which is present at
atmospheric pressure all the way from S2' to the target and further on
to S3 was He for the 158$A$ GeV and CO$_2$ for the 40$A$ GeV runs.
The signal from a small angle calorimeter (VCAL), which measured the
energy carried by the projectile spectators, was used to make off-line
centrality selections. The geometrical acceptance of the VCAL
calorimeter was adjusted for each energy in order to cover the
projectile spectator region by a proper setting of a collimator (COLL)
\cite{na49_nim,veto}. The NA49 coordinate system is defined as 
right handed with the positive $z$-axis along the beam direction and 
the $x$-axis in the horizontal and the $y$-axis in the vertical plane.

\section{Data}
\label{s:data}

The data on Pb+Pb collisions at 40$A$ GeV were collected within
the energy scan program at the CERN SPS~\cite{NA49-ADD1}. As part of
this program Pb+Pb collisions at 20, 30, 40 and 80$A$ GeV were
recorded by the NA49 detector during heavy ion runs in 1999, 2000 and
2002.  Due to limited beam time, minimum bias data required for a flow
analysis were not taken at 80 $A$ GeV, and the 20 and 30
$A$ GeV data from 2002 have not yet been analyzed. The
corresponding data at the top SPS energy (158$A$ GeV) were taken
from runs in 1996 and 2000.

\subsection{Data sets}
\label{s:datasets}

The data sets used in this analysis were recorded at 158$A$ GeV
and 40$A$ GeV with a minimum bias trigger allowing for a study of
centrality and energy dependence.  Since central collisions have a
small weight in such a selection of events, their number was augmented
by data from central trigger runs at 158$A$ GeV. The final
results for the 40$A$ GeV beam minimum bias data were obtained
from 350 k minimum bias events for the standard method and 310 k
events for the cumulant method. For the 158$A$ GeV results, the
minimum bias events used were 410 k for the standard method and 280 k
for the cumulant method. The 12.5\% most central events added in were
130 k for the standard method and 670 for the cumulant method. In
addition, for the integrated cumulant results, 280 k events from
another run triggered on 20\% most central collisions was added. 
These numbers refer to events which fulfill the selection criteria. 
After verifying that the
analysis of events recorded with opposite field polarities give
compatible results the corresponding data sets were combined and
processed together.  Full coverage of the forward hemisphere for pions
and protons is achieved by using the tracks combined from both the
Vertex and Main TPCs.

\subsection{Selections and particle identification}
\label{s:selections}

The sample of events provided by the hardware trigger is contaminated
by non-target interactions which are removed by a simultaneous cut on
the minimum number of tracks connected to the reconstructed primary
vertex (10) and on the deviation from its nominal position in space
(0.5 cm in all dimensions).  Quality criteria ensured that only
reliably reconstructed tracks were processed. This acceptance was
defined by selecting tracks in each TPC if the number of potential
points in that TPC based on the geometry of the track was at least 20
in the vertex TPCs and 30 in the main TPC. In order to avoid split
tracks the number of fit points for the whole track had to be greater
than 0.55 times the number of potential points for that track. The 
$\chi^2$ per degree of freedom of the fit had to be less than 10. 
Tracks with transverse momentum ($p_t$) up to 2 GeV/$c$ were considered.
The fraction of tracks of particles from weak decays or other secondary
vertices was reduced by cutting on the track distance from the
reconstructed event vertex in the target plane ($\pm$3 cm in the
bending and $\pm$0.5 cm in the non-bending direction).

The binning of the event samples in centrality was done on the basis
of the energy measurement in the forward calorimeter (VCAL). Its
distribution was divided into 6 bins with varying widths. Each bin has
a mean energy ($E^0$) and corresponds in a Glauber-like picture to an
impact parameter range ($b$) with an appropriate mean, a mean number
of wounded nucleons $\mean{N_{WN}}$, a mean number of participants
$\mean{N_{part}}$, and a cross section fraction $\sigma/\sigma_T$ with
$\sigma_T$ being the total hadronic inelastic cross section of Pb+Pb
collisions which has been estimated to be 7.1~b at both
energies. Details of the binning are given in
Table~\ref{tbl:centrality}. In the graphs ``central'' refers to bins 1
plus 2, ``mid-central'' to bins 3 plus 4, and ``peripheral'' to bins 5
plus 6. When we integrate over the first five centrality bins to
present ``minimum bias'' results, we believe we have integrated out to
impact parameters of about 10 fm corresponding to
$\sigma/\sigma_T$=0.435.

In the standard method of flow analysis the determination of the event
plane is required (see below). The uncertainty of its azimuthal angle
in the laboratory coordinate system depends not only on the total
number of particles used but also on the size and sign of the flow
signal of these particles, which are in general different for
different types of flow and different phase space regions. To optimize
the resolution of the event plane orientation the following selection
of tracks used for the event plane determination was made using $y$ in
the center of mass and $p_t$.  For the first harmonic: $0 < p_t < 1$
GeV/$c$ (centrality bins 3--6), $0 < p_t < 0.3$ GeV/$c$ (centrality
bin 1), $0 < p_t < 0.6$ GeV/$c$ (centrality bin 2), $ 1.1 < y < 3.1$
for 158$A$ GeV data, and $0 < p_t < 1$ GeV/$c$ (centrality bins 1--6),
$ 0.8 < y < 2.8$ for 40$A$ GeV data.  For the second harmonic: $0 <
p_t < 1$ GeV/$c$, $ -0.5 < y < 2.1$ for 158$A$ GeV data, and $0 < p_t
< 1$ GeV/$c$, $ -0.4 < y < 1.8$ for 40$A$ GeV data. In comparison to
these selections for good event plane resolution, the differential
data to be presented go to higher $p_t$ but lower maximum $y$ values.

Particle identification is based on energy loss measurements ($dE/dx$)
in the time projection chambers. An enriched sample of pions is
obtained by removing those particles which are obviously not pions by
appropriate cuts in the lab momentum - $dE/dx$ plane. The remaining
contamination amounts to less than 5\% for negatively charged
pions. For positively charged pions it is less than 20\% between 2 and
20~GeV/c momentum in the laboratory. Outside this range the
contamination increases up to 35\% for lower momenta. For higher
momentum the contribution to the measured flow is small due to the
vanishing cross section of pions in this region.  Although the
fraction of misidentified particles is substantial, the effect on the
results will be small, since $v_2$ for pions and protons is comparable
and depends on rapidity and $p_t$ in a similar way. The kaons are
expected to follow the same trend. To examine the influence of the
contamination, $v_1$ and $v_2$ of all charged particles were analyzed
and compared to results for pions. The small differences are included
in the quoted systematic errors.  The proton identification is
restricted to laboratory momenta above 3 GeV/$c$ and thus to the
region of the relativistic rise of the specific energy loss. Tight
upper limits of $dE/dx$ remove almost quantitatively all lighter
particles. The remaining contamination amounts to less than 5\% kaons
and pions for 158$A$ GeV data and less than 8\% for 40$A$ GeV data.
   
\begin{table}[hbt!]
\caption {Listed for the two beam energies and six centralities are: 
 $E^0/E^0_{beam}$, the forward calorimeter energy divided by the beam
 energy; $\mean{E^0/E^0_{beam}}$, the mean value; $\sigma/\sigma_T$,
 the fraction of the total cross section in that bin; the integral of
 $\sigma/\sigma_T$; $b$, the estimated range of impact parameters;
 $\mean{b}$, the estimated mean impact parameter; $\mean{N_{WN}}$, the
 estimated number of wounded nucleons, and $\mean{N_{part}}$, the
 estimated mean number of participants. The 6 bins in centrality have
 different values of $E^0/E^0_{beam}$ at 40 and 158$A$ GeV since
 the acceptance of the forward calorimeter depends on beam energy.}
\begin{center}
\begin{ruledtabular}
\begin{tabular}{l|c|c|c|c|c|c}
Centrality &
\multicolumn{2}{c|}{Central}    & 
\multicolumn{2}{c|}{Mid-Central}        & 
\multicolumn{2}{c}{Peripheral} \\ \hline
& \multicolumn{5}{c|}{Minimum Bias} \\ \hline
Centrality bin    & 1     & 2     & 3     & 4     & 5     & 6     \\ \hline
158$A$ GeV = 32.86 TeV    & & & & & & \\ \cline{1-1}
$E^0/E^0_{beam}$ & 0 - 0.251 & 0.251 - 0.399 & 0.399 - 0.576 & 0.576 -
0.709 & 0.709 - 0.797 & 0.797 - $\infty$ \\
$\mean{E^0/E^0_{beam}}$ & 0.19 & 0.32 & 0.49 & 0.65 & 0.75 & 0.86 \\ \hline
40$A$ GeV = 8.32 TeV    & & & & & & \\ \cline{1-1}
$E^0/E^0_{beam}$ & 0 - 0.169 & 0.169 - 0.314 & 0.314 - 0.509 & 0.509 -
0.66 & 0.66 - 0.778 & 0.778 - $\infty$ \\
$\mean{E^0/E^0_{beam}}$ & 0.12 & 0.24 & 0.41 & 0.58 & 0.71 & 0.86 \\ \hline
both energies & & & & & & \\ \cline{1-1}
$\sigma/\sigma_T$ in each bin& 0.050 & 0.075 & 0.11  & 0.10  & 0.10  & 0.57\\
${\rm Sum}~\sigma/\sigma_T$&   0.050 & 0.125 & 0.235 & 0.335 & 0.435 & 1.00  \\
$b$ (fm) & 0 - 3.4 & 3.4 - 5.3 & 5.3 - 7.4 & 7.4 - 9.1 & 9.1 - 10.2 & 10.2 - $\infty$\\
$\mean{b}$ (fm)         & 2.4   & 4.6   & 6.5   & 8.3   & 9.6  & 11.5 \\
$\mean{N_{WN}}$         & 352   & 281   & 204   & 134   &  88  & 42 \\
$\mean{N_{part}}$       & 366   & 309   & 242   & 178   & 132  & 85 \\
\end{tabular}
\end{ruledtabular}
\end{center}
\label{tbl:centrality}
\end{table}

\subsection{Acceptance}
\label{s:acceptance}

\begin{figure}[hbt!]
\includegraphics*[width=0.51\textwidth]{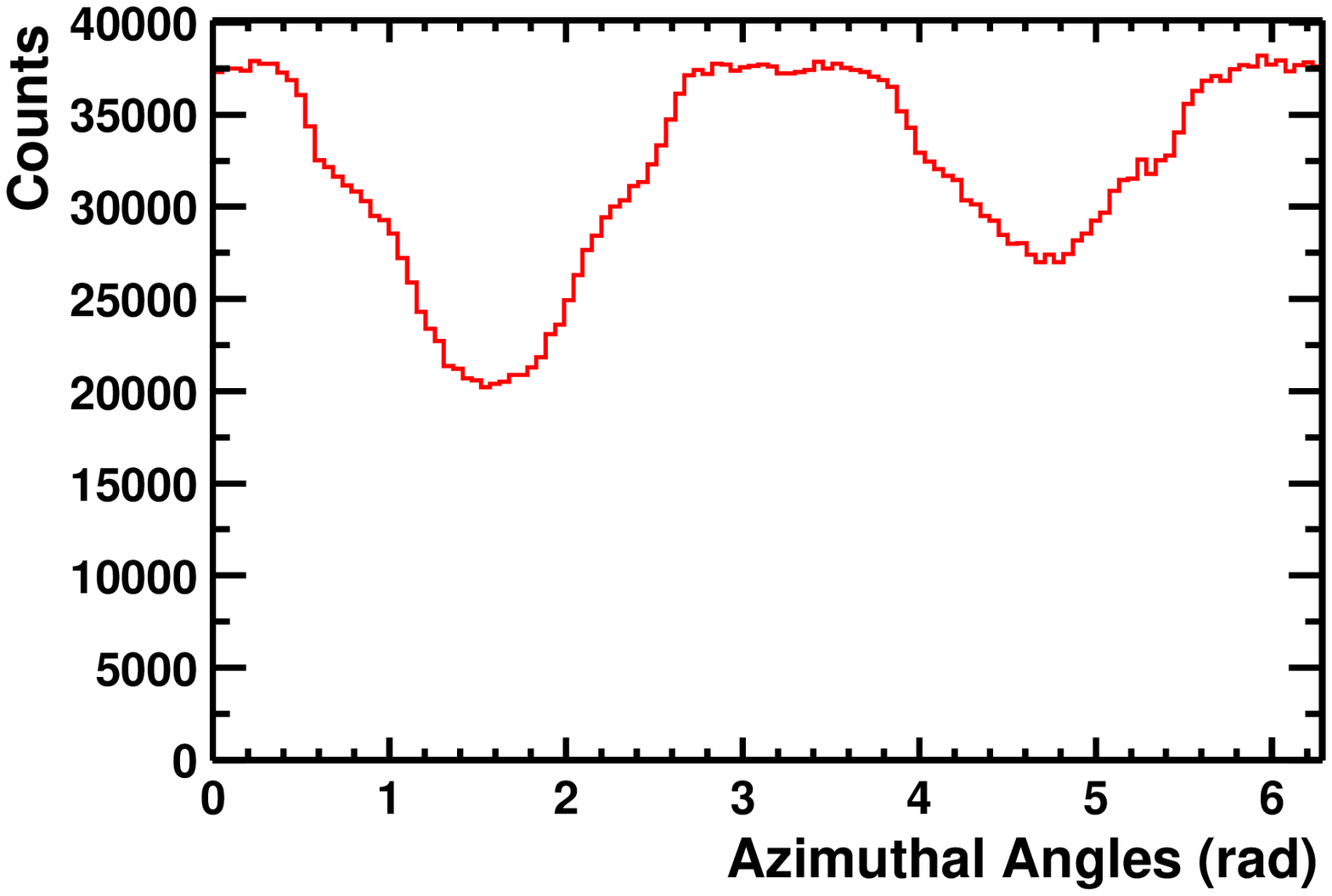}
\includegraphics*[width=0.51\textwidth]{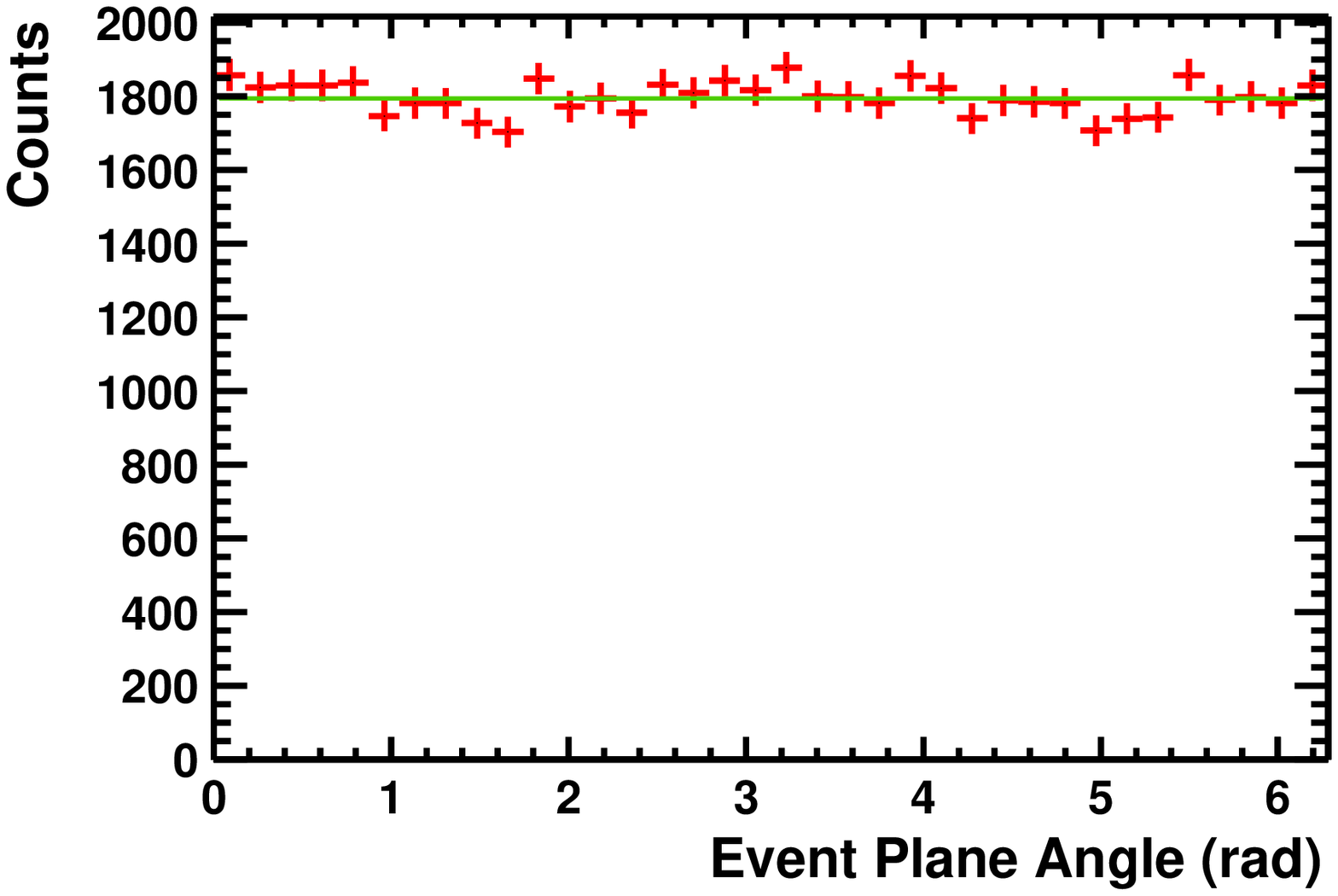}
\caption{\label{fig:phi-psi} 
(Color online) 
On top is a typical azimuthal distribution of
 particles (pions, third centrality bin) at 40$A$ GeV. On the
 bottom is the azimuthal distribution of the first harmonic event
 planes after correction for the laboratory azimuthal anisotropies.}
\end{figure}

The NA49 detector was designed for large acceptance in the forward
hemisphere of the center-of-mass frame. The resulting acceptance is
illustrated by the density distributions of protons and pions as a
function of rapidity and transverse momentum as seen in
Ref.~\cite{NA49-energy}. The scaling of the magnetic field strength
with beam energy ensures similar distributions at both energies. The
NA49 detector employs two dipole magnets with main field components
perpendicular to the beam axis. This breaking of rotational symmetry
together with the rectangular TPC shapes introduces azimuthal
anisotropies which are more pronounced at the lower beam energy. The
Lorentz boost focuses the tracks of all particles forward of
midrapidity into cones of approximately 5 and 10 degrees at 158 and
40$A$ GeV, respectively. Acceptance losses occur for particles at
large angles with respect to the bending plane. A typical inclusive
azimuthal angle ($\phi$) distribution is shown in
Fig.~\ref{fig:phi-psi} top for pions, which in an ideal detector would
be flat.

In the standard method, event plane determinations from such a
distribution would obviously lead to acceptance biased event
planes. As long as each bin in the acceptance is populated
significantly, the bias can be removed as described below by
recentering the particles in the plane perpendicular to the beam for
each rapidity and $p_t$ bin, in such a way that the event plane
distribution becomes flat.  The result of this procedure is
exemplified for the first harmonic in Fig.~\ref{fig:phi-psi} bottom.

In the cumulant method, the main contributions of the detector
inefficiency, that is, spurious correlations which have nothing to do
with physical (flow or nonflow) correlations, are automatically
removed.  The only required acceptance corrections amount to a global
multiplicative factor which depends solely on the specific detector
under study, and can thus be calculated separately from the flow
analysis \cite{Borghini:2001vi}.  Further details will be given in
Sec.~\ref{s:cumulant}.

\section{Methods}
\label{s:methods}

In this section, we recall the principles of the standard 
(Sec.~\ref{s:standard}) and cumulant (Sec.~\ref{s:cumulant})
methods of flow analysis.

\subsection{Standard method}
\label{s:standard}

The standard method~\cite{Ollitrault:1997di,Poskanzer:1998yz} correlates 
the azimuthal angles of particles, $\phi$, with an estimated event plane 
to obtain the observed coefficients in a Fourier expansion in the plane
transverse to the beam. The observed coefficients are then divided by
the resolution of the event plane obtained from the correlation of the
estimated event planes of two random subevents. The estimated event
plane angle, $\Phi_n$, are obtained from the azimuthal angles of the
$Q_n$ vectors whose $x$ and $y$ components are defined by:
\begin{eqnarray}   
Q_n \cos n\Phi_n &=& \sum_i w_i \left ( \cos n\phi_i - \mean{\cos
n\phi} \right ), \nonumber 
\\     
Q_n \sin n\Phi_n & =& \sum_i w_i \left (\sin n\phi_i - \mean{\sin
n\phi} \right ) \,,      
\label{eq:Q_n}    
\end{eqnarray}   
where $n$ is the harmonic order and the sum is taken over the $M$
particles in the event. In this work the weights, $w_i$, have been
taken to be $p_t$ for the second harmonic and $y$ in the center of
mass for the first harmonic. To make the event plane isotropic in the
laboratory in order to avoid acceptance correlations, we have used the
recentering method~\cite{Poskanzer:1998yz}. The mean $\mean{\sin
n\phi}$ and $\mean{\cos n\phi}$ values in the above equation were
calculated as a function of $p_t$ and $y$ for all particles in all
events in a first pass through the data, and then used in a second
pass to recenter the $Q_n$ vector to be isotropic as shown in
Fig.~\ref{fig:phi-psi} bottom. The mean sin and cos values were stored
in a matrix of 20 $p_t$ values and 50 $y$ values for each
harmonic. Particles were only used for the event plane determination
if the absolute value of the mean sin and cos values for that bin were
less than 0.2. Then the flow values are calculated by
\begin{equation}
\label{eq:v}
v_n = \frac{\mean{\cos(n(\phi_i - \Phi_n))}} {\mean{\cos(n(\Phi_n - \Phi_{RP}))}}.
\end{equation} 
If particle $i$ was used also for the event plane determination, its
contribution to $Q_n$ is subtracted before calculating $\Phi_n$, so as
to avoid autocorrelations. The denominator is called the resolution
and corrects for the difference between the estimated event plane and
the real reaction plane, $\Phi_{RP}$. It is obtained from the
resolution of the subevent event planes, which is
$\sqrt{\mean{\cos(n(\Phi_a - \Phi_b))}}$. The resolution of the full
event plane, for small resolution, is approximately $\sqrt{2}$ larger,
but the actual equation in Ref.~\cite{Poskanzer:1998yz} was used for
this calculation. It was found to be more accurate to calculate $v_2$
relative to the second harmonic event plane, $\Phi_2$, although the
sign of $v_2$ was determined to be positive by correlation with the
first harmonic event plane $\Phi_1$. The sign of $v_1$ was set so that
protons at high rapidity have positive $v_1$ as described below. The
software used in this analysis was derived from that used for the
STAR experiment~\cite{Ackermann:2000tr}.

Equation (\ref{eq:v}) is the most general form to determine $v_n$. In
the case of the NA49 experiment the main losses are concentrated
around 90 deg and 270 deg in the up and down directions. (See
Fig.~\ref{fig:phi-psi}~top.) In order to limit the analysis to the
regions of more uniform acceptance a cut on the particle azimuthal
angle was applied: Particles with $\cos(2\phi) < 0$ are cut out.  This
however requires large acceptance corrections if Eq.~\ref{eq:v} is
used for $v_2$ determination.  The correlation term may be modified in
Eq.~(\ref{eq:v}) in order to select azimuthal regions with small
distortions. Its numerator may be rewritten in the form of a sum of
products instead of a difference of angles:
\begin{equation}
\label{eq:vprime}
  \mean{\cos(n\phi_i) \cdot \cos(n\Phi_n)} 
+ \mean{\sin(n\phi_i) \cdot \sin(n\Phi_n)}. 
\end{equation}
Since the (ideal) inclusive azimuthal distributions in {\it A+A}
collisions are flat by definition, both terms must be of equal
magnitude and $v_n$ can be calculated by 2 times either the first or
second term.  The cut applied on the azimuthal angle ($\cos(2\phi) <
0$) leads to large corrections to the cosine term in
Eq.~(\ref{eq:vprime}).  Therefore, for the second harmonic at 40$A$
GeV the numerator in the $v_2$ calculation was done using the
expression
\begin{equation}
\label{eq:vprimeprime}
  2\mean{\sin(n\phi_i) \cdot \sin(n\Phi_n)}.
\end{equation} 
The same line of argument applies to the denominator of
Eq.~(\ref{eq:v}) where again only the sin terms were used and the
subevent plane resolution was increased by a factor of $\sqrt{2}$:
\begin{equation}
\label{eq:vmod}
v_n = \frac{2 \cdot \mean{\sin(n\phi_i) \cdot \sin(n\Phi_n)}}
        {\sqrt{2} \cdot \mean{\sin(n\Phi_n) \cdot \sin(n\Phi_{RP})}} 
\end{equation}
Determination of $v_2$ by Eq.~\ref{eq:vmod} requires the acceptance
correction only for losses in the selected angular ranges.

\begin{figure}[hbt!]
\includegraphics*[width=0.51\textwidth]{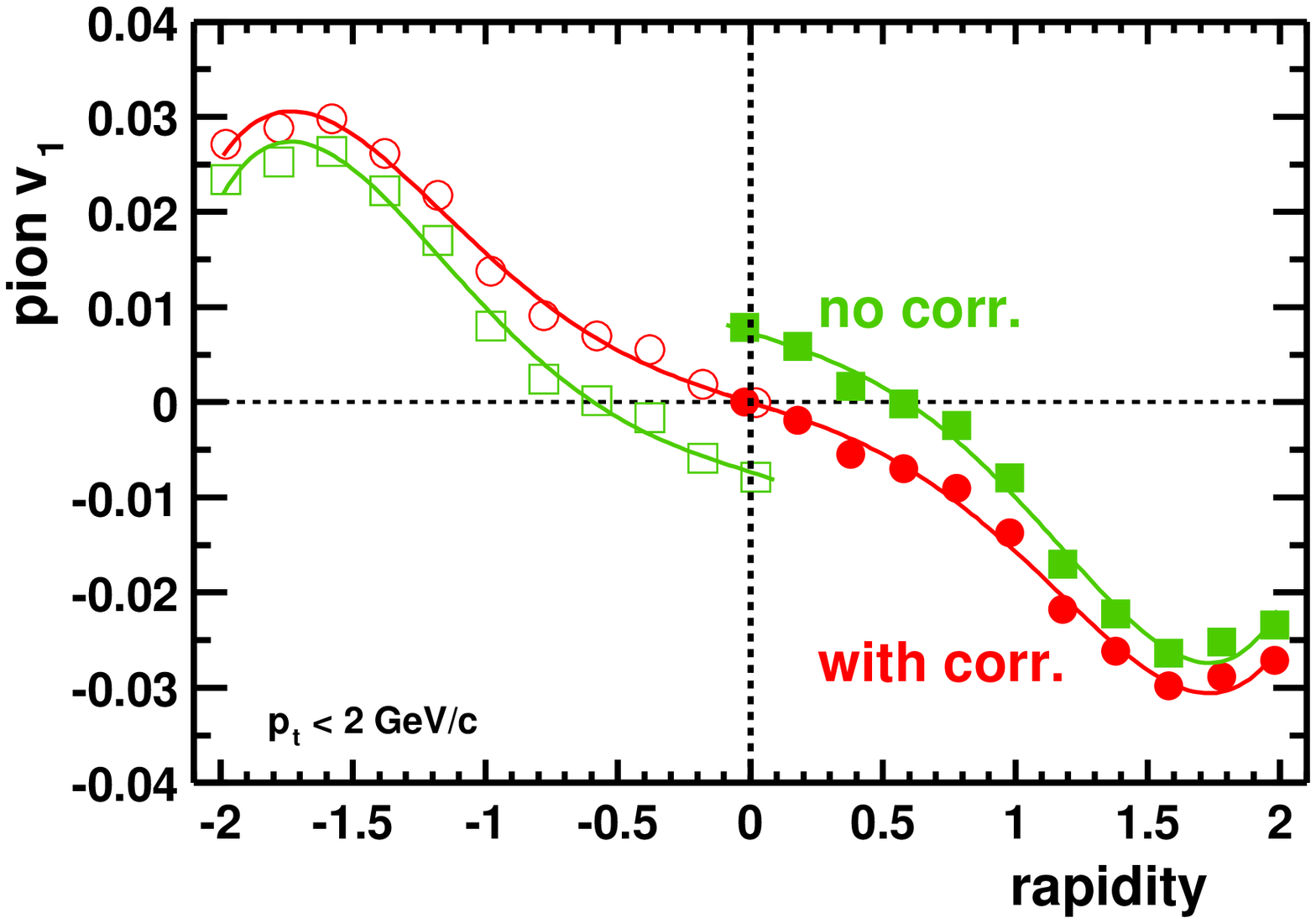}
\includegraphics*[width=0.51\textwidth]{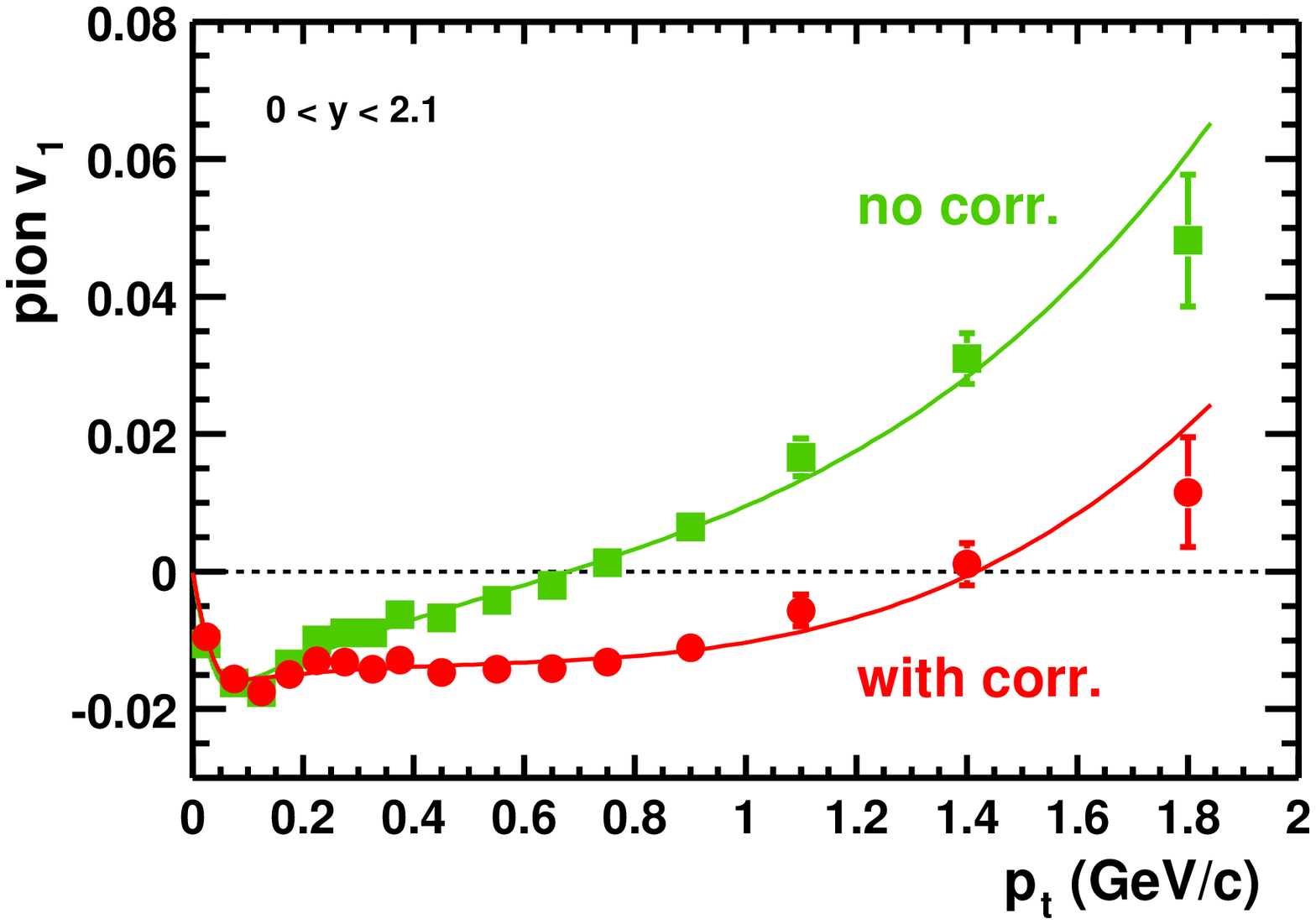}
\caption{\label{fig:momCons} (Color online) Charged pion directed flow
as a 
function of rapidity (top) and $p_t$ (bottom) for minimum bias 158$A$
GeV Pb + Pb.  Shown are $v_1$ before (squares) and after (circles)
correction for momentum conservation. The open points in the top graph
have been reflected about midrapidity. The solid lines are polynomial
fits.}
\end{figure}

A momentum conservation correction for the first harmonic was made as
described in Ref.~\cite{Borghini:2002mv}. The correction is made to
the observed differential flow values before they are divided by the
event plane resolution. Figure~\ref{fig:momCons} shows that this
correction, without any adjustable parameter, makes the directed flow
curve cross zero at midrapidity. The figure also shows that because
this correction is proportional to $p_t$ there is a large effect at
high $p_t$. There is no effect on the elliptic flow because it is
calculated relative to the second harmonic event
plane. Table~\ref{tbl:momcons} shows the parameters used in making
this correction.

\begin{table}[hbt!]
\caption {Listed for the two beam energies and six centralities are: 
$N$, the estimated total multiplicity of charged plus neutral hadrons
over all phase space; $\mean{p_t^2}$, the estimated mean $p_t^2$ of
all hadrons over all phase space; $M$, the mean multiplicity of
particles used for the event plane determination; $f$, the fraction
defined by Eq.~(17) in Ref.~\cite{Borghini:2002mv}, which controls the
correction for momentum conservation; $\chi_1$, the first harmonic
resolution parameter for the full event plane (it follows the
convention of Ref.~\cite{Ollitrault:ba} and is $\sqrt{2}$ smaller than
defined in Ref.~\cite{Poskanzer:1998yz}); the resolution of the first
harmonic full event plane; the percent increase in the resolution due
to momentum conservation, and the resolution of the second harmonic
event plane. The first two centralities at 158$A$ GeV had more
restrictive $p_t$ cuts. The first centrality bin at 40$A$ GeV did
not have sufficient statistics to determine the second harmonic event
plane.}
\begin{center}
\begin{ruledtabular}
\begin{tabular}{l|c|c|c|c|c|c}
Centrality              & 1     & 2     & 3     & 4     & 5     & 6     \\ \hline
158$A$ GeV              & & & & & & \\ \cline{1-1}
$N$                     & 2402  & 1971  & 1471  & 1028  & 717   & 457   \\
$\mean{p_t^2}$ (GeV$^2/c^2$)  
                        & 0.32  & 0.32  & 0.31  & 0.30  & 0.28  & 0.27  \\
$M$                     & 119   & 181   & 154   & 110   & 78    & 46    \\
$f$                     & 0.07  & 0.14  & 0.17  & 0.17  & 0.18  & 0.17  \\
$\chi_1$                & 0.25  & 0.27  & 0.40  & 0.44  & 0.48  & 0.47  \\
resolution 1st          & 0.22  & 0.24  & 0.34  & 0.38  & 0.40  & 0.40  \\
res. increase, \%       & 4     & 16    & 11    & 9     & 7     & 7   \\ 
resolution 2nd          & 0.23  & 0.28  & 0.36  & 0.40  & 0.37  & 0.29
\\ \hline
40$A$ GeV               & & & & & & \\ \cline{1-1}
$N$                     & 1473  & 1215  & 913   & 643   & 453   & 290   \\
$\mean{p_t^2}$ (GeV$^2/c^2$)  
                        & 0.39  & 0.38  & 0.38  & 0.36  & 0.34  & 0.32  \\
$M$                     & 89.3  & 76.4  & 59.5  & 42.2  & 29.7  & 17.2  \\
$f$                     & 0.11  & 0.11  & 0.11  & 0.12  & 0.12  & 0.11\\
$\chi_1$                & 0.23  & 0.23  & 0.30  & 0.34  & 0.38  & 0.40 \\
resolution 1st          & 0.20  & 0.20  & 0.26  & 0.29  & 0.33  & 0.34 \\
res. increase, \%       & 14    & 16    & 8     & 6     & 5     & 4  \\
resolution 2nd          & -     & 0.13  & 0.25  & 0.31  & 0.25  & 0.21 \\
\end{tabular}
\end{ruledtabular}
\end{center}
\label{tbl:momcons}
\end{table}

\subsection{Cumulant method}
\label{s:cumulant}

In this section we first recall the motivations for developing
alternative methods of flow analysis. We then explain the principle of
the method. Unlike the standard method, the cumulant method yields in
principle several independent estimates of directed and elliptic flow,
which will be defined below.  Finally, we describe the practical
implementation of the method.

At the core of the standard method outlined in Sec.~\ref{s:standard}
lies a study of two-particle correlations: one correlates either {\em
two} subevents, to derive the event plane resolution, or one particle
with (a second particle belonging to) the $Q_n$ vector. The basic
assumption is that the correlation between two arbitrary particles is
mainly due to the correlation of each single particle with the
reaction plane, that is, due to flow. However, there exist other
sources of two-body correlations, that do not depend on the reaction
plane; for instance, physical correlations arising from quantum (HBT)
effects, global momentum conservation, resonance decays, or jets. At
SPS energies, it turns out that these ``nonflow'' two-particle
correlations are {\em a priori} of the same magnitude as the
correlations due to flow~\cite{Dinh:1999mn,Borghini:2000cm}, at least
in some phase space regions. While some of these correlations can be
taken into account in the standard method with a minimal modeling of
the collisions (see end of Sec.~\ref{s:standard}), others cannot be
estimated as reliably.

This observation motivated the elaboration of new methods of flow
analysis, which are much less biased by nonflow correlations than the
standard method~\cite{Borghini:2000sa}.  The basic idea of the methods
is to extract flow from multiparticle azimuthal correlations, instead
of using the correlation between two particles only.  Naturally, the
measured $k$-body correlations also consist of contributions due to
flow and to nonflow effects.  Nevertheless, by performing a cumulant
expansion of the measured correlations, it is possible to disentangle
the flow contribution from the other, unwanted sources of
correlations.  Thus, at the level of four-particle correlations, one
can remove {\em all} nonflow two- and three-particle correlations,
keeping only the correlation due to flow, plus a systematic
uncertainty arising from genuine nonflow {\em four}-particle
correlations, which is expected to be small.

The cumulant method not only minimizes the influence of nonflow
correlations; it also provides several independent estimates of $v_1$
and $v_2$, which will be labeled by the order $k$ at which the
cumulant expansion is performed: for instance, $v_2\{4\}$ denotes our
estimate of $v_2$ using cumulants of $4$-particle correlations, etc.
Generally speaking, the systematic error due to nonflow correlations
decreases as the order $k$ increases, at the expense of an increased
statistical error.

To be more specific, we first consider a simplified situation where
one wishes to measure the average value of the flow $v_n$ over the
detector acceptance, which is assumed to have perfect azimuthal
symmetry.  The lowest order estimate of $v_n$ from two-particle
correlations, $v_n\{2\}$, is then defined by
\begin{equation}
\label{eq:vn2}
v_n\{2\}^2\equiv \mean{e^{in(\phi_1-\phi_2)}}
\end{equation}
where brackets denote an average value over pairs of particles emitted
in a collision, and over events.  Please note that $v_n\{2\}$ is {\em
a priori} consistent with the value given by the standard method,
Eq.~(\ref{eq:v}), at least if the cuts in phase space are identical in
both analyses.

Higher order estimates are obtained from two complementary
multiparticle methods.  The first one~\cite{Borghini:2001vi} measures
the flow harmonics separately, either $v_1$ or $v_2$.  For instance,
the four-particle estimate $v_n\{4\}$ is defined by
\begin{eqnarray}
\label{eq:vn4}
-v_n\{4\}^4 &\equiv& \mean{e^{in(\phi_1+\phi_2-\phi_3-\phi_4)}} \cr
 & &-\mean{e^{in(\phi_1-\phi_3)}}\mean{e^{in(\phi_2-\phi_4)}} \cr
 & &-\mean{e^{in(\phi_1-\phi_4)}}\mean{e^{in(\phi_2-\phi_3)}},
\end{eqnarray}
where the average runs over quadruplets of particles emitted in the
collision, and over events.  The right-hand side defines the cumulant
of the four-particle correlations.  This can be generalized to an
arbitrary even number of particles, which yields higher order
estimates $v_n\{6\}$, $v_n\{8\}$, etc.

The second multiparticle method~\cite{Borghini:2002vp} was used to
analyze directed flow $v_1$. It relies on a study of {\rm
three}-particle correlations which involve both $v_1$ and $v_2$:
\begin{equation}
\label{v1/2idea}
\mean{e^{i(\phi_1+\phi_2-2\phi_3)}}\simeq (v_1)^2 v_2.
\end{equation}
In the case of NA49, we shall see that the first multiparticle method
provides reliable estimates of $v_2$ (the most reliable was found to
be $v_2\{2\}$ at 40$A$ GeV and $v_2\{4\}$ at 158$A$ GeV, as
will be discussed later).  Then, the above equation can be used to
obtain an estimate of $v_1$, which is denoted by $v_1\{3\}$ since it
involves a $3$-particle correlation.  As shown in Ref.\
\cite{Borghini:2002vp}, and will be seen below in Sec.~\ref{s:v1},
$v_1\{3\}$ offers the best compromise between statistical errors
(which prevent obtaining $v_1\{4\}$ with the first method) and
systematic errors from nonflow correlations, which plague the lowest
order estimate $v_1\{2\}$.  In particular, among other nonflow
correlations, $v_1\{3\}$ is insensitive to the correlation due to
momentum conservation, so that one need not compute it explicitly as
in the case of the standard method.  As a matter of fact, a
straightforward calculation using the {\em three}-particle correlation
due to momentum conservation, given by Eq.~(12) in
Ref.~\cite{Borghini:2003ur}, shows that the contribution of $p_t$
conservation to the average $\mean{e^{i(\phi_1+\phi_2-2\phi_3)}}$ is
$\mean{p_t}^2 / (N^2
\mean{p_t^2})$, roughly smaller than $1/N^2$.  (Please note that this
three-particle correlation is positive, while the two-particle one is
negative, back-to-back.)  With the values of $N$ listed in
Table~\ref{tbl:momcons}, this ranges from $0.2\times 10^{-6}$ to
$4.8\times 10^{-6}$ at 158$A$ GeV: for the various centrality
bins, this is a factor of 10 smaller than the $(v_1)^2 v_2$ values we
shall find. Therefore, the contamination of correlations due to
transverse momentum conservation in our derivation of the estimate
$v_1\{3\}$ is indeed negligible.

The flow analysis with either multiparticle method consists of two
successive steps. The first step is to estimate the average value of
$v_1$ and $v_2$ over phase space (in practice, these are weighted
averages, as we shall see shortly), which we call ``integrated
flow''. This is done using Eq.~(\ref{eq:vn2}), Eq.~(\ref{eq:vn4}) or
Eq.~(\ref{v1/2idea}), which yield $v_n\{2\}$, $v_n\{4\}$ and
$v_1\{3\}$, respectively.  The second step is to analyze differential
flow, $v_n(p_t,y)$, in a narrow $(p_t,y)$ window.  For this purpose,
one performs averages as in Eqs.~(\ref{eq:vn2}-\ref{v1/2idea}), where
the particle with angle $\phi_1$ belongs to the $(p_t,y)$ window under
study, while the average over $\phi_2$, $\phi_3$, $\phi_4$ is taken
over all detected particles. The left-hand sides of
Eqs.~(\ref{eq:vn2},\ref{eq:vn4}) and the right-hand side of
Eq.~(\ref{v1/2idea}) are then replaced by $v_n\{2\}(p_t,y)\times
v_n\{2\}$, $v_n\{4\}(p_t,y)\times v_n\{4\}^3$, $v_1\{3\}(p_t,y) \times
v_1\{3\} v_2$, respectively.  This defines the estimates of
differential flow from 2, 4 and 3 particle correlations. Note that
they can be obtained only once the integrated flow is known.  In order
to reduce the computing time, the analysis was performed over 20 $p_t$
bins of 0.1 GeV/$c$ and 20 $y$ bins of 0.3 rapidity units (instead of
50 $y$ bins in the standard method).

In the case of NA49, the use of higher order cumulants was limited by
statistical errors, in particular for differential flow.  In practice,
up to four estimates of integrated elliptic flow were obtained, namely
$v_2\{2\}$, $v_2\{4\}$, $v_2\{6\}$ and $v_2\{8\}$, but at most two
($v_2\{2\}$ and $v_2\{4\}$) for differential flow.  In the case of
directed flow, at most three estimates were obtained for integrated
flow ($v_1\{2\}$, $v_1\{3\}$ and $v_1\{4\}$) and two for differential
flow ($v_1\{2\}$ and $v_1\{3\}$).

The practical implementation of these multiparticle methods is
described in detail in Refs.~\cite{Borghini:2001vi,Borghini:2002vp}.
In order to illustrate the procedure, we recall here how estimates of
integrated (directed or elliptic) flow are obtained from the first
multiparticle method outlined above.  One first defines the generating
function
\begin{equation}
\label{genfunc}
\mean{G_n(z)} = \mean{\prod_{j=1}^M 
\left( 1+ \frac{w_n(j)}{M} (z\,e^{-in\phi_j} + z^*\,e^{in\phi_j}) \right)},
\end{equation}
where $z$ is a complex variable, and $z^*$ its complex conjugate.  The
product runs over particles detected in a single event, and
$w_n(j)\equiv w_n({p_t}_j, y_j)$ is the weight attributed to the
$j$-th particle with azimuthal angle $\phi_j$. Angular brackets denote
an average over events.  A similar generating function for the
analysis of directed flow from three-particle correlations can be
found in Ref.~\cite{Borghini:2002vp}. Weights in Eq.~(\ref{genfunc})
are identical to the weights used in the standard method (see Eq.\
(\ref{eq:Q_n})), namely $y$ in the center of mass for $v_1$, and $p_t$
for $v_2$.  As in the standard method, they are introduced in order to
reduce the statistical error.

In Eq.~(\ref{eq:Q_n}), we use the same value of $M$ for all
events in a given centrality bin. This value has been fixed to 80\% of
the average event multiplicity in the bin. The small fraction of
events having multiplicity less than $M$ are rejected. For the events
having multiplicity greater than $M$, the $M$ particles required to
construct the generating function are chosen randomly. Alternatively,
one could have chosen for $M$ in Eq.~(\ref{eq:Q_n}) the total event
multiplicity. We have checked on a few examples that results are the
same within statistical errors. Note that the value of $M$ is much
larger for the cumulant method (Table~\ref{tbl:cum}) than for the
standard method (Table~\ref{tbl:momcons}). In the standard method, we
have seen that cuts were performed in order to minimize the azimuthal
asymmetry of the detector, resulting in a lower value of $M$. In the
cumulant method, such detector effects are taken into account, as will
be explained below, so that cuts are not necessary. Furthermore,
statistical errors are extremely sensitive to $M$ for higher-order
estimates, so that it is important to use as many particles as
possible.

The cumulants of $2k$-particle correlations, $c_n\{2k\}$ are 
then obtained by expanding in power series the generating function of 
cumulants, ${\cal C}_n(z)$, defined as 
\begin{eqnarray}
\label{gencum}
{\cal C}_n(z) &=& M\left( \mean{G_n(z)}^{1/M} -1 \right) \cr
 & \equiv & \sum_{k=0}^{+\infty} c_n\{2k\}\mean{w_n^2}^k \frac{|z|^{2k}}{(k!)^2}.
\end{eqnarray}
The average value of the weight squared, $\mean{w_n^2}$, has been
introduced so that the cumulants $c_n\{2k\}$ are dimensionless.  In
practice, the cumulants are obtained from the generating function
using interpolation formulas given in Appendix \ref{s:interpolation}.
Finally, each cumulant yields an independent estimate of the
integrated flow $v_n$:
\begin{eqnarray}
\label{flow&cumul-int}
c_n\{2\} = v_n\{2\}^2 &,& c_n\{4\} = -v_n\{4\}^4 \,,  \cr
c_n\{6\} = 4\,v_n\{6\}^6 &,& c_n\{8\} = -33\,v_n\{8\}^8 \ldots
\end{eqnarray} 
A similar procedure holds for differential flow.  If the cumulant
extracted from the data comes out with the wrong sign (for instance, a
positive number for $c_n\{4\}$), one cannot obtain the corresponding
flow estimate ($v_n\{4\}$).  As we shall see in Sec.~\ref{s:results},
this does occur, most often for central collisions where the flow is
small.  There are two reasons for this: statistical fluctuations,
which may be large for multiparticle cumulants
\cite{Borghini:2001vi}; nonflow correlations, in particular 
for two-particle cumulants, which may be opposite to the correlations
due to flow (see the effect of momentum conservation 
on directed flow in Sec.\ref{s:v1}).

In the last equation, $v_n$ denotes the weighted integrated flow, 
defined as 
\begin{equation}
\label{weight-int}
v_n \equiv \frac{\langle w_n e^{in(\phi-\Phi_R)}\rangle}
{\sqrt{\langle w_n^2 \rangle}}. 
\end{equation}
It is dimensionless, as it should be.  We decided to normalize with
$\sqrt{\langle w_n^2\rangle}$ rather than $\langle w_n \rangle$ since
the weight $w_n$ can be negative: for a perfect detector,
$\mean{w_1}=\mean{y}$ would vanish!  One should note that this
integrated flow differs from that obtained in the standard method,
which integrates the doubly differential flow without weights.  Since
the average values in Eq.~(\ref{weight-int}) are taken over the whole
detector acceptance, the integrated $v_n$ is a strongly
detector-dependent quantity, whose absolute value has little physical
significance.  It is essentially an intermediate step: as explained
above, one can analyze differential flow only once integrated flow is
known.  However, we shall see in Sec.~\ref{s:centrality} that the
centrality dependence of $v_n$ is meaningful.  The magnitude of
integrated flow also determines the magnitude of statistical errors
through the resolution parameter $\chi_n=v_n\sqrt{M}$, which is
essentially the same quantity as for the standard method.  Using
weights increases $\chi_n$ roughly by a factor of 1.2.  In
Table~\ref{tbl:cum} are presented the multiplicity used in the
cumulant method and the corresponding $\chi$ parameters.
\begin{table}[hbt!]
\caption{Listed for the two beam energies and six centralities are:
  $M$, multiplicity used in the cumulant method, Eq. \
  (\ref{genfunc}), for the reconstruction of the integrated flow;
  $\chi_n=v_n\sqrt{M}$, the resolution parameter, where $v_n$ is a
  (rescaled) weighted flow, see Eq.\ (\ref{weight-int}). The
  multiplicities are larger for the cumulant method than for the
  standard method because all particles were used, including particles
  in the backward hemisphere.}
\begin{center}
\begin{ruledtabular}
\begin{tabular}{l|c|c|c|c|c|c}
Centrality              & 1    & 2    & 3    & 4    & 5    & 6    \\ \hline
158$A$ GeV              & & & & & & \\ \cline{1-1}
$M$                     & 591  & 528  & 419  & 301  & 209  & 109   \\
$\chi_1$                & -    & 0.27 & 0.33 & 0.39 & 0.41 & 0.35  \\
$\chi_2$                & 0.27 & 0.42 & 0.63 & 0.67 & 0.62 & 0.44  \\ \hline
40$A$ GeV              & & & & & & \\ \cline{1-1}
$M$                     & 318  & 257   & 185  & 120  & 80   & 42   \\
$\chi_1$                & -    & 0.003 & 0.19 & 0.23 & 0.16 & 0.26  \\
$\chi_2$                & -    & 0.45  & 0.43 & 0.42 & 0.40 & 0.23  \\
\end{tabular}
\end{ruledtabular}
\end{center}
\label{tbl:cum}
\end{table}

Although the formalism may at first sight look complicated, its
various features make it the simplest to use in practice, for several
reasons.  First, the several estimates are obtained from a single
generating function. Second, the generating function automatically
involves all possible $k$-tuplets of particles in the construction of
the $k$-particle cumulants.  Last but not least, the formalism can be
used even if the detector does not have perfect azimuthal symmetry. In
this case, Eqs.~(\ref{eq:vn2}-\ref{v1/2idea}) no longer hold.  Other
terms must be added in order to remove the spurious, nonphysical
correlations arising from detector inefficiencies, and the number of
these terms increases tremendously as the order of the cumulant
increases.  With the generating-function formalism, they are
automatically included and require no additional work.

When the azimuthal coverage of the detector is strongly asymmetric,
further acceptance corrections must be made, which amount to a global
multiplicative factor~\footnote{In general, $c_n\{k\}$ depends not
only on $v_n$ but also on other harmonics $v_p$ with $p\neq
n$. However for most detectors, these interferences are negligible and
this is indeed the case for the NA49 acceptance.} in the relations
Eqs. (\ref{flow&cumul-int}) between the cumulants $c_n\{k\}$ and the
flow estimates $v_n\{k\}$~\cite{Borghini:2001vi,Borghini:2002vp}.  In
the case of the NA49 acceptance, the corrections are negligible at
158$A$ GeV but can become significant at 40$A$ GeV.  We present the
range of the corrections factors on the reconstructed flow values in
Table~\ref{tbl:cum&acc}.
\begin{table}[hbt!]
\caption { Acceptance correction factors
  on the reconstructed values of the flow $v_n\{k\}$.  For integrated
  flow, the range corresponds to the various centrality bins.  For
  differential flow, the correction factors are calculated in
  centrality bin 4 and the range corresponds to the various ($p_t,y$)
  bins. The larger corrections are for the highest $p_t$ or $y$
  values.}
\begin{center}
\begin{ruledtabular}
\begin{tabular}{l|c|c|c|c||c|c|c|c|c}
158$A$ GeV & $v_1\{2\}$ & $v_1\{3\}$ & $v_2\{2\}$ & $v_2\{4\}$ &
40$A$ GeV & $v_1\{2\}$ & $v_1\{3\}$ & $v_2\{2\}$ & $v_2\{4\}$ \\
\hline 
Integrated & 1.00 & 1.01 & 1.00 & 1.00 & 
Integrated & 1.03 -- 1.04 & 1.07 -- 1.09 & 1.03 -- 1.04 & 1.01
\\ \hline 
Differential & 1.00 -- 1.08 & 1.01 & 1.00 -- 1.02 & 1.01 --
1.06 & 
Differential & 1.07 -- 1.23 & 1.10 -- 1.17 & 1.07 -- 1.29 & 1.01 --
1.78 \\ 
\end{tabular}
\end{ruledtabular}
\end{center}
\label{tbl:cum&acc}
\end{table}

\subsection{Systematic uncertainties}
\label{s:sys}

Various measures are introduced above in order to quantify azimuthal
correlations of particles produced in heavy ion collisions.  These
measures are used in the data analysis with the aim to extract
information on directed and elliptic flow of primary charged pions and
protons emitted in the interaction of two nuclei.  The measured
correlations consist, however, not only of the genuine flow
correlations, but also from other physical correlations of primary
hadrons (nonflow physical correlations) as well as correlations
introduced by the imperfectness of the measuring methods.  The issue
of nonflow physical correlations is addressed in the description of
the methods in Secs.~\ref{s:standard}, ~\ref{s:cumulant} and, in
particular, in Sec.~\ref{s:nonflow}.  In the following we discuss
various sources of detector induced correlations, as well as
corrections and cuts used to reduce their influence on the results and
the systematic uncertainties.

The geometrical acceptance of the detector is not uniform in azimuthal
angle as seen in Fig.~\ref{fig:phi-psi} top. This effect is corrected
for in both methods.  As the geometrical acceptance of the detector
can be probed to high accuracy by the particle yields, the systematic
uncertainty caused by non-uniform acceptance is small except as noted
below.  The majority of the events selected by the hardware trigger
and off--line event cuts (see Sec.~\ref{s:selections}) are Pb+Pb
collisions.  However, there is a small ($<5\%$) contamination in low
multiplicity events from collisions of the Pb beam with the material
surrounding the target foil.  A possible bias caused by this
contamination was estimated by varying off--line selection cuts on the
primary vertex position. No influence on the magnitude of the $v_1$
and $v_2$ was observed.

\begin{figure}[hbt!]
\includegraphics*[height=.40\textheight]{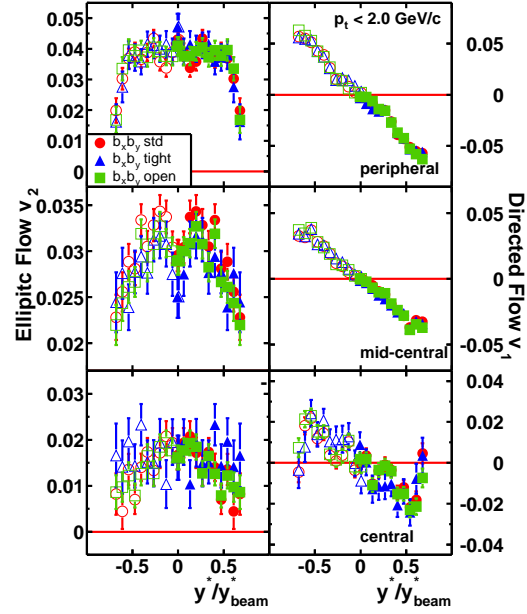}
\caption{\label{fig:systBxBy} (Color online) Rapidity dependence of
pion elliptic 
flow for different cuts on the track impact parameter. The sets of the
cuts were chosen to be 1.0 cm for $b_x$ and 0.2 cm for $b_y$ for the
tight cut. For the open case no cut was used. The standard cut is 3.0
cm for $b_x$ and 0.5 cm for $b_y$.}
\end{figure}

About 90\% of tracks selected by the track selection cuts (see
Sec.~\ref{s:selections}) are tracks of primary hadrons coming from the
main interaction vertex.  The remaining fraction of tracks originates
predominately from weak decays and secondary interactions occurring in
the detector material.  In order to estimate a possible bias due to
this contamination the cuts on track distance from the reconstructed
event vertex ($b_x$,$b_y$) were varied as shown in
Fig.~\ref{fig:systBxBy} with little effect on the results.

\begin{figure}[hbt!]
\includegraphics*[height=.40\textheight]{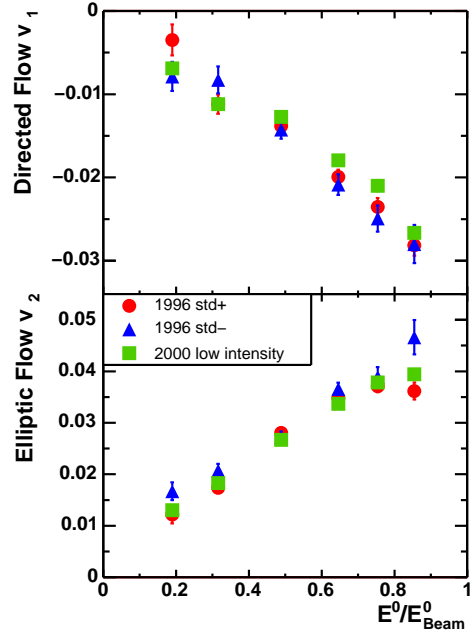}
\caption{\label{fig:systB} (Color online)  Centrality dependence of
results for 
different data sets taken with the same beam energy, same centrality
definition, but two different field polarities and beam
intensities. The most recent data set was taken after some changes to
the detector and analysis procedures.}
\end{figure}

The efficiency of track reconstruction and track selection cuts
depends on track density in the detector and this efficiency is the
lowest ($\approx 80\%$) for central Pb+Pb collisions at 158$A$ GeV at
midrapidity.  The systematic uncertainty due to track losses was
estimated by varying track selection cuts (see
Sec.~\ref{s:selections}).  Additionally, data taken at the two
magnetic field polarities and during two running periods were analyzed
separately and the results compared in Fig.~\ref{fig:systB}. The
systematic variations are largest in near central collisions for $v_1$
($\Delta v_1=0.005$ ) and in peripheral collisions for $v_2$ ($\Delta
v_2 = 0.01$).

The influence of the particle identification procedures on the flow
values was probed by changing the energy loss criteria within
reasonable limits. The resulting variation of $v_1$ and $v_2$ are
below 0.005. The integrations over $p_t$ and rapidity involve a
weighting procedure on the basis of differential cross sections (see
Sec~\ref{s:results}). These cross sections are available at
40$A$ GeV only for pions in central collisions, for other cases
they were estimated based on data systematics. Variation of the cross
section weights within reasonable limits results in variations below
0.001 for $v_2$ and 0.005 for $v_1$.

The azimuthal coverage of the NA49 TPCs is significantly reduced at
40$A$ GeV as compared to 158$A$ GeV (see
Sec.~\ref{s:acceptance}). Modifications to the standard analysis
method were necessary to reduce the systematic errors. The validity of
the results from the modified method were scrutinized by applying it
to the 158$A$ GeV data. The results of this test are stable and in
agreement with those of the standard method for mid-central
collisions. In near central and peripheral collisions large relative
differences are taken as estimate of systematic uncertainties. These
are significant, if low event multiplicity or low flow values make the
event plane determination unreliable.

The results of our study of systematic uncertainties can be summarized
as follows. The systematic error of $v_2(p_t,y)$ and $v_1(p_t,y)$ for
pions in mid-central and peripheral collisions is 0.005 and 0.002
respectively. For central collisions the numbers increase to 0.01 for
both. The error for protons for mid-central and peripheral collisions
is 0.005 for $v_2$ and 0.01 for $v_1$. For central collisions it
increases to about 0.03 for both. At 40$A$ GeV these errors
could be 50\% larger than those at 158$A$ GeV. Note that the errors 
plotted in all figures are statistical ones only.

\section{Results}
\label{s:results}

Both methods outlined in Sec.~\ref{s:methods} have been applied at
40$A$ GeV and 158$A$ GeV.  The double differential flow values
$v_n(p_t,y)$ for each harmonic as obtained from the methods were
tabulated as a function of $p_t$, $y$, and centrality. Integration of
$v_n(p_t,y)$ to obtain $v_n(p_t)$ or $v_n(y)$ values was done by
averaging over the integration variable using the cross sections of
the particles as weights. The cross section values at 158$A$ GeV had
been parametrized~\cite{crosssections} and were available as a
macro. Since no cross sections were available at 40$A$ GeV for
non-central collisions, the width of the pion Gaussian rapidity
distribution and the separation of the two proton rapidity Gaussian
distributions were scaled down by the ratio of the beam rapidities at
40 and 158$A$ GeV. Since we chose larger $y$ bins in the cumulant
method than in the standard method, the integration over rapidity was
not performed over exactly the same $y$ range.  More precisely, the
upper limit is always smaller for the cumulant method because the
results for proton flow did not seem to be very stable with respect to
integration up to high rapidity values. In the cumulant $v(p_t)$
graphs the indicated $y$ ranges refer to the cumulant results; the
reproduced standard method results in these graphs have the $y$ ranges
indicated in the preceding standard method plots. The results are
presented for three centrality bins (two successive bins have been
combined, weighted with the known cross sections and the fraction of
events in each bin, see Table~\ref{tbl:centrality}) and also
integrated over the first five centrality bins (weighted with the
known cross sections and the fraction of the geometric cross section
for each bin given in Table~\ref{tbl:centrality}), which we call
minimum bias. We present elliptic flow (Sec.~\ref{s:v2}), directed
flow (Sec.~\ref{s:v1}), minimum bias results (Sec.~\ref{s:minbias}),
centrality dependence (Sec.~\ref{s:centrality}), nonflow effects
(Sec.~\ref{s:nonflow}), and beam energy dependence
(Sec.~\ref{s:energy}).  The standard method $v_1$ values have been
corrected for momentum conservation but the cumulant method $v_1\{2\}$
values have not.

In the graphs of flow as a function of rapidity the points have been
reflected about midrapidity and fitted with polynomial curves to guide
the eye.  Please note that we always use the rapidity in the center of
mass, and to calculate this the nominal laboratory rapidity of the
center of mass was taken to be 2.92 at 158$A$ GeV and 2.24 at 40$A$
GeV.  In the graphs of flow as a function of $p_t$ the smooth curves
shown to guide the eye were obtained by fitting to a simple
hydrodynamic motivated Blast Wave model as described in
Ref.~\cite{Snellings:2001,pasi2} but generalized to also describe
$v_1$:
\begin{equation}
  v_n(p_t)=
  \frac{  \int_0^{2\pi} d\phi_b
    \cos(n \phi_b)
    I_n(\alpha_t) K_1 (\beta_t)
    (1+2 s_n \cos(n\phi_b))}
  {  \int_0^{2\pi} d\phi_b
    I_0(\alpha_t) K_1 (\beta_t)
    (1+2 s_n \cos(n\phi_b))},
  \label{modblastwave}
\end{equation}
where the harmonic $n$ can be either one or two, where $I_0$, $I_n$,
and $K_1$ are modified Bessel functions, and where
$\alpha_t(\phi_b)=(p_t/T_f)\sinh(\rho(\phi_b))$ and
$\beta_t(\phi_b)=(m_t/T_f)\cosh(\rho(\phi_b))$.  The basic assumptions
of this model are boost-invariant longitudinal expansion and
freeze-out at constant temperature $T_f$ on a thin shell, which
expands with a transverse rapidity exhibiting a first or second
harmonic azimuthal modulation given by $\rho(\phi_b)=\rho_0 +\rho_a
\cos(n \phi_b)$.  In this equation, $\phi_b = \phi - \Phi_{RP}$ is the
azimuthal angle (measured with respect to the reaction plane) of the
boost of the source element on the freeze-out
hyper-surface~\cite{pasi2}, and $\rho_0$ and $\rho_a$ are the mean
transverse expansion rapidity [$v_0 = \tanh(\rho_0)$] and the
amplitude of its azimuthal variation, respectively. The parameters in
this model are $T_f$, the temperature, $\rho_0$, the transverse flow
rapidity, $\rho_a$, the azimuthal flow rapidity, and $s$, the surface
emission parameter. The $T_f$ parameter was fixed and the $s$
parameter was allowed to be non-zero only for pions at 158$A$ GeV. The
values of the fit parameters themselves are not very meaningful
because the flow values derived from two-particle correlations contain
nonflow effects and the values from many-particle analyses have poor
statistics, but the fits do provide the curves to guide the eye shown
in the $p_t$ graphs. The data are clearly not boost invariant, but
since we use the blast wave model only to fit the $p_t$ dependence of
the flow, it was felt that a more sophisticated model was not
warranted.  When the fits would not converge the points were just
connected, giving rise to the jagged lines in some graphs.

The error bars shown for the standard method are the standard
deviation of the data.  For the cumulant method, they are calculated
analytically following the formulas given in
Refs.~\cite{Borghini:2001vi,Borghini:2002vp}. Tables of the data
can be found on the web at http://na49info.cern.ch/na49/Archives/Data/FlowWebPage/.

\subsection{Elliptic flow}
\label{s:v2}

\subsubsection{158$A$ GeV}

\begin{figure}[hbt!]
\includegraphics*[height=.40\textheight]{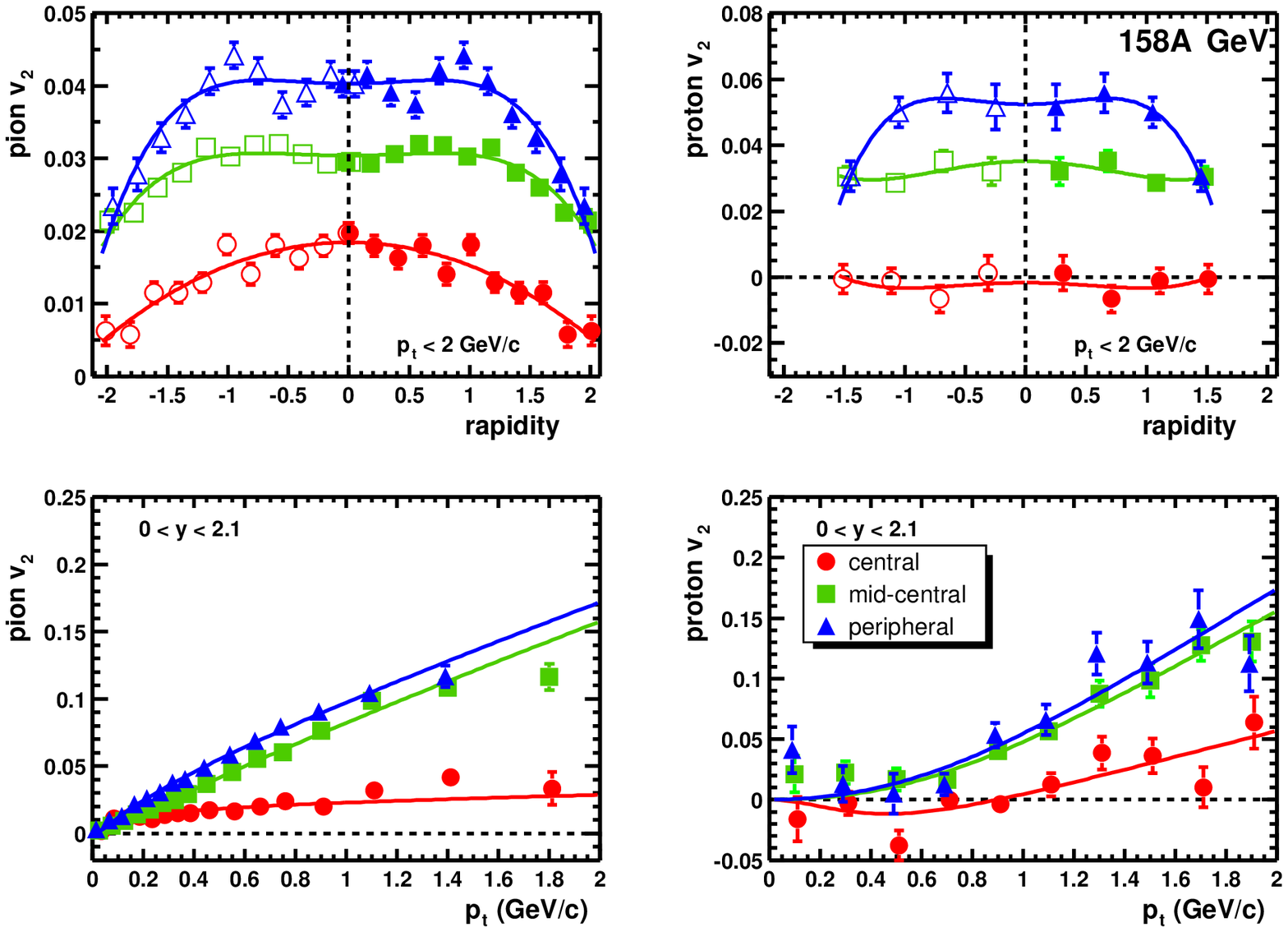}
\caption{\label{fig:158_v2_all} (Color online) Elliptic flow obtained
 from the standard  
 method as a function of rapidity (top) and transverse momentum
 (bottom) for charged pions (left) and protons (right) from 158$A$ GeV
 Pb + Pb. Three centrality bins are shown. The open points in the top
 graphs have been reflected about midrapidity. Solid lines are
 polynomial fits (top) and Blast Wave model fits (bottom).}
\end{figure}

\begin{figure}[hbt!]
\includegraphics*[height=.40\textheight]{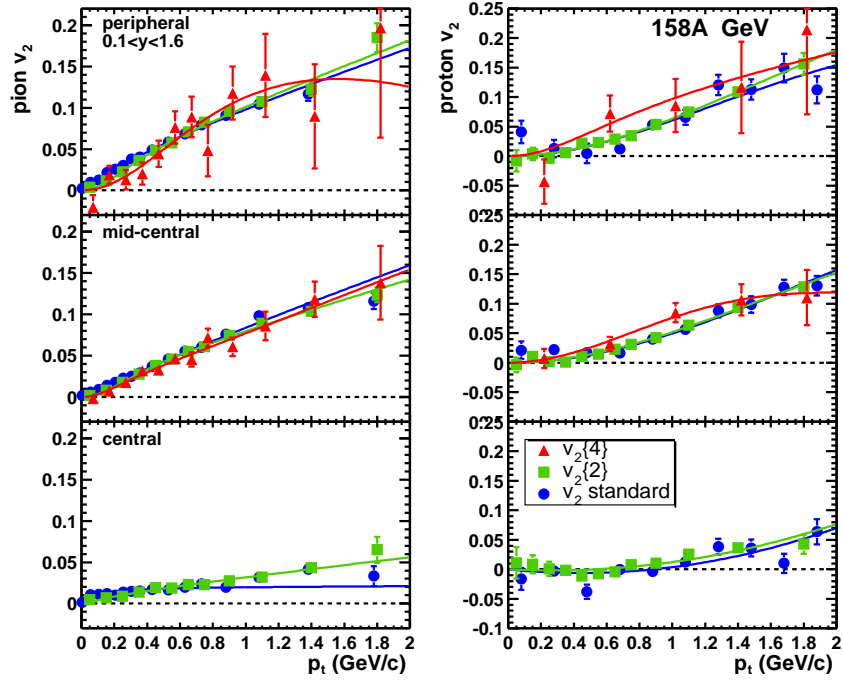}
\caption{\label{fig:158_V2vsPt} (Color online) Elliptic flow of
  charged pions (left) and 
  protons (right) from the cumulant method as a function of transverse
  momentum in 158$A$ GeV Pb + Pb collisions. Three centrality
  bins are shown with results from the standard method, from cumulants
  for two-particle correlations ($v_2\{2\}$), and from cumulants for
  four-particle correlations ($v_2\{4\}$). Solid lines are from Blast
  Wave model fits.}
\end{figure}

\begin{figure}[hbt!]
\includegraphics*[height=.40\textheight]{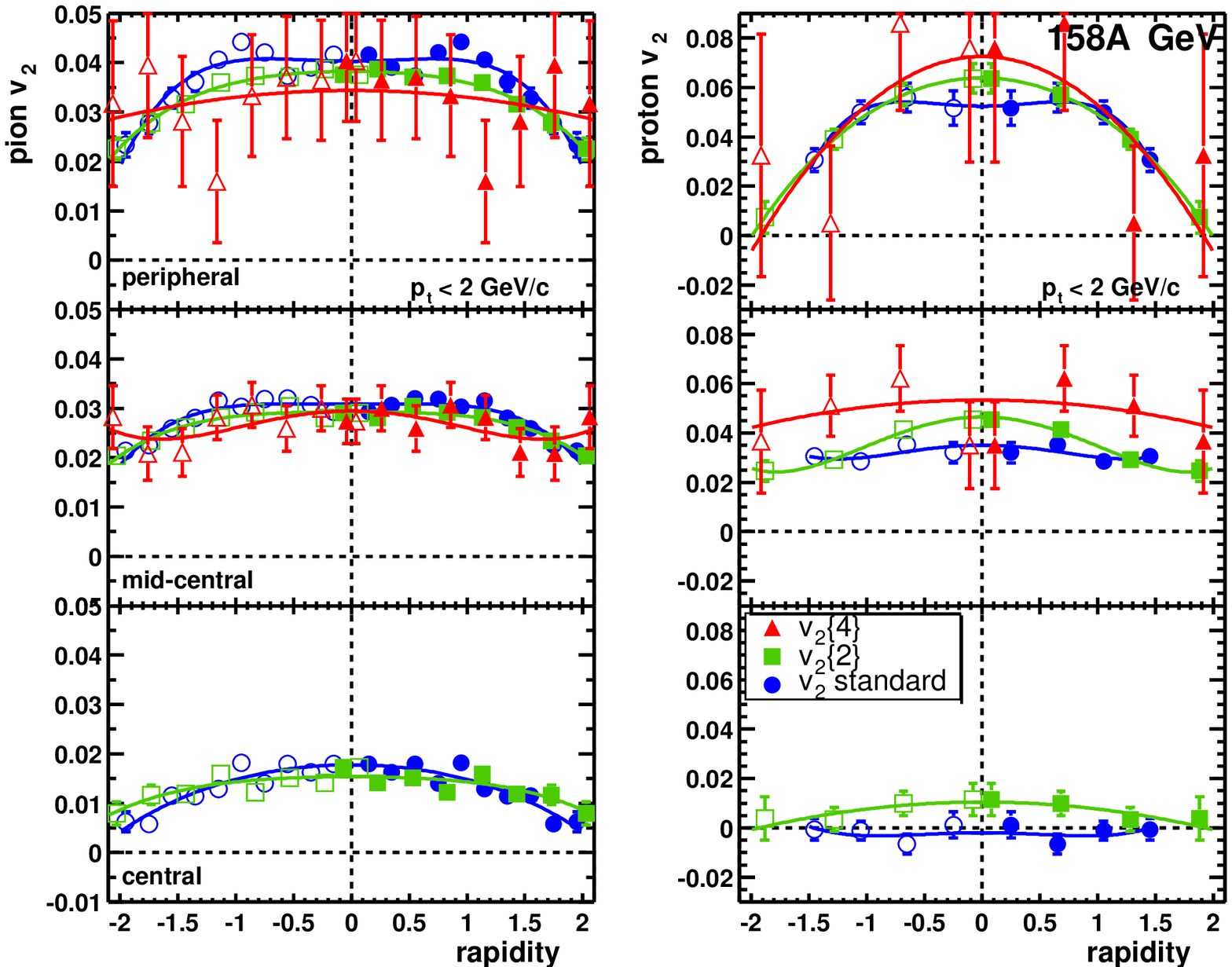}
\caption{\label{fig:158_V2vsY} (Color online) Same as
  Fig.~\ref{fig:158_V2vsPt}, as a  
  function of rapidity. The open points have been reflected about
  midrapidity. Solid lines are from polynomial fits. Please note the
  different vertical scales.}
\end{figure}

Results from 158$A$ GeV collisions are displayed in
Fig.~\ref{fig:158_v2_all} (from the standard method) and are compared
to the results from the cumulant method in Figs.~\ref{fig:158_V2vsPt}
and \ref{fig:158_V2vsY}. 

We first discuss results for mid-central collisions
(Figs.~\ref{fig:158_V2vsPt} and \ref{fig:158_V2vsY}, middle) for
the following reason: as usual in flow analyses, they have smaller
errors than peripheral collisions (Figs.~\ref{fig:158_V2vsPt} and
\ref{fig:158_V2vsY}, top) due to the larger multiplicity, and
also smaller than central collisions (Figs.~\ref{fig:158_V2vsPt} and
\ref{fig:158_V2vsY}, bottom) due to the larger value of the flow.  By
errors, we mean both the statistical error, shown in the figures, and
the uncertainty of the contribution of nonflow correlations,
which is not known and not shown in the figures.  For mid-central
collisions, $v_2$ is positive over all phase space for pions and
protons.  As a function of $p_t$, it rises linearly for pions up to
2~GeV/c (Fig.~\ref{fig:158_V2vsPt} middle left).  For protons
(Fig.~\ref{fig:158_V2vsPt} middle right), the rise is slower at low
$p_t$ (quadratic rather than linear up to 1~GeV/c), but interestingly,
$v_2$ reaches the same value as for pions at 2~GeV/c.  All three
methods give compatible results within statistical error bars.  As a
function of rapidity, the pion $v_2$ exhibits the usual bell shape
(Fig.~\ref{fig:158_V2vsY} middle left), with a maximum at
mid-rapidity. This maximum, however, is not very pronounced, and $v_2$
remains essentially of the same magnitude, between 2\% and 3\%, over
four units of rapidities.  For protons (Fig.~\ref{fig:158_V2vsY}
middle right), the rapidity dependence is similar. Note that $v_2(y)$
is slightly larger than for pions, although $v_2(p_t)$ was
smaller. This can be explained simply: $v_2(y)$ is integrated over
$p_t$, protons have higher average $p_t$ than pions, and $v_2$
increases with $p_t$.  For protons, a small discrepancy appears around
$y=0$ between the estimate from the two-particle cumulants
($v_2\{2\}$) and the standard reaction plane estimate, which we do not
understand, and consider part of our systematic error.

For peripheral collisions, $v_2$ is somewhat larger than for
mid-central collisions at low $p_t$ (Fig.~\ref{fig:158_v2_all}
bottom), but comparable at high $p_t$. $v_2(y)$ is dominated by the
low $p_t$ region where the yield is larger, hence it is also larger
for peripheral than for mid-central collisions
(Fig.~\ref{fig:158_v2_all} top).  A small discrepancy can be seen
around $y=1$ for pions between $v_2\{2\}$ and the standard $v_2$
(Fig.~\ref{fig:158_V2vsY} top left).

For central collisions, elliptic flow is much smaller
(Fig.~\ref{fig:158_v2_all}).  As a consequence, four-particle
cumulants could not be used, due to large statistical fluctuations.
Contrary to more peripheral collisions, $v_2(y)$ is larger for pions
than for protons (Figs.~\ref{fig:158_v2_all}, top, and
\ref{fig:158_V2vsY}, bottom), for which it is consistent with zero in
the available rapidity range.

We have also compared these results with those of the earlier analysis
published in Ref.~\cite{NA49PRL} and updated on the collaboration web
page~\cite{NA49Web}.  Note that the previous analysis used narrower
cuts in $p_t$ and $y$, and had much larger statistical errors.
Results are compatible for protons, but not for pions.  In particular
$v_2(p_t)$ is significantly different at low $p_t$, where the increase
is now much smoother.  The earlier analysis was biased by
Bose-Einstein correlations~\cite{Dinh:1999mn}, whose contribution is
reduced in the present analysis thanks to the $w_i=p_t$ weights used
in Eq.~(\ref{eq:Q_n}), rather than the unit weights used in the
previous analysis.

\subsubsection{40$A$ GeV}

\begin{figure}[hbt!]
\includegraphics*[height=.40\textheight]{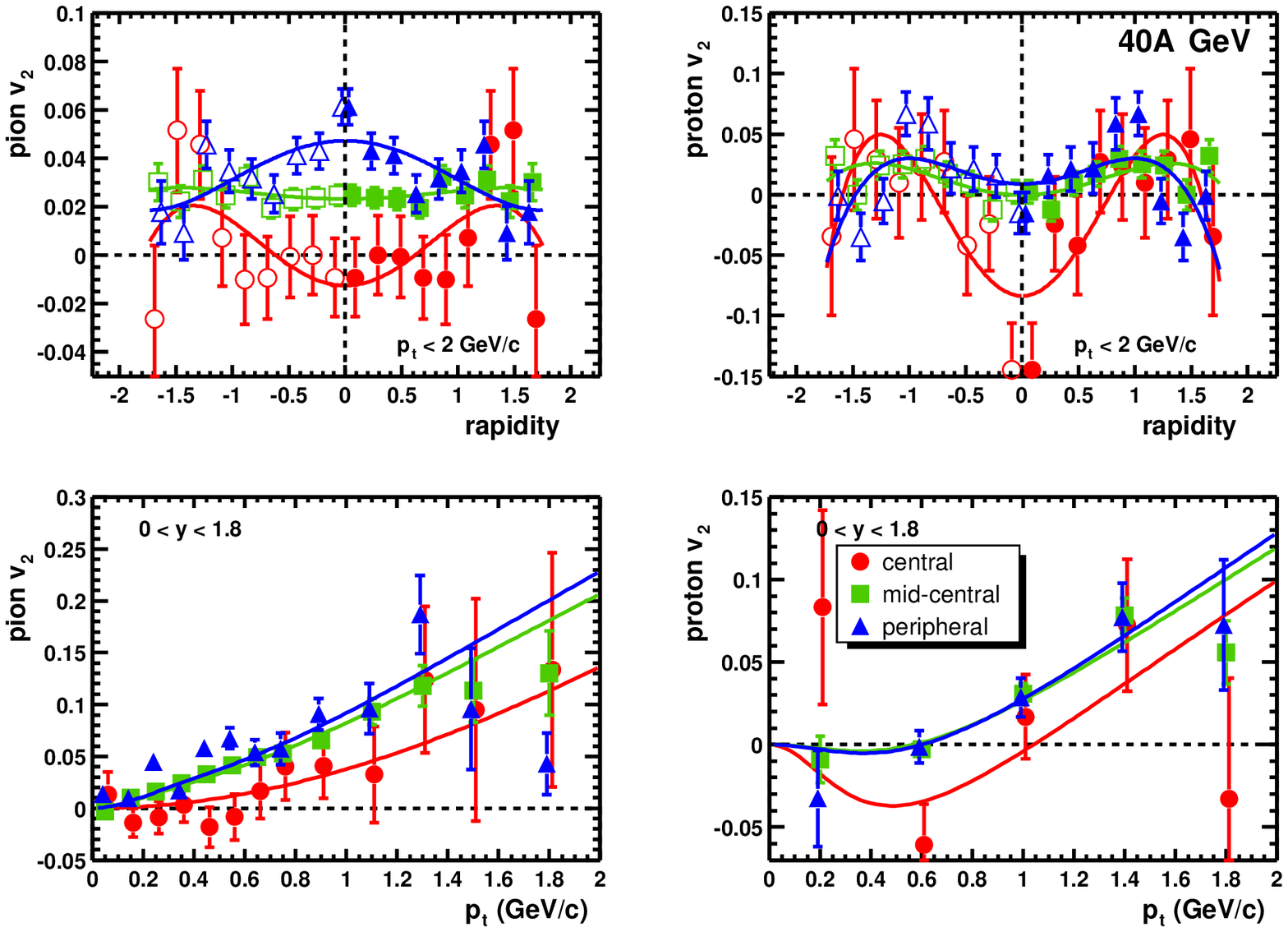}
\caption{\label{fig:40_v2_all} (Color online) Elliptic flow obtained
 from the standard  
 method as a function of rapidity (top) and transverse momentum
 (bottom) for charged pions (left) and protons (right) from 40$A$ GeV
 Pb + Pb. Three centrality bins are shown (``central'' corresponds to
 centrality bin 2 only, see Table~\ref{tbl:centrality}). The open
 points in the top graphs have been reflected about midrapidity. Solid
 lines are polynomial fits (top) and Blast Wave model fits (bottom).}
\end{figure}

\begin{figure}[hbt!]
\includegraphics*[height=.40\textheight]{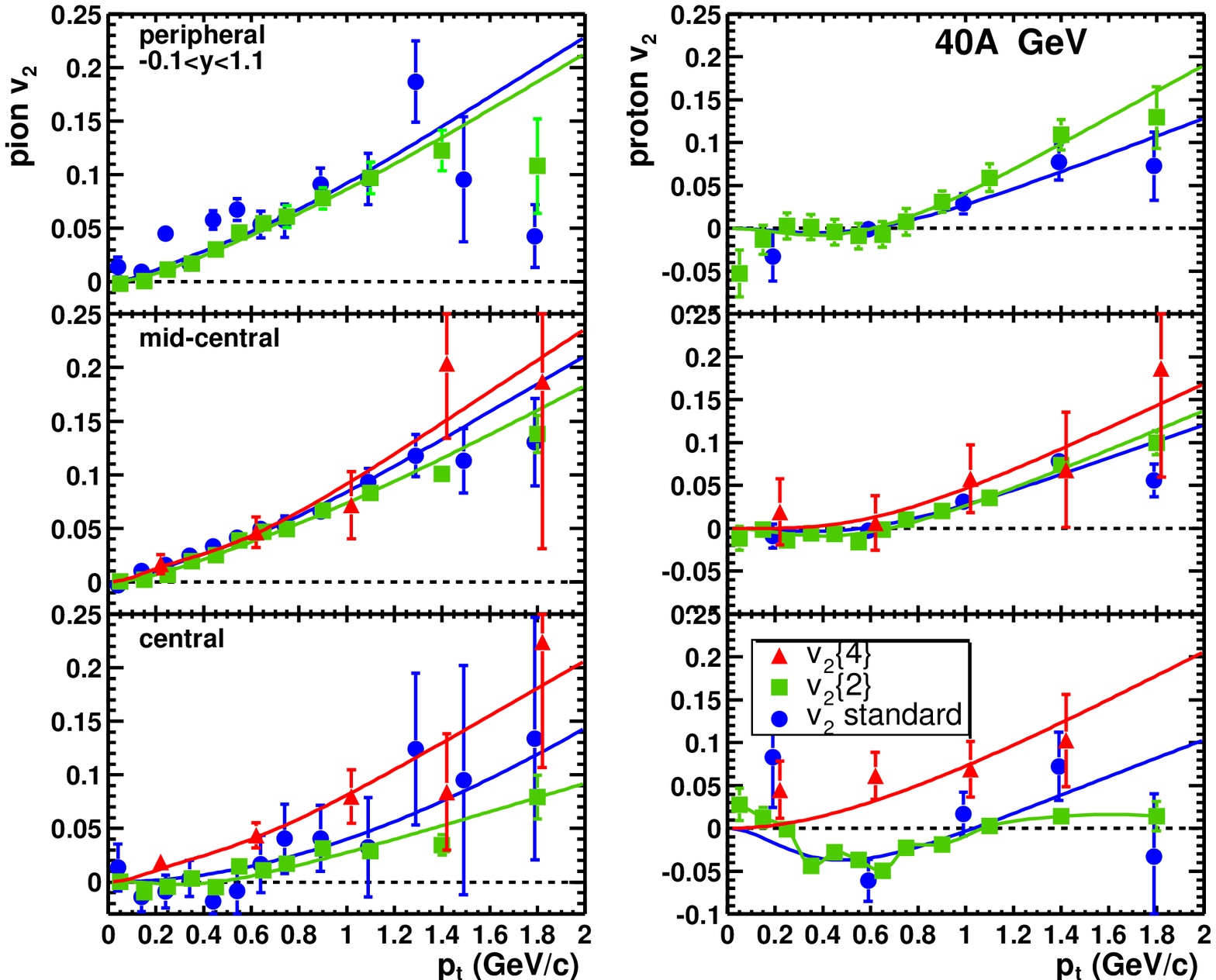}
\caption{\label{fig:40_V2vsPt} (Color online) Elliptic flow of charged
  pions (left) and 
  protons (right) obtained from the cumulant method as a function of
  transverse momentum from 40$A$ GeV Pb + Pb. Three centrality
  bins are shown.  Here, ``central'' (bottom) corresponds only to
  centrality bin 2 (see Table~\ref{tbl:centrality}). Results are shown
  from the standard method, from cumulants for two-particle
  correlations ($v_2\{2\}$), and from cumulants for four-particle
  correlations ($v_2\{4\}$). The smooth solid lines are from
  Blast Wave model fits.}
\end{figure}

\begin{figure}[hbt!]
\includegraphics*[height=.40\textheight]{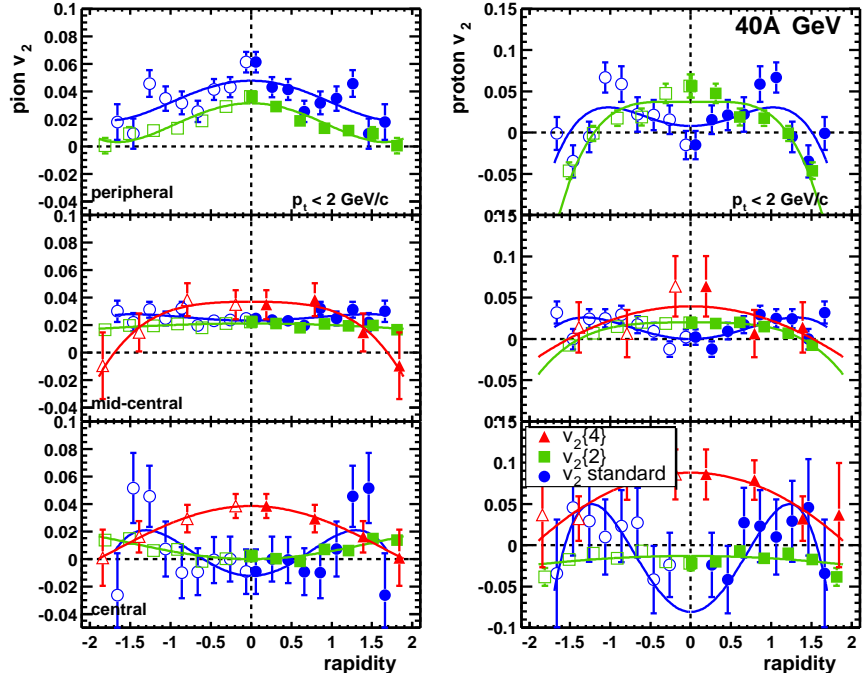}
\caption{\label{fig:40_V2vsY} (Color online) Same as
  Fig.~\ref{fig:40_V2vsPt}, as a  
  function of rapidity. The open points have been reflected about
  midrapidity. Solid lines are from polynomial fits.}
\end{figure}

Results from 40$A$ GeV collisions are displayed in
Figs.~\ref{fig:40_v2_all} (from the standard method),
\ref{fig:40_V2vsPt}, and \ref{fig:40_V2vsY} (from the cumulant
method). Both the values of $v_2$ and the multiplicities are
smaller than at 158$A$ GeV, which results in larger errors.

In mid-central collisions, $v_2(p_t)$ (Fig.~\ref{fig:40_V2vsPt}
middle) is roughly the same as at 158$A$ GeV
(Fig.~\ref{fig:158_V2vsPt} middle), both in shape and magnitude, and
all three methods give consistent results.  The value of $v_2$ at a
fixed $p_t$ is in fact very similar up to the highest RHIC energy,
as pointed out by Snellings~\cite{Snellings:2003mh}.  Since the
average $p_t$ is smaller at 40$A$ GeV than at 158$A$ GeV,
however, the corresponding $v_2(y)$ is smaller
(Fig.~\ref{fig:40_V2vsY} middle).  As a function of rapidity, both the
standard $v_2$ and $v_2\{2\}$ of pions are remarkably flat. On the
other hand, $v_2\{4\}$ is bell-shaped, as at 158$A$ GeV, but
statistical error bars prevent any definite conclusion.  For protons
(Fig.~\ref{fig:40_V2vsY} middle right), a discrepancy appears between
$v_2\{2\}$, which is bell-shaped, and the standard $v_2$, which has a
dip at mid-rapidity. 

For peripheral collisions, $v_2$ is slightly larger, but 
similar to mid-central collisions  (Fig.~\ref{fig:40_v2_all}). 
The four-particle cumulant result $v_2\{4\}$ could not be 
obtained due to large statistical errors.

At this energy, statistics did not allow the determination of the
second harmonic event plane, nor the derivation of estimates from two-
or four-particle cumulants for the most central bin.  Thus, the
``central'' bin corresponds to bin 2 only in
Figs.~\ref{fig:40_v2_all} to \ref{fig:40_V2vsY}.  A striking
discrepancy appears between the two-particle estimates (standard $v_2$
and $v_2\{2\}$) and the four-particle cumulant result $v_2\{4\}$
in the $p_t$ dependence (Fig.~\ref{fig:40_V2vsPt} bottom). It is
closely related to that observed in the rapidity dependence: For
pions (Fig.~\ref{fig:40_V2vsY}, bottom left), the standard $v_2$ and
$v_2\{2\}$ both seem to show a dip at midrapidity, where they are
compatible with zero; for protons (Fig.~\ref{fig:40_V2vsY},
bottom right), the standard $v_2$ and $v_2\{2\}$ are even negative at
midrapidity.  By contrast, the four-particle cumulant result,
$v_2\{4\}$, is positive for both pions and protons and has the usual
bell shape, with a maximum value at midrapidity.

Since the difference between two-particle estimates and $v_2\{4\}$ is
much beyond statistical error bars, this is a hint that the
two-particle estimates for pions are affected by nonflow effects
(recall that $v_2\{4\}$ is expected to be free from two-particle
nonflow effects). This could be due to correlations from $\rho$
decays, since $\rho$ mesons are more concentrated near midrapidity,
and produce a negative correlation in the second
harmonic~\cite{Borghini:2000cm}, which tends to lower the two-particle
$v_2$ estimates.  This effect, however, cannot explain the difference
for protons.

While the four-particle cumulant result $v_2\{4\}$ for central
collisions looks very reasonable in shape (bell shape in rapidity, 
regular increase with $p_t$ for both pions and protons), its 
magnitude is unusual: it is as large as for mid-central collisions, 
while we would have expected a significantly smaller value 
following the observations at 158$A$  GeV.

\subsection{Directed flow}
\label{s:v1}

\subsubsection{158$A$ GeV}

\begin{figure}[hbt!]
\includegraphics*[height=.40\textheight]{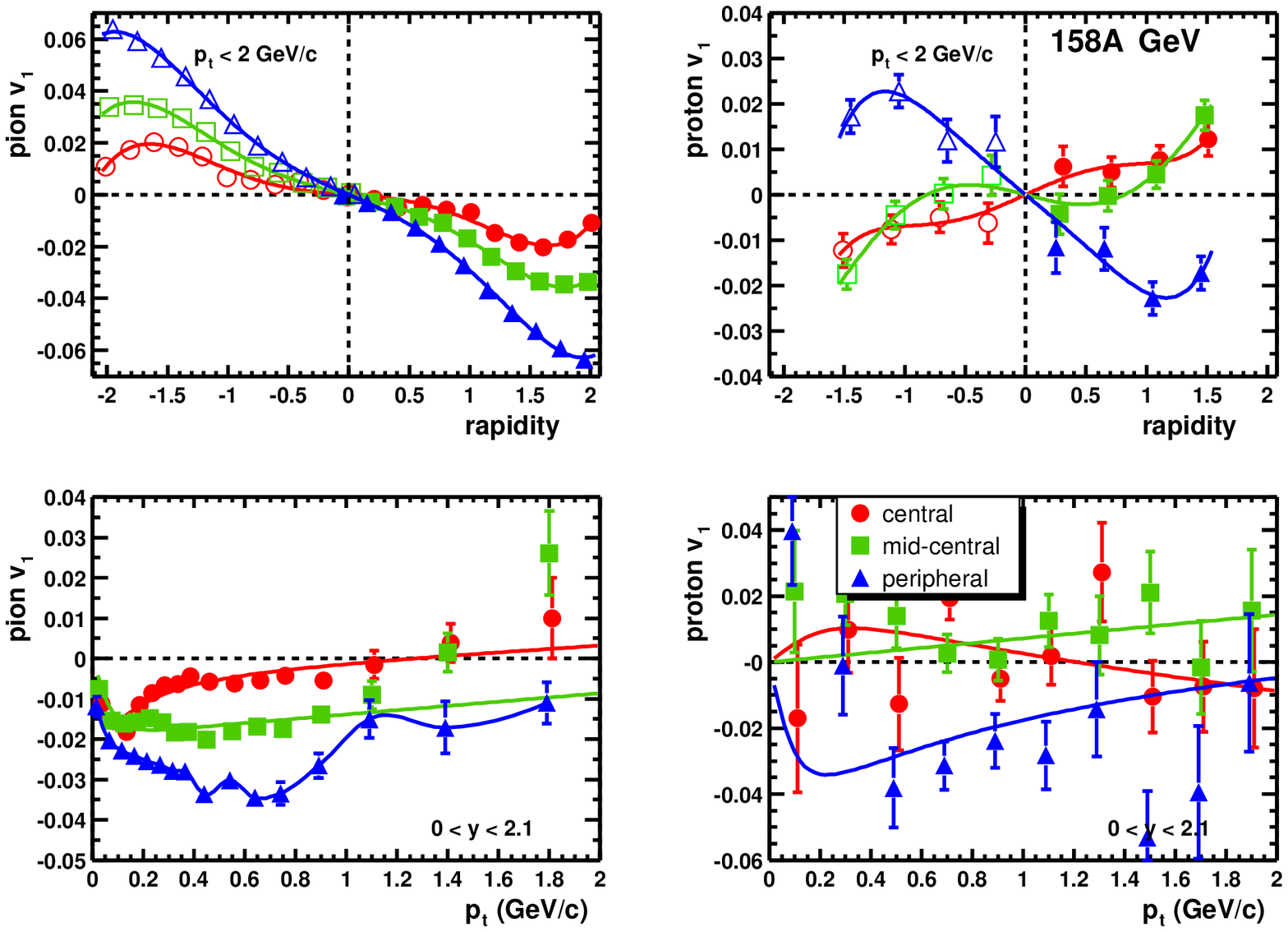}
\caption{\label{fig:158_v1_all} (Color online) Standard  directed
 flow as a function of rapidity (top) and transverse momentum (bottom)
 for charged pions (left) and protons (right) from 158$A$ GeV Pb
 + Pb. Three centrality bins are shown. The open points in the top
 graphs have been reflected about midrapidity. On the top, solid
 lines are polynomial fits. On the bottom, smooth solid lines are
 Blast Wave model fits.}
\end{figure}

\begin{figure}[hbt!]
\includegraphics*[height=.40\textheight]{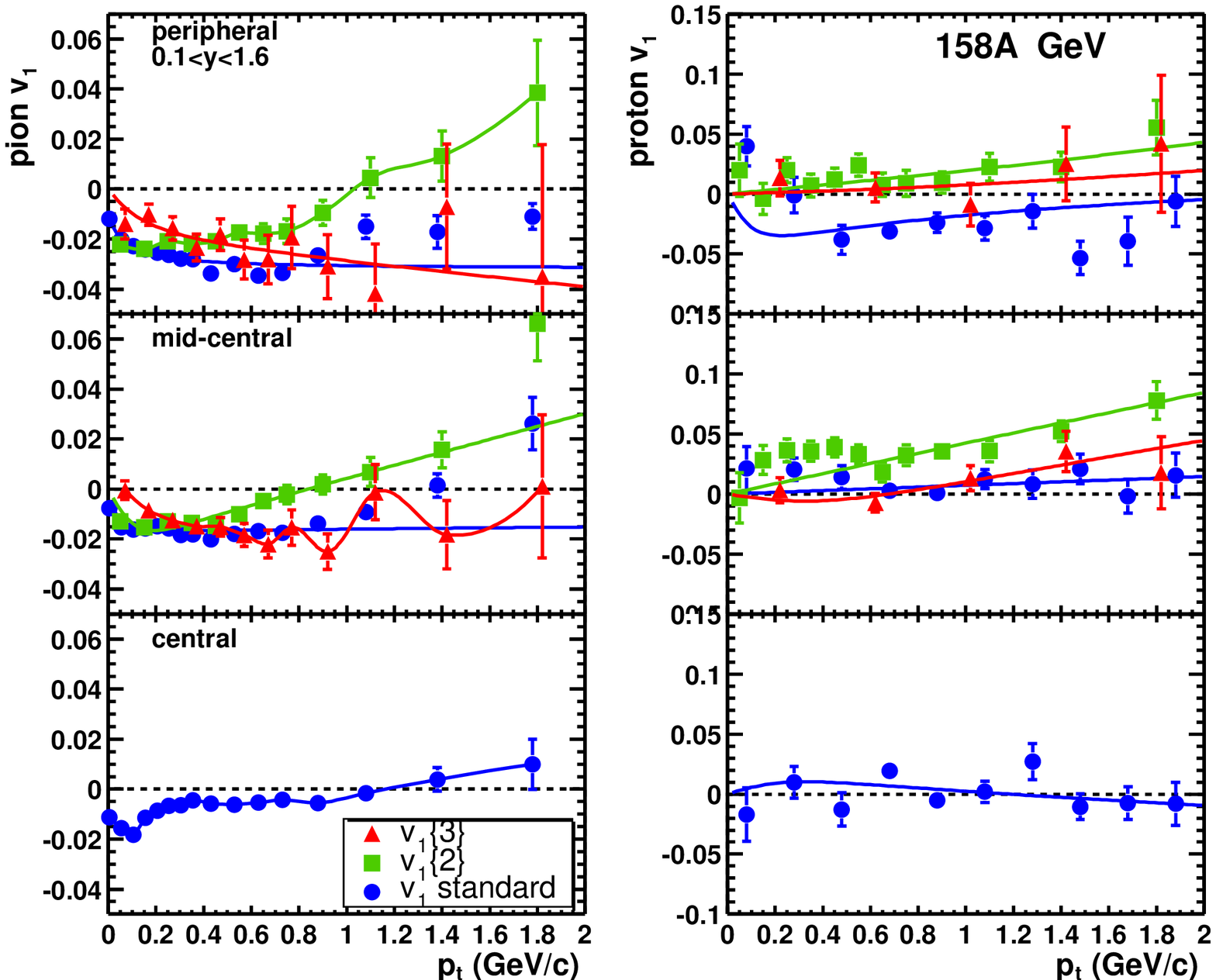}
\caption{\label{fig:158_V1vsPt} (Color online) Directed flow of
 charged pions (left) and 
 protons (right) from cumulant method, as a function of transverse
 momentum in 158$A$ GeV Pb + Pb. Three centrality bins are shown.
 Here, ``central'' (bottom) corresponds to centrality bin 2 (see
 Table~\ref{tbl:centrality}) only. Results are shown from the standard
 method, from cumulants for two-particle correlations ($v_1\{2\}$),
 and from cumulants for three-particle correlations ($v_1\{3\}$). The
 $v_1\{2\}$ estimates are {\em not} corrected for momentum
 conservation, while the standard method results are. Smooth
 solid lines are Blast Wave model fits. Please note the different
 vertical scales.}
\end{figure}

\begin{figure}[hbt!]
\includegraphics*[height=.40\textheight]{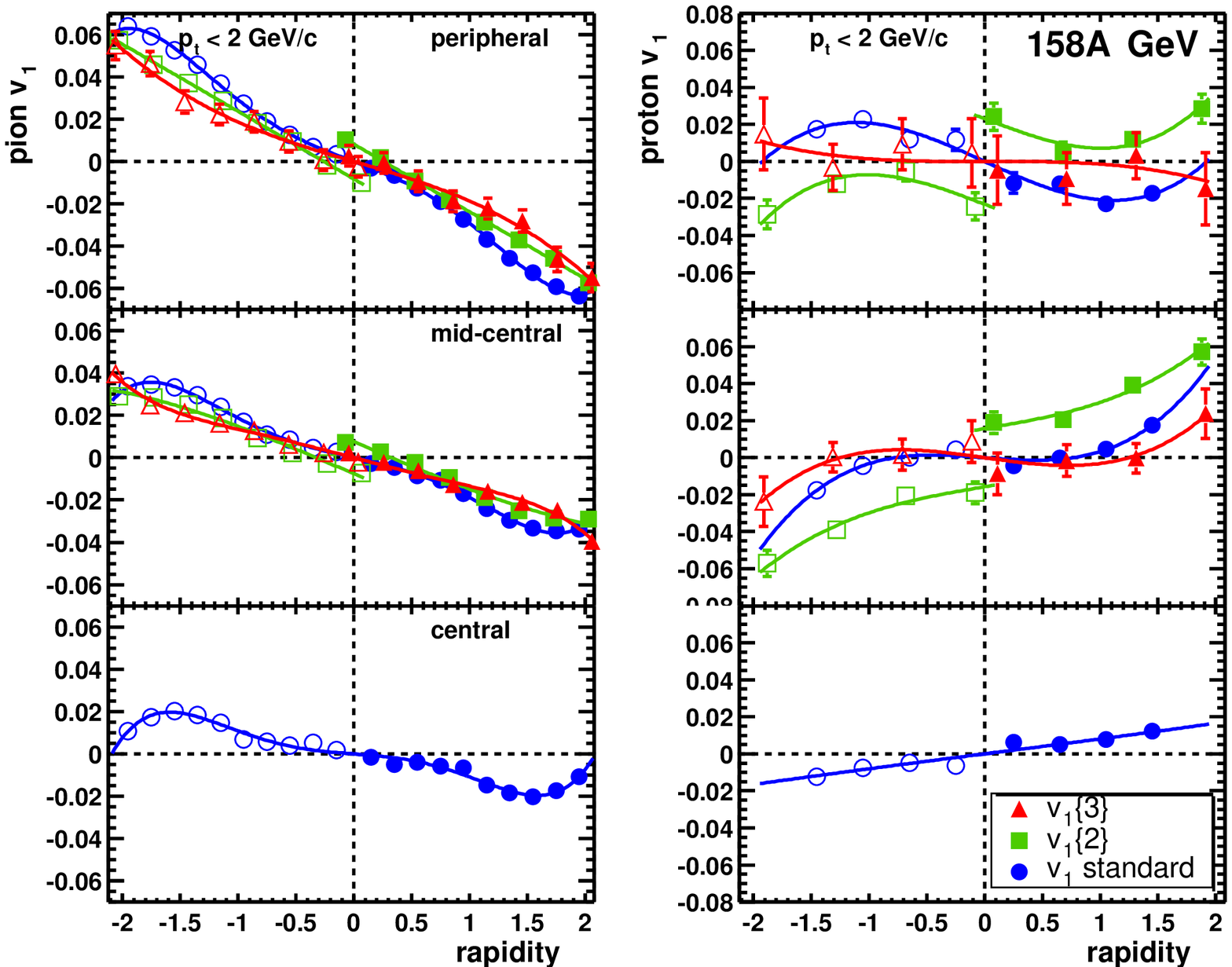}
\caption{\label{fig:158_V1vsY} (Color online) Same as
  Fig.~\ref{fig:158_V1vsPt}, as a  
  function of rapidity. The open points have been reflected about 
  midrapidity. Solid lines are polynomial fits.}
\end{figure}

Results from collisions at 158$A$ GeV are shown in
Figs.~\ref{fig:158_v1_all} (from the standard method),
~\ref{fig:158_V1vsPt}, and \ref{fig:158_V1vsY} (from the cumulant
method).  It can be seen in Fig.~\ref{fig:158_v1_all} top left that
the curves for the different centralities all cross zero at
midrapidity, indicating that the correction for global momentum
conservation in the standard method, shown in
Fig.~\ref{fig:momCons} for minimum bias data, also works for the
individual centralities.  One clearly sees the magnitude of the
correction for momentum conservation in Figs.~\ref{fig:158_V1vsPt},
and \ref{fig:158_V1vsY}, by comparing the results from the
two-particle cumulants, $v_1\{2\}$ (squares), which are not corrected,
with the results of the standard reaction plane analysis (circles),
which are corrected (see Sec.~\ref{s:standard}).  The difference
between the two results rises linearly with $p_t$, and is larger for
peripheral collisions (Fig.~\ref{fig:158_V1vsPt} top) than for
mid-central collisions (Fig.~\ref{fig:158_V1vsPt} middle), as expected
from the discussion in Ref.~\cite{Borghini:2002mv}.  For central
collisions, the negative correlations due to momentum conservation
become larger in absolute value than the positive correlations due to
flow (because flow is much smaller in central collisions), so that the
cumulant $c_1\{2\}$ in Eq.~(\ref{flow&cumul-int}) becomes negative and
the flow estimate $v_1\{2\}$ could not be obtained.  As a function of
rapidity, the difference between $v_1\{2\}$ and the standard $v_1$ is
approximately constant (Fig.~\ref{fig:158_V1vsY} top and
middle). Unlike the standard $v_1$, $v_1\{2\}$ does not cross zero at
midrapidity, while the true directed flow should.  This is a direct
indication that $v_1\{2\}$ is biased by global momentum conservation
(see Fig.~\ref{fig:momCons}, top).  It is shown merely as an
illustration of this effect, and will not be discussed further.

We first discuss the directed flow of pions for mid-central
collisions.  Its transverse momentum dependence
(Fig.~\ref{fig:158_V1vsPt} middle left) is peculiar.  The standard
$v_1$ (circles) is negative at low $p_t$, but then increases and
becomes positive above 1.4~GeV/c.  The result from the mixed
three-particle correlation method [Eq.~(\ref{v1/2idea})], $v_1\{3\}$,
which is expected to be free from all nonflow effects including
momentum conservation, is also shown (triangles).  Above 0.6~GeV/c, it
is compatible with the standard reaction plane estimate, but not with
the two-particle estimate $v_1\{2\}$.  This suggests that nonflow
effects are dominated by momentum conservation in this region.
Unfortunately, statistical error bars on $v_1\{3\}$ are too large at
high $p_t$ to confirm the change of sign observed in the standard
$v_1$.  Below 0.2~GeV/c, where momentum conservation is negligible,
both the standard analysis value and $v_1\{2\}$ seem to intercept at a
finite value when $p_t$ goes to zero.  This is suggestive of nonflow
effects arising from quantum Bose-Einstein effects between identical
pions~\cite{Dinh:1999mn}.  The three-particle estimate $v_1\{3\}$,
which is free of nonflow effects, smoothly vanishes at $p_t=0$.  This
discussion shows that results on directed flow must be interpreted
with care, and are more biased by nonflow effects than results on
elliptic flow. This is mostly due to the smaller value of directed
flow (typically 2\% in absolute value, instead of 3\% for elliptic
flow).

The rapidity dependence of the pion $v_1$ for mid-central collisions
is displayed in Fig.~\ref{fig:158_V1vsY} middle left.  One notes that
$v_1\{3\}$ vanishes at midrapidity (unlike $v_1\{2\}$), which confirms
that it is automatically corrected for momentum conservation effects.
Both $v_1\{3\}$ and the standard $v_1$ exhibit a smooth, almost linear
rapidity dependence. They are in close agreement near mid-rapidity,
where we observe clear evidence that the slope is negative. At the
more forward rapidities, however, the standard $v_1$ becomes larger in
absolute value than $v_1\{3\}$ and seems to saturate, while the slope
becomes steeper for $v_1\{3\}$. This small discrepancy can be
attributed to the above mentioned Bose-Einstein effects which bias the
standard $v_1$ at low $p_t$. This bias is only present over the phase
space used to determine the event plane~\cite{Dinh:1999mn}, i.e. above
center-of-mass rapidity $y=1.08$.  Note that the value of $v_1\{3\}$
is only 4\% in absolute value at $y=2$. This is smaller by at least a
factor of 5 than the value obtained by the WA98 collaboration in the
target fragmentation region (center-of-mass rapidity around -3), where
$v_1$ reaches 20\% for pions~\cite{Aggarwal:1998vg}.  There is no
overlap between their rapidity coverage and ours, so that we cannot
check whether the two analyses are consistent.  Nevertheless, this
comparison suggests that the slope becomes much steeper toward beam
rapidity, a trend already seen on our $v_1\{3\}$ result.

In peripheral collisions, the pion $v_1$ (Fig.~\ref{fig:158_V1vsPt}
and Fig.~\ref{fig:158_V1vsY} top left) has a $p_t$ and $y$ dependence
very similar to the one in mid-central collisions.  Its magnitude,
however, is larger, and the increase is more significant than for
elliptic flow: $v_1\{3\}$ is 50\% larger at the most forward
rapidities (-6\%, instead of -4\% for mid-central collisions).  In
central collisions, $v_1\{3\}$ could not be obtained due to large
statistical errors.  The standard $v_1$ (Fig.~\ref{fig:158_V1vsPt} and
Fig.~\ref{fig:158_V1vsY} top left) is largest at low $p_t$ (below
0.3~GeV/c) and forward rapidities (above $y=1$) where it is biased by
Bose-Einstein effects as explained above.  This bias is even more
important for the centrality bin 1, where a tighter $p_t$ cut (below
0.3~GeV/c) was chosen for the event plane determination.

We now discuss the directed flow of protons in mid-central collisions.
As a function of $p_t$ (Fig.~\ref{fig:158_V1vsPt} middle right) both
the standard $v_1$ and $v_1\{3\}$ are generally positive but almost
consistent with zero.  The rapidity dependence
(Fig.~\ref{fig:158_V1vsY} middle right) is more interesting. $v_1$ is
flat near midrapidity (error bars are large, but we can safely state
that it is flatter than for pions). Only one point, at the most
forward rapidity, clearly deviates from zero.  This is an essential
point, since we use it to resolve the overall sign ambiguity of the
$v_1$ analysis: at high energies, $v_1$ is assumed to be positive for
protons at forward rapidities, and this determines the pion flow to be
negative in this region (Fig.~\ref{fig:158_V1vsY} middle left).  In
peripheral and central collisions where the sign of the proton $v_1$
could not be clearly determined, the pion $v_1$ was assumed to be
negative for sake of consistency.  This sign will be established more
firmly at 40$A$ GeV (see below).  As already noted for pions, the
slope of $v_1$ above $y=2$ must be very steep in order to match WA98
results, which give a proton $v_1$ of the order of 20\% around
$y=3$~\cite{Aggarwal:1998vg}.

In peripheral collisions, there is a discrepancy between the standard
$v_1$ and $v_1\{3\}$ for protons, which is clearly seen on the $p_t$
graph (Fig.~\ref{fig:158_V1vsPt} top right).  This difference may be
due to nonflow correlations from $\Delta$ decays into protons and
pions \cite{Borghini:2000cm}, which are automatically corrected for in
the three-particle analysis, but not in the reaction plane analysis.
This contamination from correlations from $\Delta$ decays would then
explain why the standard $v_1$ has a negative sign at forward
rapidities (Fig.~\ref{fig:158_V1vsY} top right).  In central
collisions, where only the reaction plane estimate is available,
$v_1(p_t)$ is compatible with zero (Fig.~\ref{fig:158_V1vsPt} bottom
right), but is positive at forward rapidities
(Fig.~\ref{fig:158_V1vsY} bottom right), as expected.

Let us briefly compare the present results with those of the earlier
analysis~\cite{NA49PRL,NA49Web}.  As in the case of elliptic flow, the
most significant differences are seen for pions, where $v_1$ was
biased in the earlier analysis by Bose-Einstein correlations and
global momentum conservation~\cite{Dinh:1999mn,Borghini:2000cm}.  The
earlier $v_1(p_t)$ is in fact similar to the present $v_1\{2\}$
(squares in Fig.~\ref{fig:158_V1vsPt}, middle left), which suffer from
the same biases.

\subsubsection{40$A$ GeV}

\begin{figure}[hbt!]
\includegraphics*[height=.40\textheight]{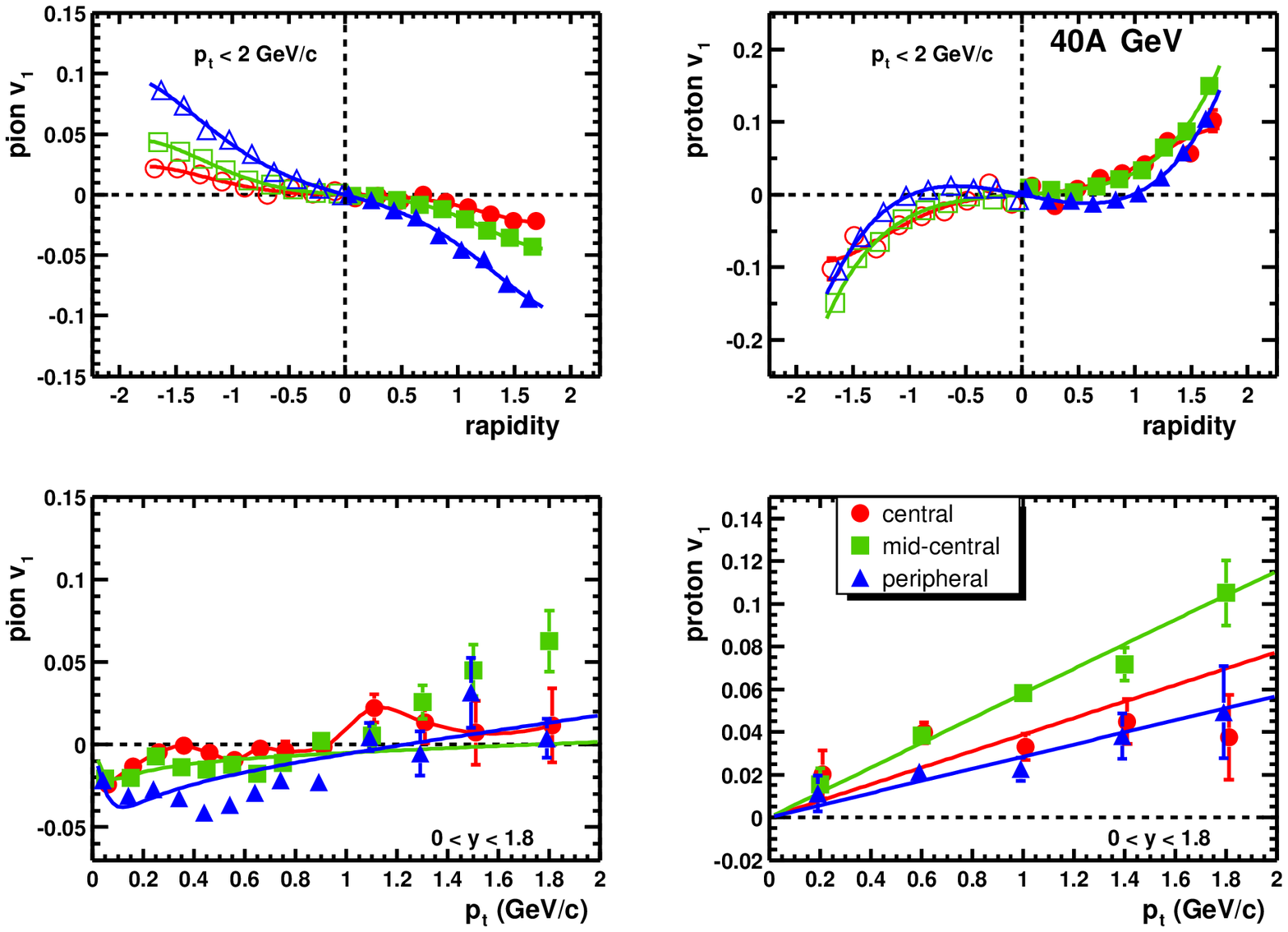}
\caption{\label{fig:40_v1_all} (Color online) Standard  directed
 flow as a function of rapidity (top) and transverse momentum (bottom)
 for charged pions (left) and protons (right) from 40$A$ GeV Pb +
 Pb. Three centrality bins are shown. The open points in the top
 graphs have been reflected about midrapidity. On the top, solid
 lines are polynomial fits. On the bottom, smooth solid lines are
 Blast Wave model fits.}
\end{figure}

\begin{figure}[hbt!]
\includegraphics*[height=.40\textheight]{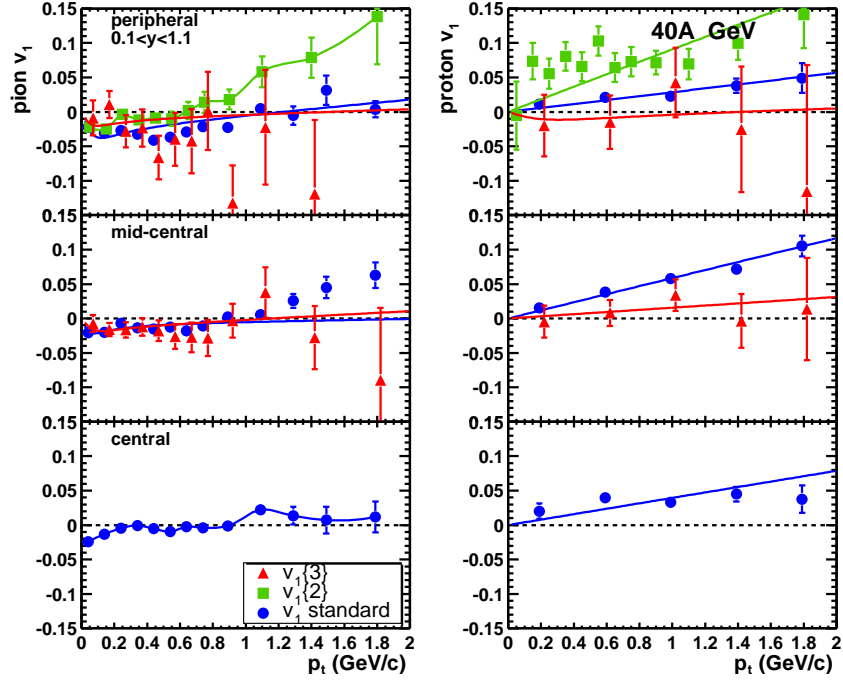}
\caption{\label{fig:40_V1vsPt} (Color online) Directed flow of charged
 pions (left) and 
 protons (right) from the cumulant method, as a function of transverse
 momentum in 40$A$ GeV Pb + Pb. Three centrality bins are
 shown. Results are shown from the standard method, from cumulants for
 two-particle correlations ($v_1\{2\}$), and from cumulants for
 three-particle correlations ($v_1\{3\}$). Please note that the
 $v_1\{2\}$ estimates are {\em not} corrected for momentum
 conservation, while the standard method results are. Smooth solid
 lines are Blast Wave model fits.}
\end{figure}

\begin{figure}[hbt!]
\includegraphics*[height=.40\textheight]{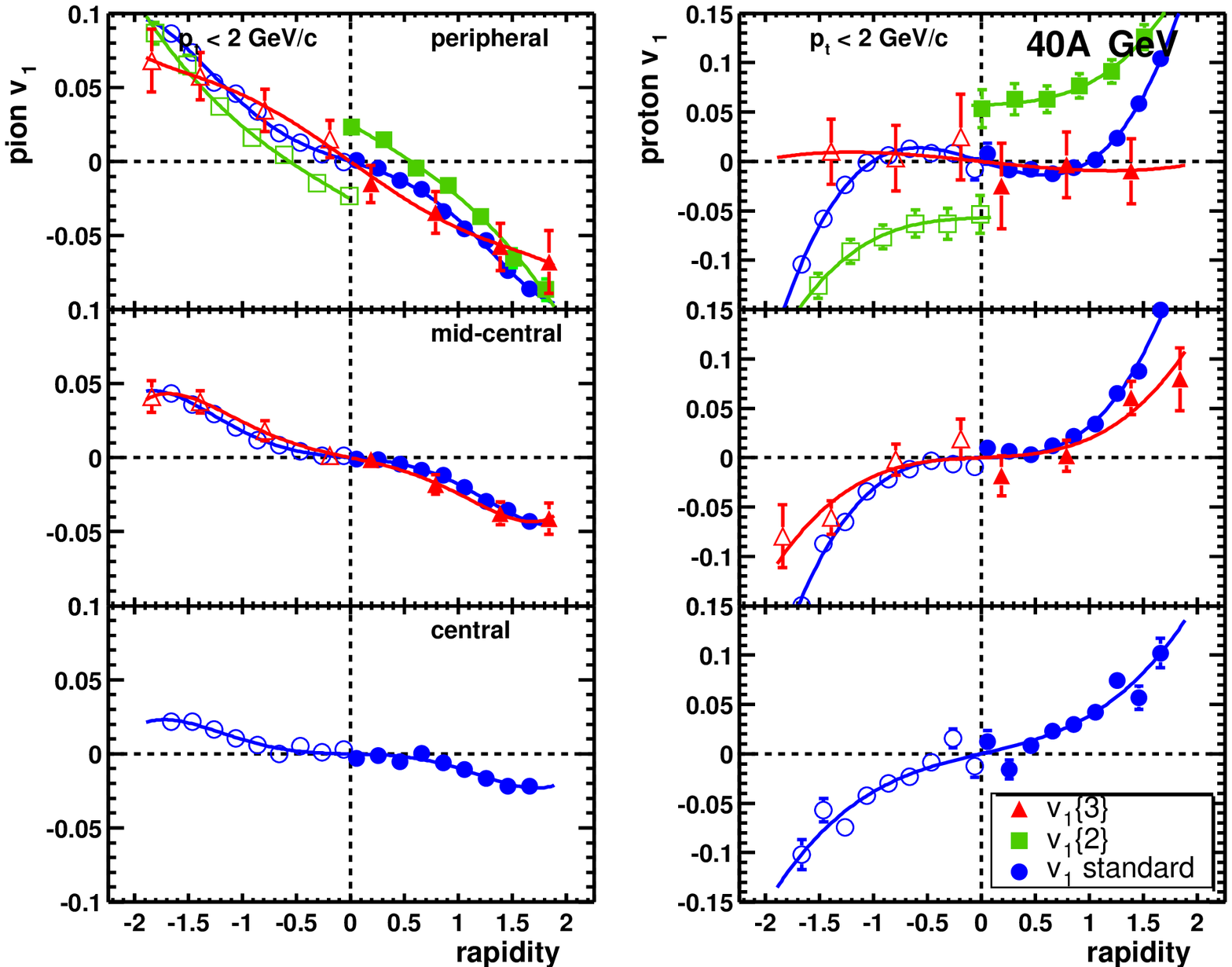}
\caption{\label{fig:40_V1vsY} (Color online) Same as
 Fig.~\ref{fig:40_V1vsPt}, as a  
 function of rapidity. The open points have been reflected about
 midrapidity. Solid lines are polynomial fits. Please note the
 different vertical scales.}
\end{figure}

Results from 40$A$ GeV collisions are shown in
Fig.~\ref{fig:40_v1_all} (from the standard method),
\ref{fig:40_V1vsPt}, and \ref{fig:40_V1vsY} (from the cumulant
method).  In Fig.~\ref{fig:40_v1_all}, top, one can see that the
standard $v_1$ crosses zero at midrapidity for all centralities, which
shows that the correlation from global momentum conservation has been
properly subtracted.  The three-particle results $v_1\{3\}$, which is
automatically free from all nonflow effects including momentum
conservation, also crosses zero at mid-rapidity for peripheral and
mid-central collisions (Fig.~\ref{fig:40_V1vsY} top and middle).  It
could not be obtained for central collisions due to large statistical
fluctuations.  The effect of momentum conservation is even larger than
at 158$A$ GeV, as can be seen by comparing $v_1\{2\}$ (not
corrected) with the standard $v_1$ (corrected) for peripheral
collisions (Figs.~\ref{fig:40_V1vsPt} and \ref{fig:40_V1vsY} top).
For mid-central and central collisions, the momentum conservation
effect is so large that $v_1\{2\}$ could not even be obtained, as for
central collisions at 158$A$ GeV.

The directed flow of pions (Figs.~\ref{fig:40_v1_all}, 
\ref{fig:40_V1vsPt} and \ref{fig:40_V1vsY} left)
is similar at 40$A$ GeV and at 158$A$ GeV, both in magnitude
and shape. The standard $v_1$ and $v_1\{3\}$ are compatible. Error
bars are larger at the lower energy due to the lower multiplicity,
especially for the three-particle result $v_1\{3\}$.  The nonzero
value of the standard $v_1$ at low $p_t$ (Fig.~\ref{fig:40_v1_all}
bottom left) probably results from nonflow Bose-Einstein correlations,
as at 158$A$ GeV.

The directed flow of protons, on the other hand, 
(Figs.~\ref{fig:40_v1_all}, 
\ref{fig:40_V1vsPt} and \ref{fig:40_V1vsY} right),
is significantly larger at 40$A$ GeV than at 158$A$ GeV.
For mid-central collisions, both the standard $v_1$ and $v_1\{3\}$
clearly differ from zero at forward rapidities
(Fig.~\ref{fig:40_V1vsY} middle right).  As already explained, the
proton $v_1$ is assumed to be positive at forward rapidities, and this
fixes the sign of the pion flow to be negative in this region
(Fig.~\ref{fig:40_V1vsY} middle left).  This is consistent with our
prescription at 158$A$ GeV.  The standard $v_1$ is larger than
$v_1\{3\}$. The discrepancy is beyond statistical error bars
(Fig.~\ref{fig:40_V1vsPt} middle right) and might be due to some
nonflow effect.

Unlike the pion $v_1$, the proton $v_1$ does not seem to be 
larger for peripheral collisions than for mid-central collisions.
In Figs.~\ref{fig:158_v1_all} and \ref{fig:40_v1_all}, top right
panels, the peripheral data seem to exhibit a ``wiggle'' such that the
proton $v_1$ has a negative excursion.  Due to the large statistical
error bars, this can neither be confirmed nor invalidated by the
three-particle cumulant results in Figs.~\ref{fig:158_V1vsY} and
\ref{fig:40_V1vsY}, top right. Nevertheless, such a behavior has been
predicted~\cite{Snellings:1999bt} due to the variation in stopping in
the impact parameter direction in peripheral collisions coupled with the
space-momentum correlations of flow~\cite{Voloshin:QM02}. This is the
first experimental observation of this phenomenon.

\subsection{Minimum Bias}
\label{s:minbias}

\begin{figure}[hbt!]
\includegraphics*[height=.40\textheight]{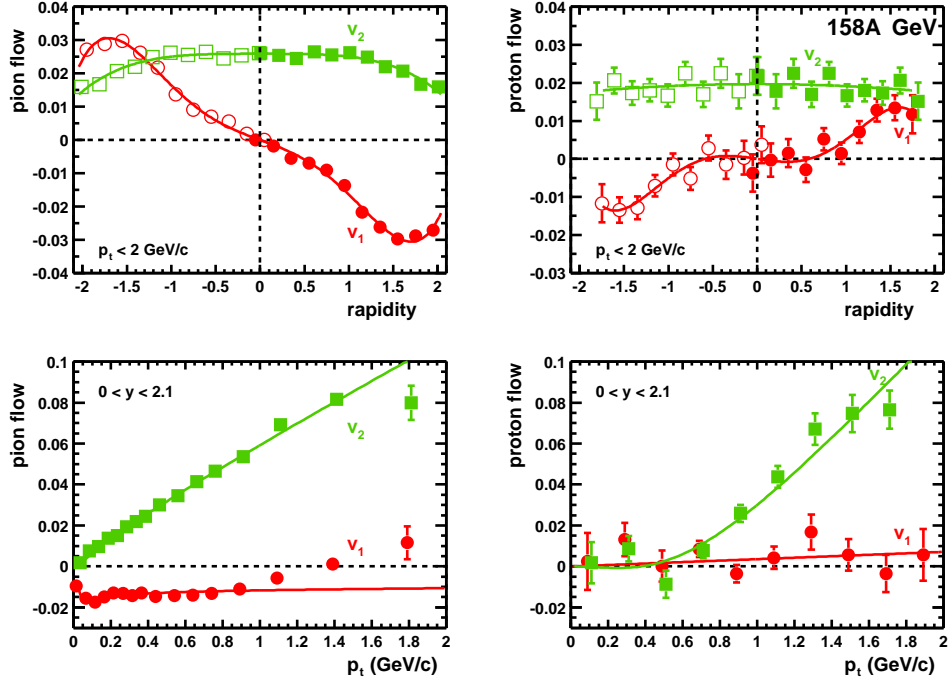}
\caption{\label{fig:158_mb} (Color online) Standard minimum bias
 directed and elliptic 
 flow as a function of rapidity (top) and transverse momentum (bottom)
 for charged pions (left) and protons (right) from 158$A$ GeV Pb
 + Pb. Shown are $v_1$ (circles) and $v_2$ (squares). The open points
 in the top graphs have been reflected about midrapidity. Solid lines
 are polynomial fits on the top and Blast Wave model fits on the
 bottom.}
\end{figure}

\begin{figure}[hbt!]
\includegraphics*[height=.40\textheight]{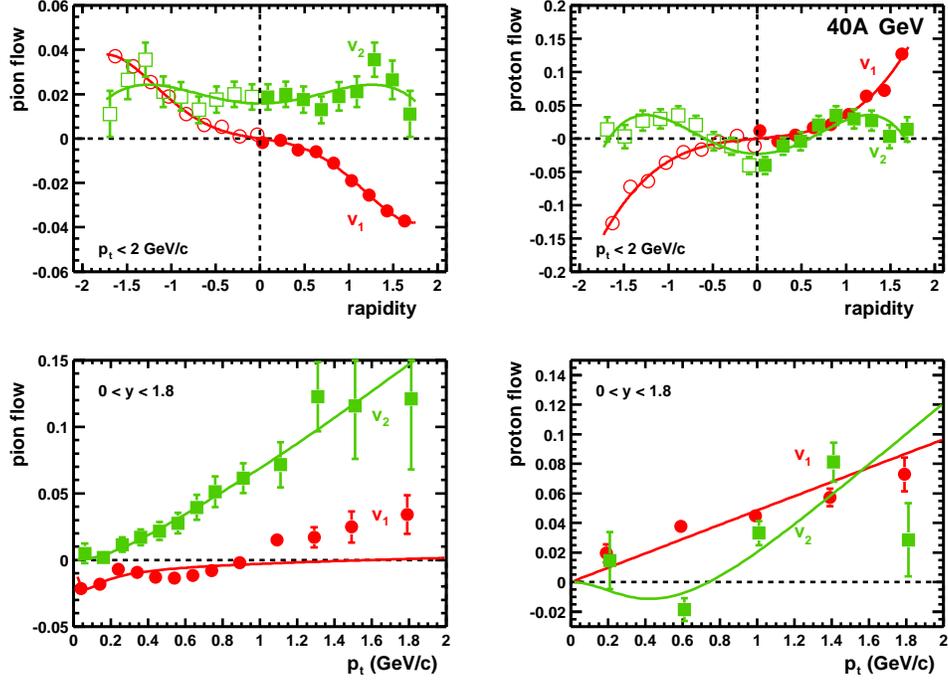}
\caption{\label{fig:40_mb} (Color online) Same as Fig.\ref{fig:158_mb}
 for 40$A$ GeV  Pb + Pb.}
\end{figure}

The results of the standard method integrated over the first five
centrality bins weighted with the fraction of the geometric cross
section for each bin given in Table~\ref{tbl:centrality} are shown in
Figs.~\ref{fig:158_mb} and \ref{fig:40_mb} for the two beam
energies. For 40$A$ GeV $v_2$ bin 1 was not included because we
had no results for it. (Also, the cumulant data could not be summed
for minimum bias graphs because too many centrality bins were
missing.) In the lower left graphs, $v_1$ for pions shows a sharp
negative excursion in the first 100 MeV/$c$ of $p_t$. This can also be
seen in the graphs for the individual centralities. To describe this
feature the Blast Wave fits require a very large $\rho_0$ parameter.
However, the physical explanation is not clear and the effect may in
fact be caused by some very low $p_t$ short range nonflow correlation,
most probably quantum correlations between
pions~\cite{Dinh:1999mn,Borghini:2000cm}.  Similarly, the positive
values of $v_1$ for pions at high transverse momenta is due to nonflow
correlations (see $v_1\{3\}$ in Fig.~\ref{fig:40_V1vsPt}).  In
Fig.~\ref{fig:158_mb} lower right, the proton $v_1$ is consistent with
zero at all $p_t$ values because of the accidental cancellation of its
positive and negative values at the different centralities.

\subsection{Centrality dependence of integrated flow values}
\label{s:centrality}

\begin{figure}[hbt!]
\includegraphics*[width=0.35\textwidth]{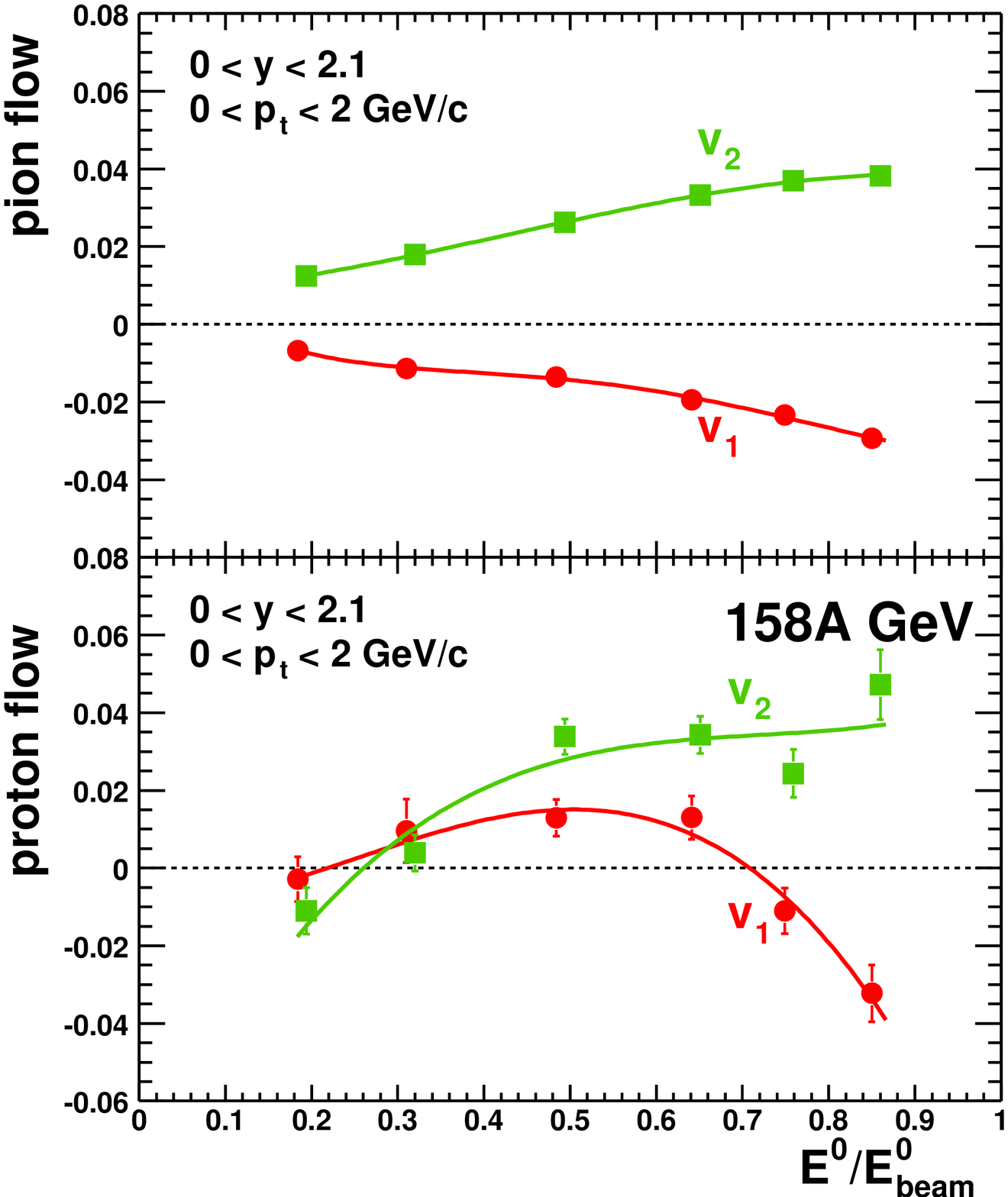}
\includegraphics*[width=0.35\textwidth]{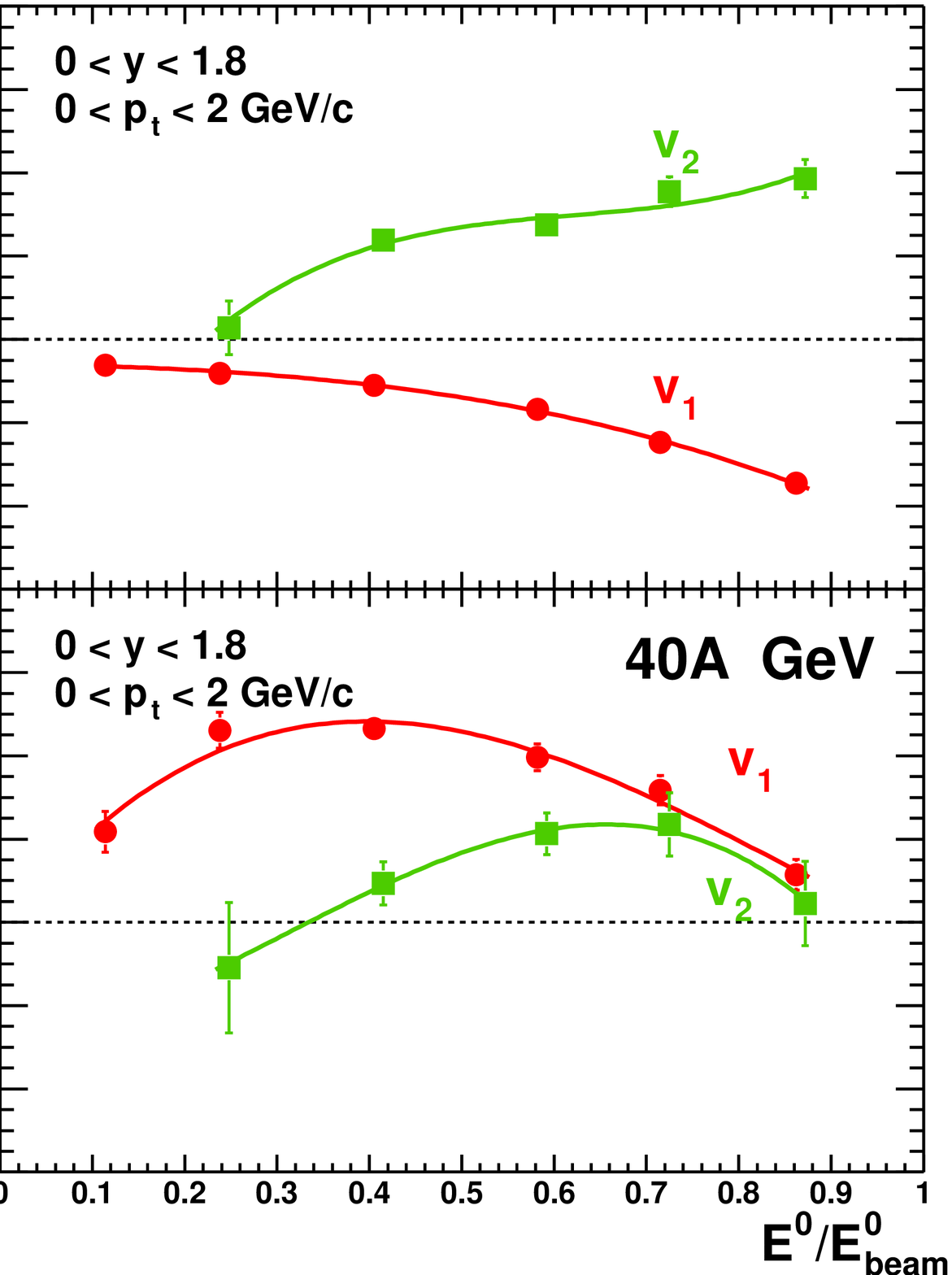}
\caption{\label{fig:v_cen} (Color online) Directed and elliptic
 flow as a function of centrality for charged pions (top) and protons
 (bottom) from 158$A$ GeV Pb + Pb (left) and 40$A$ GeV Pb +
 Pb (right) from the standard analysis. Shown are $v_1$ (circles) and
 $v_2$ (squares). The more central collisions are on the left side of
 each graph. Solid lines are polynomial fits.}
\end{figure}

Results from the standard method for the doubly-integrated $v_n$ as a
function of centrality are shown in Fig.~\ref{fig:v_cen}.  As the
differential flow values $v_n(p_t)$ or $v_n(y)$, these results were
obtained by averaging the tabulated $v_n(p_t,y)$ values, here over
both transverse momentum and rapidity, using the cross sections as
weights.  For pions on top, the values generally increase in absolute
magnitude in going from central to peripheral collisions. However, for
protons on bottom, the values appear to peak at mid-centrality.

\begin{figure}[hbt!]
\includegraphics*[height=.35\textheight]{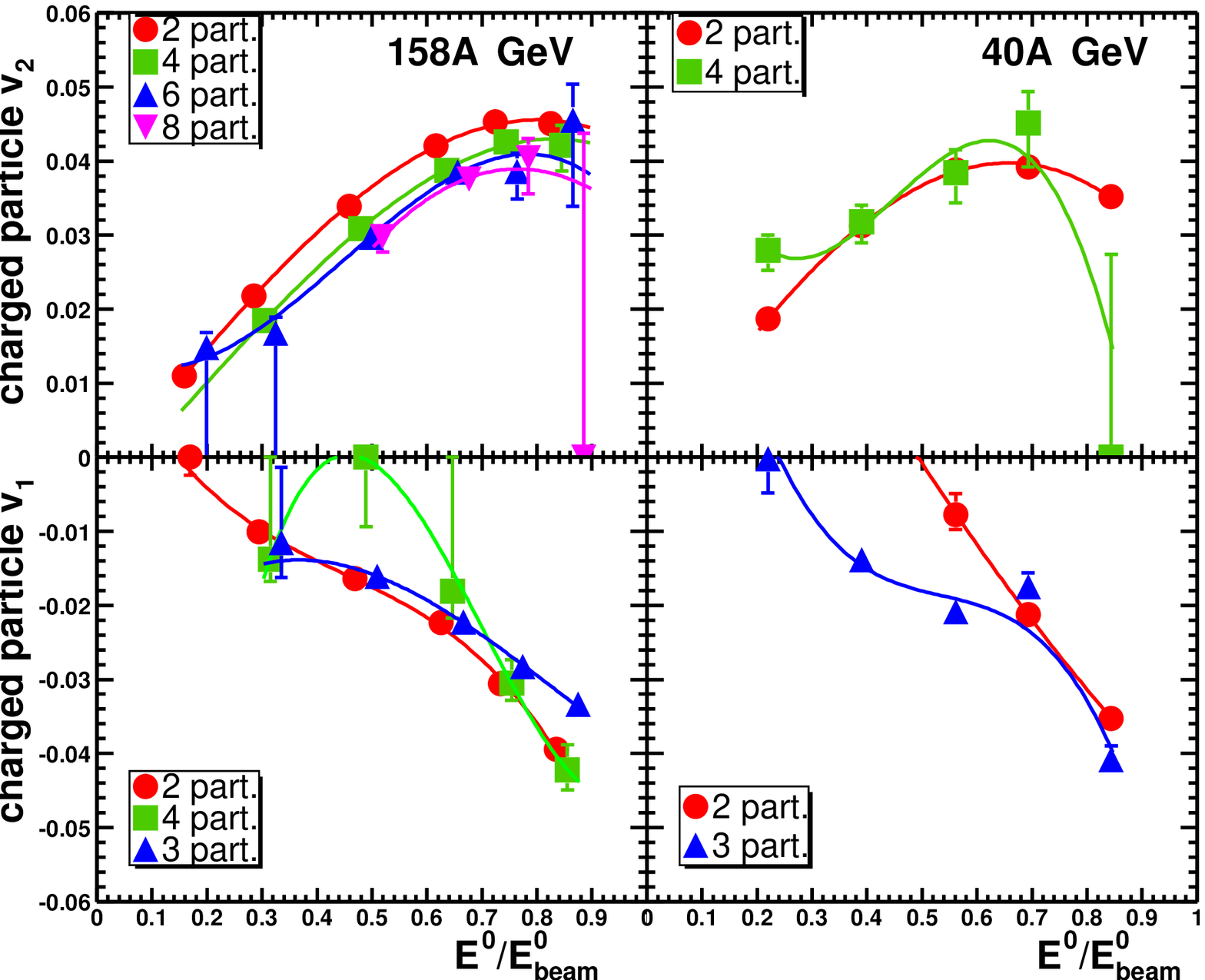}
\caption{\label{fig:Vn_int_cum} (Color online) Weighted flow 
  $\langle w_n e^{in(\phi-\Phi_{RP})}\rangle / \sqrt{\mean{{w_n}^2}}$,
  Eq.~(\ref{weight-int}), with $w_1=y$ in the center of mass frame and
  $w_2=p_t$, from the cumulant method as a function of centrality in
  158$A$ GeV Pb + Pb (left) and 40$A$ GeV Pb+Pb (right). The
  more central collisions are on the left side of each graph. The
  lines are polynomial fits. The unplotted points could not be
  obtained or had error bars which were too large.}
\end{figure}

Figure \ref{fig:Vn_int_cum} shows the weighted integrated flow values
for charged particles, Eq.\ (\ref{weight-int}), from the
cumulant method as a function of centrality, up to eight-particle
correlations.  Contrary to the standard method results, these values
were not obtained by averaging the differential values {\em a
posteriori}.  As mentioned in Sec.~\ref{s:cumulant}, integrated flow
is the first outcome of the cumulant method, which is then used to
obtain differential values.  In particular, this explains why for
$v_2$ we could derive integrated values from up to eight-particle
cumulants, while only four-particle results are available for
differential flow.  Also for $v_1$ we could obtain integrated values
from up to four-particle cumulants while only three-particle results
are available for differential flow with reasonable error bars.

Please note that the plotted quantity is not $v_n$, but a weighted
flow $\mean{w_n e^{in(\phi-\Phi_{RP})}} / \sqrt{\mean{{w_n}^2}}$.
Since it is only intended as a reference value in the cumulant method,
the weights were chosen so as to maximize it, and indeed it is about a
factor of 1.2 larger than the standard analysis value because of the
weighting.

A striking feature is the consistency between the flow estimates using
more than two particles.  Thus, in Fig.~\ref{fig:Vn_int_cum} top left
for 158$A$ GeV, $v_2\{4\}$, $v_2\{6\}$, and $v_2\{8\}$ are in
agreement within statistical error bars.  This is a clear signal that
these estimates indeed measure the ``true'' $v_2$, and are all arising
from a collective motion.  In addition, it is interesting to note that
these three estimates differ significantly from the two-particle
estimate $v_2\{2\}$, which suggests that the latter is contaminated by
nonflow effects.  This will be further examined in
Sec.~\ref{s:nonflow}.  Apart from that, these charged particle values
follow the usual trend for pions: the absolute value of $v_1$
increases in going to more peripheral collisions; $v_2$ also increases
in going from central to semi-central collisions, but then starts to
decrease for the most peripheral bins, both at 40 and 158$A$ GeV.

\subsection{Nonflow effects}
\label{s:nonflow}

The purpose of the cumulant method is to remove automatically nonflow
effects, so as to isolate flow. Nevertheless, one may wonder whether
this removal is really necessary, and if nonflow effects are significant.

\begin{figure}[hbt!]
\includegraphics*[height=.30\textheight]{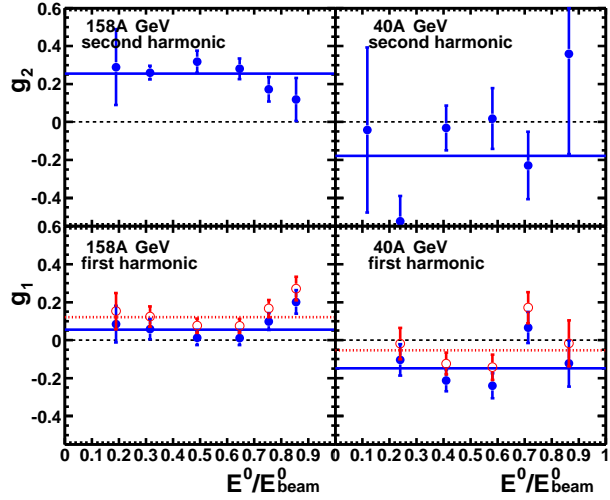}
\caption{\label{fig:nonflow} (Color online) Nonflow azimuthal
 correlations from  
 Eq.(\ref{eq:gn}), for the first, $g_1$, (bottom) and second, $g_2$,
 (top) Fourier harmonics, from 158$A$ GeV (left) and 40$A$ GeV (right)
 Pb+Pb collisions. For $g_1$, the solid points represent all nonflow
 effects, while the open points are corrected for momentum
 conservation. The horizontal lines are at the mean values.}
\end{figure}

A first way to estimate the contribution of nonflow effects, which was
proposed in Ref.~\cite{Borghini:2002vp}, consists in plotting the
quantity $N\cdot(v_n\{2\}^2-v_n\{k\}^2)$ as a function of centrality,
where $N$ is the mean total number of particles per collision in a
given centrality bin, and $k>2$.  The reason is that the two-particle
estimate $v_n\{2\}$ is contaminated by nonflow effects, while the
estimate from more particles $v_n\{k\}$ is not, hence their difference
should be due to the nonflow correlations.  More precisely, the
two-particle cumulant reads
\begin{eqnarray}
\label{nonflow}
c_n\{2\} &\equiv& v_n\{2\}^2\, \cr
&=& (v_n)^2 + {\rm nonflow}\, \cr
&=& (v_n)^2 + \frac{g_n}{N},
\end{eqnarray}
where we have recalled the definition of the two-particle flow
estimate $v_n\{2\}$, see Eq.~(\ref{flow&cumul-int}), and the nonflow
term scales as $1/N$.  Since it is hoped that the multi-particle
estimates reflect only flow, $v_n \simeq v_n\{k\}$, a straightforward
rearrangement shows that 
\begin{eqnarray}
\label{eq:gn}
  N\cdot(v_n\{2\}^2-v_n\{k\}^2) = g_n
\end{eqnarray}
should be approximately constant. Note
that $g_1$ and $g_2$ were calculated directly from the cumulants, not
from the values of the flow estimates:
\begin{eqnarray}
\label{g1-g2}
g_1 &=& N \cdot \left( c_1\{2\} - c_1\{3\}/v_2 \right), \cr 
g_2 &=& N \cdot \left( c_2\{2\} - \sqrt{-c_2\{4\}} \right)
\end{eqnarray}
($c_1\{3\}$ denotes the left-hand side of Eq.~(\ref{v1/2idea}), 
and the other cumulants have been defined in Sec.~\ref{s:cumulant}).
This explains why we can display values for almost all centrality
bins, while there might be no corresponding flow values because of a
wrong sign for the cumulant~\footnote{When $c_2\{4\}>0$, which occurs
for centrality bin 1 at 158$A$ GeV, we set $c_2\{4\}=0$ in
Eq.~(\ref{g1-g2}).}.  In Fig.~\ref{fig:nonflow} are shown the
coefficients $g_1$ and $g_2$ as a function of centrality, for both 158
and 40$A$ GeV, where the values of $N$ are taken from
Table~\ref{tbl:momcons}.  The magnitude of the statistical error
follows the analytical formulas given in Appendix
\ref{s:nonflow-error}.

As could be expected from the fact that in Fig.~\ref{fig:Vn_int_cum}
top left the two-particle points are above the multi-particle points,
a clear nonflow signal can be observed in Fig.~\ref{fig:nonflow} top
left for the second harmonic in full energy collisions.  Moreover,
$g_2$ is approximately constant, as expected for nonflow correlations:
one has approximately $v_2\{2\}^2 \simeq v_2\{4\}^2 + 0.26/N$, at
least for the four more central bins.

For the first harmonic, two types of results are displayed in
Fig.~\ref{fig:nonflow} bottom: the solid points correspond to all
nonflow effects in Eq.\ (\ref{eq:gn}), while the open points are
corrected for momentum conservation. The uncorrected points are
approximately constant and close to zero at 158$A$ GeV for all
but the most peripheral bin. This is quite surprising since we have
seen above that the two-particle $v_1\{2\}$ is strongly contaminated
by correlations arising from momentum conservation.  Actually the
contribution of this nonflow effect can be explicitly calculated using
Eq.\ (4) of Ref.~\cite{Borghini:2002mv}:
\begin{equation}
\label{g1^mom-cons}
g_1^{\rm mom-cons} = - \frac{\mean{y \, p_t}^2}{\mean{y^2}
  \mean{{p_t}^2}_{\rm all}} = - \frac{N}{M} f^2\,,
\end{equation}
where averages are calculated over the phase space covered by the
detector, except for the subscript ``all'' which means that all
particles are taken into account, including those that fall outside
the detector acceptance.  The corresponding quantity
$\mean{{p_t}^2}_{\rm all}$ is estimated from a model calculation, see
Table~\ref{tbl:momcons}.  In the second version of the equation, $N$
is the multiplicity over all phase space, $M$ is the multiplicity of
detected particles, and $f$ is defined by Eq.\ (19) of
Ref.~\cite{Borghini:2002mv}. Note that for a perfect acceptance, or an
acceptance symmetric about midrapidity, $g_1^{\rm mom-cons}$
vanishes. As expected, $g_1^{\rm mom-cons}$ depends weakly on
centrality and its mean value is about $-0.067$ at 158$A$ GeV, and
$-0.10$ at 40$A$ GeV. Since the acceptance is better at 158$A$ GeV
than at 40$A$ GeV, we expect to get a lower value at that energy, and
this is the case indeed. Finally, subtracting this contribution to
$g_1$ (which amounts to an addition, since $g_1^{\rm mom-cons} < 0$),
all values are shifted upward and we get a clear positive value at
158$A$ GeV for all nonflow effects except momentum conservation.

The important point is that the whole $g_1$ is
only an integrated quantity, averaged over phase space and summed over
different particle types, and variations of opposite sign may
cancel. As a matter of fact, we have already noted in Sec.~\ref{s:v1}
that for pions the three-particle estimate $v_1\{3\}$ is lower than
$v_1\{2\}$ at large $p_t$, but larger at low $p_t$
(Fig.~\ref{fig:158_V1vsPt}, left): ascribing the large-momentum
discrepancy to momentum conservation, and the low-$p_t$ one to
short-range correlations, it turns out that they compensate when one
averages over $p_t$. Similarly, we have already mentioned that the
difference seen in the directed flow of protons
(Fig.~\ref{fig:158_V1vsPt}, right) between the standard analysis value
(corrected for $p_t$ conservation) and the three-particle estimate at
high momentum could be due to correlations from $\Delta$ decays, an
effect that may explain the open points in Fig.\ref{fig:nonflow},
bottom.

On the contrary, the nonflow correlations in the second harmonic,
which are seen in Fig.~\ref{fig:nonflow}, cannot be localized in any
definite region of phase space in Figs.~\ref{fig:158_V2vsPt} or
\ref{fig:158_V2vsY}, and thus there is no clue regarding the actual
effects which contribute.  One may simply notice that the order of
magnitude of $g_2$ is the same as the estimate for the contribution
from $\rho$-meson decays~\cite{Borghini:2000cm}, although resonance
decays may only explain part of $g_2$.  Nevertheless, one can safely
conclude that nonflow effects are significant at 158$A$ GeV and
play a role in the analysis of both $v_1$ and $v_2$.

At 40$A$ GeV, Fig.~\ref{fig:nonflow} is less conclusive than at
158$A$ GeV, but we have seen in Sec.~\ref{s:results} clear
indications of nonflow correlations.  First, we recall that the
importance of nonflow effects, and especially of momentum conservation,
in the first harmonic is such that it is not possible to obtain
directed flow from the two-particle cumulant, $v_1\{2\}$, while the
standard method (including correction for $p_t$ conservation) or the
three-particle cumulant do give results. Apart from this large effect,
we also mentioned the possible presence of short-range correlations at
low $p_t$, which could explain why the standard method value for $v_1$
does not go smoothly to zero at vanishing momentum, while the
three-particle result $v_1\{3\}$ does.  In the second harmonic, we
have seen in Sec.~\ref{s:v2} large discrepancies between the two- and
four-particle estimates, in central collisions.  As suggested above,
the observed differences may arise from two-particle correlations due
to $\rho$-meson decays.

\subsection{Beam energy dependence}
\label{s:energy}

\begin{figure}[hbt!]
\includegraphics*[height=.30\textheight]{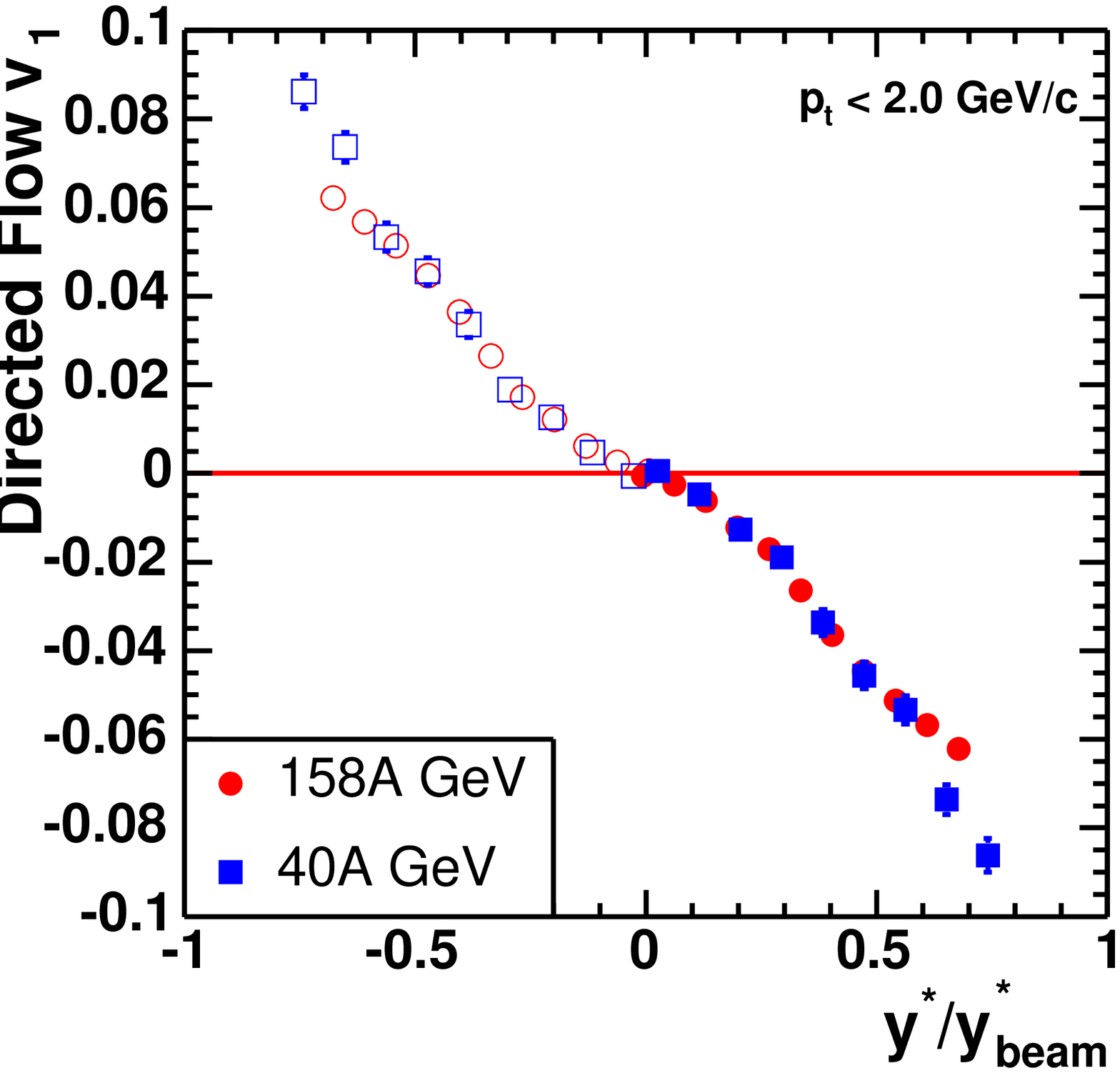}
\includegraphics*[height=.30\textheight]{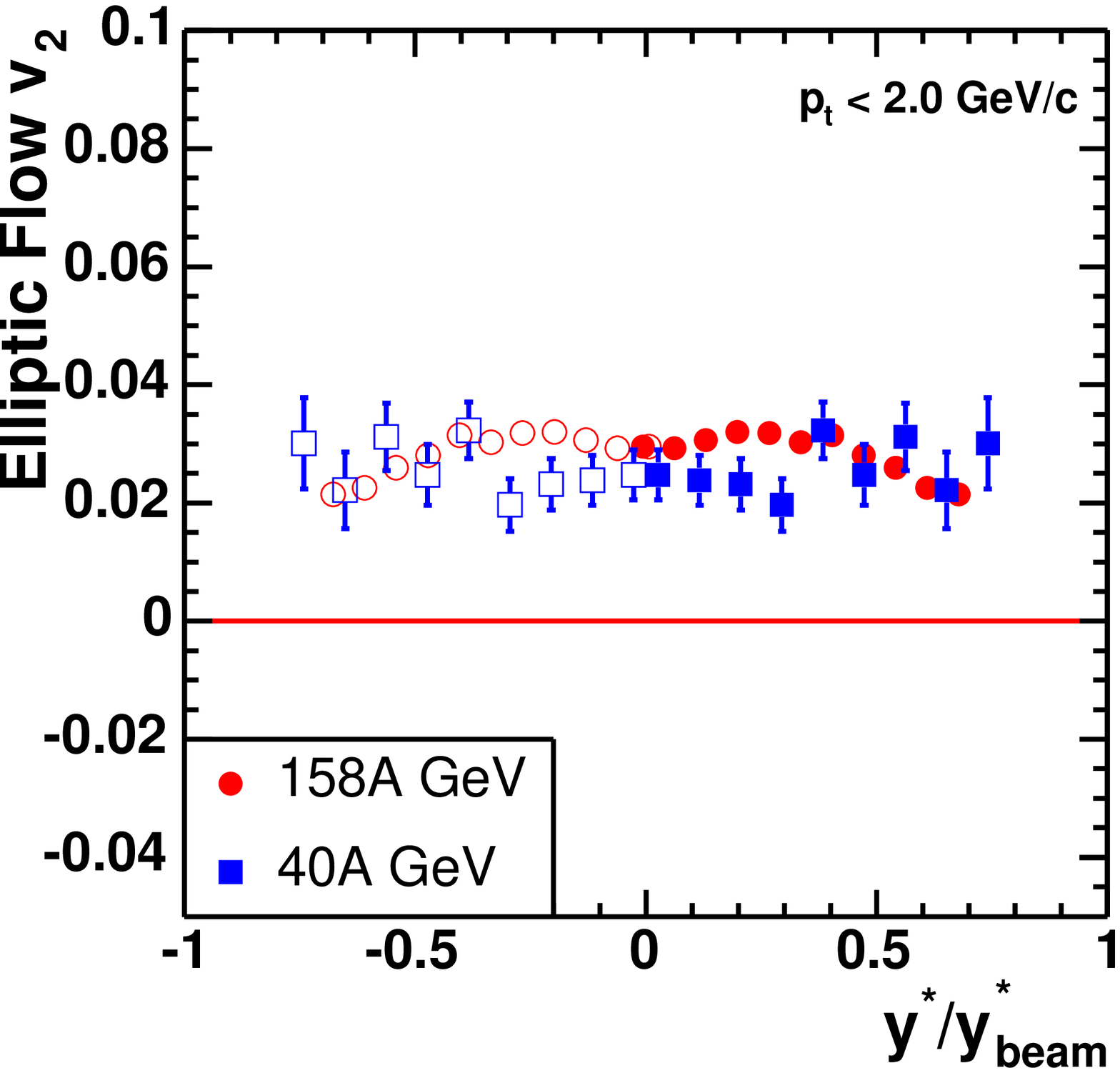}
\caption{\label{fig:v2_scaled_eComp} (Color online) A comparison of
directed and 
elliptic flow of charged pions at 40 and 158$A$ GeV using the
standard method is shown as a function of scaled rapidity. The
left graph for $v_1$ is for peripheral collisions and the right graph
for $v_2$ is for mid-central collisions. The open points have been
reflected about midrapidity.}
\end{figure}

A direct comparison between flow at 40 and 158$A$ GeV is presented in
Fig.~\ref{fig:v2_scaled_eComp}, where $v_1$ and $v_2$ for pions are
plotted as a function of scaled rapidity $y/y_{beam}$.  The use of
scaled rapidity is justified by the fact that the width of the
rapidity distribution of produced hadrons increases approximately as
$y_{beam}$~\cite{NA49-energy} and that the shapes of the transverse
momentum distributions are similar at 40 and 158$A$
GeV~\cite{NA49-energy}.  Dependence of $v_1$ and $v_2$ on scaled
rapidity is similar for both energies, however there is an indication
that $v_2$ is slightly larger at 158$A$ GeV than at 40$A$ GeV,
although $v_1$ appears to be the same at the two beam energies. For
$v_1$ at the other centralities the agreement is not as close, but the
$v_1\{3\}$ integrated results, which are free from nonflow effects,
agree at the two beam energies.

\begin{figure}[hbt!]
\includegraphics*[height=.40\textheight]{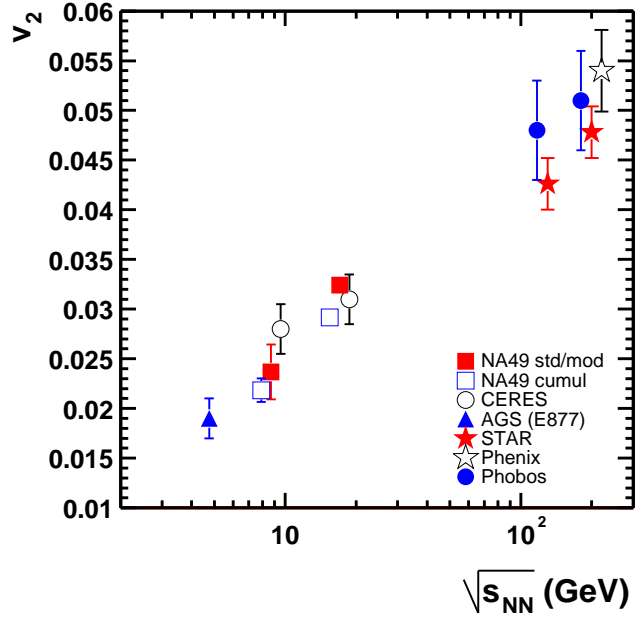}
\caption{\label{fig:v2_edep} (Color online) Energy dependence of $v_2$
near mid-rapidity 
($0 < y < 0.6$ for 40$A$ GeV and $0 < y < 0.8$ for 158$A$ GeV)
for mid-central collisions (approx. 12 to 34 \% of geometrical cross 
section). The results of NA49 pion $v_2$ are compared to charged 
particle $v_2$ measured by E877, STAR, PHENIX and PHOBOS.}
\end{figure}

Elliptic flow was recently measured at midrapidity in Au+Au at RHIC
energies~\cite{Ray:QM02,Esumi:QM2002,Manly:QM2002}.  These
measurements together with the corresponding measurements presented in
this paper and other SPS measurements~\cite{CERES:INPC01,CERES:QM02}
(there are newer measurements not taken into account yet, as they were
taken at different centralities~\cite{CERES:YUN02}), as well as the
measurement at the AGS~\cite{Voloshin:1999gs}, allow us to establish
the energy dependence of $v_2$ in a broad energy range.  The STAR data
have been scaled down to take into account the low $p_t$ cut-offs of
150 MeV/c at $\sqrt{s_{NN}}=200$~GeV and 75 MeV/c at
$\sqrt{s_{NN}}=130$~GeV. The correction factors have been
obtained by extrapolating the charged particle yield to zero $p_t$ and
assuming a linear dependence of $v_2(p_t)$ at small transverse
momenta. The factors were 1.14 for the $\sqrt{s_{NN}}=200$~GeV data
and 1.06 for the $\sqrt{s_{NN}}=130$~GeV data. The SPS data do not
have a low $p_t$ cutoff.  In Fig.~\ref{fig:v2_edep} $v_2$ at
midrapidity is plotted as a function of center of mass energy per
nucleon--nucleon pair for mid-central collisions. The rise with beam
energy is rather smooth.

\begin{figure}[htb!]
\includegraphics*[height=.40\textheight]{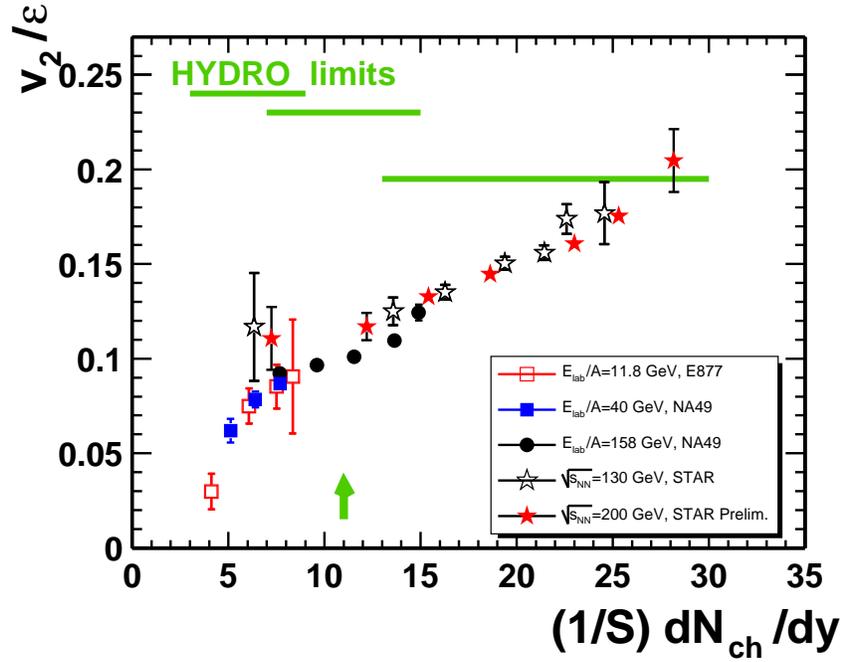}
\caption{\label{fig:fscaling}  (Color online) $v_2/\epsilon$ as a
function of particle 
density. The $v_2$ values are for near mid-rapidity ($0 < y < 0.6$ for
40$A$ GeV and $0 < y < 0.8$ for 158$A$ GeV). The results of
NA49 pion $v_2$ are compared to charged particle $v_2$ measured by
E877 and STAR. The meaning of the horizontal lines (hydro limits) and
of the arrow will be discussed in Sec.~\ref{s:models}.}
\end{figure}

$v_2$ divided by the initial eccentricity of the overlap region, 
$\epsilon$, is free from geometric contributions~\cite{Sorge:1998mk}. 
It is useful to plot this quantity versus the particle density as 
estimated by $dN/dy$ of charged particles divided by the area of 
the overlap region, $S$~\cite{Voloshin:1999gs,Voloshin:QM02}.
The initial spatial eccentricity is calculated for a Woods-Saxon
distribution with a wounded nucleon model from
\begin{equation}
\epsilon = \frac{ \langle y^2 \rangle - \langle x^2 \rangle }      
            { \langle y^2 \rangle + \langle x^2 \rangle }\,
\end{equation}
where $x$ and $y$ are coordinates in the plane perpendicular to the
beam and $x$ denotes the in-plane direction. Fig.~\ref{fig:fscaling}
shows our results for $v_2\{2\}$ together with the recent STAR results
on elliptic flow from 4-particle cumulants~\cite{STAR:130, Ray:QM02},
and E877 results~\cite{Voloshin:1999gs} from the standard method. The
RHIC data have been corrected for their low $p_t$ cut-off as described
for Fig.~\ref{fig:v2_edep}.  Only statistical errors are shown.

The NA49 results shown in the figure were obtained with the 2-particle
cumulant method. The total systematic uncertainties for the points are
unfortunately rather large and amount to about 30\% of the presented
values.  The systematic errors for the most central collisions are
even larger and taking into account also the larger statistical errors
we do not plot them in the figure. The results obtained for NA49 with
the standard method as well as from 4-particle cumulants have larger
statistical errors but are within the systematic errors quoted.  There
are also possible systematic errors of the order of 10-20\% due to
uncertainties in the centrality measurements from one experiment to
the other. Note that the most peripheral RHIC data are slightly above
the slightly more central data from the SPS, but would agree within
these systematic errors. This slight non-scaling could be related to
the use of $\epsilon$ calculated with a weight proportional to the
number of wounded nucleons.  A different weight could give somewhat
different relative values for spatial eccentricities of central and
peripheral collisions.  In general the figure shows a nice scaling of
the elliptic flow divided by the initial spatial eccentricity when
plotted against the produced particle density in the transverse
plane. The physical interpretation will be discussed in
Sec.~\ref{s:models}.

The energy dependence of directed flow is also instructive.  Directed
flow has not yet been seen at RHIC, but it has been extensively
studied at lower energies at the AGS.  The most striking difference
between the present results and AGS results is in the rapidity
dependence of the proton directed flow. Up to the top AGS energy, the
proton $v_1(y)$ follows the well-known S-shape, with a maximum slope
at midrapidity~\cite{Liu:2000am}. By contrast, at SPS, already at
40$A$ GeV, the slope at midrapidity is consistent with zero within our
errors. This is compatible with a smooth extrapolation of AGS results,
which already show a significant decrease of the slope at midrapidity
between 2 and 11~$A$ GeV.

While the proton $v_1$ near midrapidity becomes much smaller as the
energy increases, the pion $v_1$ near midrapidity remains of
comparable magnitude.  The main difference, compared to AGS energies,
is that $v_1$ remains negative until very high values of $p_t$, while
at AGS it becomes positive typically above
500~MeV/c~\cite{Barrette:1997pt}, which was interpreted as an effect
of the sidewards motion of the source.

\section{Model comparisons}
\label{s:models}

Here we review theoretical predictions for elliptic and directed flow
at SPS energies and compare them with our results. Note however that
detailed predictions have only been made for 158$A$ GeV
collisions.

\subsection{Elliptic flow}

Elliptic flow at ultrarelativistic energies is interpreted as an
effect of pressure in the interaction region.  In the transverse
plane, particles are created where the two incoming nuclei
overlap. This defines a lens-shaped region for non-central
collisions. Subsequent interactions between the particles drive
collective motion along the pressure gradient, which is larger
parallel to the smallest dimension of the lens. This creates in-plane,
positive elliptic flow~\cite{OL92}.  At early times, however,
spectator nucleons tend to produce negative elliptic
flow~\cite{Sorge:1996pc}: while this effect dominates at energies
below 5~GeV per nucleon \cite{Pinkenburg:1999ya}, one expects it to be
negligible at SPS energies, except close to the projectile rapidity,
which is not covered by the present experiment.

As it is essentially a pressure effect, elliptic flow is sensitive
both to the equation of state of nuclear matter (i.e., the relation
between pressure and density), and the degree of thermalization
reached in the system.  The equation of state enters mainly through
the velocity of sound, $c_s\equiv (dP/d\epsilon)^{1/2}$, which
controls the magnitude of pressure gradients: naturally, a ``softer''
equation of state, with smaller $c_s$, produces smaller elliptic flow,
and this is true in particular in the presence of a first-order phase
transition where $c_s$ vanishes~\cite{OL92}.  From a microscopic point
of view, elliptic flow increases with the number of collisions per
particle: very generally, one expects a linear increase, followed by a
saturation when the system reaches thermal
equilibrium~\cite{Molnar:2001ux}.  Disentangling both effects
(equation of state vs. degree of thermalization), however, is far from
obvious.

Various predictions have been made for elliptic flow at SPS
energies. They are based either on hydrodynamic models, where full
thermal equilibrium is assumed, or on transport models, where
collisions between particles are treated at a microscopic level.  We
briefly review these predictions and discuss how they compare with the
present data.

Let us start with the transverse momentum dependence of $v_2$.  As can
be seen in Figs.~\ref{fig:158_V2vsPt} and \ref{fig:40_V2vsPt}, $v_2$
vanishes at low $p_t$ and then increases regularly.  The dependence is
almost linear for pions, but rather quadratic below 1~GeV/$c$ for
protons: therefore $v_2$ is smaller for protons than for pions at low
$p_t$, while they become of comparable magnitude at 2~GeV/$c$.  These
nontrivial features are qualitatively reproduced by hydrodynamical
models~\cite{Kolb:00,Kolb:2000fh}, where the dependence on the hadron
mass is well understood~\cite{pasi2}.  Unfortunately, there do not
seem to be any predictions from transport models for $v_2(p_t)$ at SPS
energies. Several such calculations are available at the higher RHIC
energies, where the same features are
observed~\cite{Bleicher:2000sx,Lin:2001zk,Humanic:2002iw,Sahu:2002sp}.
Coming to quantitative comparisons, elliptic flow at 158$A$ GeV
seems to be overestimated by hydrodynamical calculations, although not
by a large factor.  Kolb {\em et al.}~\cite{Kolb:00} predict
$v_2(p_t=1\,{\rm GeV}/c)\simeq 12\%$ for pions in mid-central
collisions, while our various estimates (see
Fig.~\ref{fig:158_V2vsPt}, middle left) are closer to $8\%$.  The
model of Teaney {\em et al.}~\cite{Teaney:2001av}, which couples
hydrodynamics to a transport model, naturally yields smaller values of
$v_2$ (about $10\%$ for pions at $p_t=1$~GeV/$c$), but still higher 
than our data. These hydrodynamical calculations use ``soft''
equations of state, which are needed in order to reproduce the
measured $p_t$ spectra.  The fact that they tend to overestimate $v_2$
suggests that thermalization is only partially achieved even at the
top SPS energy.

The rapidity dependence of $v_2$ is also instructive.  Due to the
strong Lorentz contraction at ultrarelativistic energies, all the
produced particles essentially originate from the same point
$z=t=0$. Therefore, the longitudinal velocity $v_z$ of a particle (or
its rapidity) is expected to be strongly correlated to its position
$z$ throughout the evolution of the system.  This means that particles
with the same rapidity are those which interacted throughout the
evolution.  Following the common interpretation of elliptic flow as
due to secondary collisions, one expects an increase of $v_2$ with the
multiplicity density $dN/dy$, up to the limiting value corresponding
to thermal equilibrium.  It is interesting to note that, at 158$A$
GeV, the rapidity dependence of $v_2$ (Fig.~\ref{fig:158_V2vsY})
follows that of the multiplicity $dN/dy$
\cite{crosssections}, although it is less pronounced.  This again
suggests only partial thermalization.  Let us now come to quantitative
comparisons with the various models. Most of the calculations were
done with the same $p_t$ cuts as in the earlier NA49
analysis~\cite{NA49PRL}, i.e. $50<p_t<350$~MeV/$c$ for pions, which
would yield values of $v_2(y)$ smaller than in the present analysis by
a factor of approximately 2 according to our measurements.  First, all
above mentioned hydrodynamical calculations assume rapidity
independence~\cite{Bjorken:1982qr}, so that they are unable to predict
the rapidity dependence of $v_2$.  Hirano~\cite{Hirano:2000eu}
performs a full three-dimensional hydro calculation, tuned to
reproduce both rapidity and transverse momentum spectra. The resulting
$v_2(y)$ is about $3\%$ at midrapidity, significantly larger than our
result in Fig.~\ref{fig:158_V2vsY}, middle left (with due attention to
the different cuts in $p_t$, as explained above), and the decrease at
higher rapidities seems to be steeper than in our data.  Various
predictions are available from microscopic models for $v_2(y)$ of
pions and protons at the top SPS energy.  Predictions of
RQMD~\cite{Liu:1998yc} and UrQMD~\cite{Soff:1999yg} seem to be in
agreement with our data, at least qualitatively.  Calculations based
on the QGSM model~\cite{Bravina:2000dg} show a dip near midrapidity
for peripheral collisions, which we do not see in the data.

The centrality dependence of $v_2$ is particularly interesting.  In
hydrodynamics, it closely follows the centrality dependence of the
initial asymmetry, i.e., $v_2$ is proportional to the eccentricity
$\epsilon$ of the lens-shaped object~\cite{OL92}.  It thus increases
with impact parameter up to very peripheral collisions if the system
is thermalized.  On the other hand, if only partial thermalization is
achieved, $v_2$ is smaller for peripheral collisions, and the maximum
of $v_2$ is shifted to less peripheral
collisions~\cite{Kolb:2000fh,Voloshin:1999gs}.  Since the eccentricity
of the lens-shaped object, $\epsilon$, increases linearly with the
number of spectator nucleons~\cite{OL92}, one would expect a linear
increase of $v_2$ with $E^0$ in Figs.~\ref{fig:v_cen} and
\ref{fig:Vn_int_cum}.  At 158$A$ GeV, the experimental curve
clearly bends over in the two most peripheral bins, and this tendency
is more pronounced with estimates from higher order cumulants which
are less sensitive to nonflow correlations.  This again suggests that
thermalization is not fully achieved.  At 40$A$ GeV, error bars are
larger, but the decrease of $v_2$ is even more pronounced in the most
peripheral bin: departure from thermalization would then be more
important than at the higher energy, which is to be expected since the
density is lower.  Finally, a peculiar centrality dependence of $v_2$
was proposed as a signature of the phase transition to the quark-gluon
plasma~\cite{Sorge:1998mk}: with increasing centrality
(i.e. decreasing impact parameter), the ratio $v_2/\epsilon$ was
predicted to first increase, due to better thermalization, then
saturate, and eventually increase again for the most central
collisions as a result of the re-hardening of the equation of state
when a quark-gluon plasma is produced.  At 158$A$ GeV, however,
Fig.~\ref{fig:Vn_int_cum} shows that the increase of $v_2$ with the
number of spectators is linear for the most central collisions, which
suggests a constant $v_2/\epsilon$. At 40$A$ GeV, on the other hand,
the flow obtained from 4-particle cumulants remains unusually high for
central collisions, which may be a hint of interesting physics.

We finally come to the dependence of $v_2$ on beam energy.  There are
very few model calculations of $v_2$ at 40$A$ GeV in the
literature.  The most natural expectation is that $v_2$ should
increase regularly as the beam energy increases, and eventually
saturate: first, because the effect of spectators (which tends to
lower $v_2$) decreases, second, because the degree of thermalization
increases.  Such a behavior is indeed observed with a smoothly varying
equation of state~\cite{Kolb:99}. If a strong first order phase
transition occurs in the system, however, an interesting structure is
predicted~\cite{Kolb:99}, where a small maximum in $v_2$ occurs
somewhere between the top AGS and the top SPS energies.  Such a
structure does not appear in Fig.~\ref{fig:v2_edep}, which shows that
the increase of $v_2$ from AGS to RHIC is smooth.

There is a nice way to combine the centrality dependence and the beam
energy dependence of $v_2$, by plotting the scaled anisotropy
$v_2/\epsilon$ versus the particle density as estimated by $dN/dy$ of
charged particles divided by the area of the overlap region,
$S$~\cite{Voloshin:1999gs,Voloshin:QM02}.
If the system is locally thermalized, and if there is no marked
structure in the equation of state (such as a strong first order phase
transition), $v_2/\epsilon$ is expected to be centrality independent
as explained above, and to depend only weakly on the beam energy
through the equation of state: this is indicated by the horizontal
lines (``hydro limits'') in Fig.~\ref{fig:fscaling}.  If the mean free
path of the particles becomes as large as the transverse size,
thermalization is not reached, and $v_2/\epsilon$ is smaller than the
hydro limit: it is expected to increase smoothly with the ratio
between the transverse size and the mean free path, and eventually
saturate at the hydro limit.  The latter ratio scales itself like the
charged particle density per unit transverse area $(1/S)dN/dy$, which
is the abscissa in Fig.~\ref{fig:fscaling}.  The data shown in this
figure are compatible with the above picture, since data taken at
different beam energies fall on the same universal curve.  It is quite
remarkable, and far from obvious, that this simple physical picture
allows us to compare quantitatively a peripheral collision at the full
RHIC energy with a central collision at the lower SPS energy.

A further interest in this plot is the
speculation~\cite{Voloshin:1999gs} whether there is flattening at an
abscissa value of about 10 which could indicate a change in the
physics of rescattering and suggest a deconfinement phase transition
in this region.  It is noteworthy that the color percolation
point~\cite{Satz:QM02}, which is shown by an arrow in the figure, is
just in the same region.  
Unfortunately, taking into account the large systematic errors of the
results, we are not able to resolve this question. Further data
analysis at 80$A$~GeV SPS and $\sqrt{s_{NN}}=20 $GeV RHIC
beam energies would be useful.

\subsection{Directed flow}

Elliptic flow at SPS is created by the interactions between the
produced particles, but is to a large extent independent of the
mechanism of particle production. By contrast, the physics of directed
flow probes a time scale set by the crossing time of the
Lorentz-contracted nuclei~\cite{Bleicher:2002xx}.  It is therefore
more subtle, and more strongly model dependent than the physics of
elliptic flow.

Directed flow defines the impact parameter as an oriented arrow in the
transverse plane. As such, it reflects the asymmetry between target
and projectile (to which elliptic flow is insensitive).  For this
reason, most model calculations only address the directed flow of
nucleons.  In addition, most papers do not deal with $v_1$, but with
the mean transverse momentum projected on the reaction plane,
$\mean{p_x}=\mean{p_t\, v_1(p_t,y)}$, as a function of rapidity, for
historical reasons~\cite{Danielewicz:hn}.

It was first expected that directed flow would be negligible at SPS
energies.  In 1991, it was predicted {\em et al.}~\cite{Amelin:cs}
that it might however be large enough to be measurable in Pb-Pb
collisions. Non-zero $v_1$ for nucleons was predicted both by a
transport model (QGSM) including re-scatterings, and by a
hydrodynamical model. Furthermore, in the hydro model, $\mean{p_x}$
depended strongly on the equation of state: as elliptic flow, directed
flow is smaller with a softer equation of state (involving for
instance a phase transition to a quark-gluon plasma).

The interest in directed flow was revived following the prediction
that the ``softest point'' of the equation of state could be directly
observed at the AGS ~\cite{Hung:1994eq}.  A deep minimum of
$\mean{p_x}$ may appear, at an energy of about
6~$A$ GeV~\cite{Rischke:1995pe}.  These predictions, however,
crucially rely on the assumption that the early stages of the
collision, when particles are produced, can be described by one-fluid
hydrodynamics.  A two-fluid model~\cite{Ivanov:2000dr} predicts no
minimum as energy increases.  In a three-fluid
model~\cite{Brachmann:1999xt}, on the other hand, a minimum occurs,
but at a higher energy, around 10-20~GeV per nucleon. It is followed
by an increase up to a maximum at 40$A$ GeV.  Unfortunately, no
quantitative estimate is provided.  A similar structure is predicted
in the transport model UrQMD~\cite{Bleicher:2002xx}, where the minimum
of $\mean{p_x}$ also appears as a consequence of the softening of the
equation of state (although no quark-gluon plasma is explicitly
incorporated in the model): $\mean{p_x}$ increases from the top AGS
energy and then saturates above 40~GeV per nucleon at a value of
$30$~MeV/$c$.  Unfortunately, an observable such as $\mean{p_x}$ is
largest in the target and projectile rapidity regions, which are not
covered by the present experiment, so that quantitative comparisons
are not possible.

We therefore concentrate on predictions for differential flow as a
function of rapidity, $v_1(y)$.  The remarkable feature of our data in
Figs.~\ref{fig:158_V1vsY} and \ref{fig:40_V1vsY} is the flatness of
$v_1$ for protons near mid-rapidity, where the slope at midrapidity
may even be negative for peripheral collisions.  This ``antiflow,'' or
``wiggle,'' was predicted at RHIC energies~\cite{Snellings:1999bt}. It
is also present at SPS energies in most fluid-dynamical
calculations~\cite{Brachmann:1999xt,Csernai:1999nf,Ivanov:2000dr}.
Transport models like UrQMD~\cite{Soff:1999yg} predict a too large
value of $v_1(y)$ for protons, although they do see some flattening
near midrapidity.  In the QGSM model \cite{Bravina:2000dg}, $v_1(y)$
for protons shows no flattening, except for peripheral collisions. The
value of $v_1$ of protons is much larger than our data: about 7\% for
central collisions and 10\% for mid-central collisions at $y=1.5$,
while we see at most 2\%.  The value of $v_1(p_t)$ in
QGSM~\cite{Zabrodin:2000xc} is also larger than our data for protons.

There are comparatively very few studies of pion directed flow at SPS
energies.  Pion directed flow was first seen in asymmetric
collisions~\cite{Gosset:1988cm} at lower energies.  The $v_1$ of pions
usually has a sign opposite to the proton $v_1$, which is understood
as an effect of shadowing and absorption on
nucleons~\cite{Bass:em,Li:tj}.  The same effect was observed at
AGS~\cite{Barrette:1996rs} and is clearly seen in all our data.
UrQMD~\cite{Soff:1999yg} correctly predicts $v_1(y)$ for pions at the
top SPS energy, although their prediction for protons is too high.  In
the QGSM model~\cite{Bravina:2000dg}, the pion $v_1(y)$ has a minimum
at about half the projectile rapidity.  In Fig.~\ref{fig:158_V1vsY},
the same tendency is seen in the standard analysis (also quantitative
agreement) but not in the three-particle estimate which shows further
decrease toward projectile rapidities.  The peculiar transverse
momentum dependence of $v_1$ for pions, which is negative at low $p_t$
and then positive above 1~GeV/$c$ (see Figs.~\ref{fig:158_V1vsPt} and
\ref{fig:40_V1vsPt}), does not seem to have been predicted.

\section{Summary}
\label{s:summary}

We have presented a true multiparticle analysis of directed and
elliptic flow which provides results on $v_1$ and $v_2$ values of
pions and protons as function of rapidity, transverse momentum,
collision centrality and beam energy (40 and 158$A$ GeV).

Two independent analyses were carried out using two different methods:
the standard method of correlating particles with an event plane, and
the cumulant method of studying genuine multiparticle
correlations. The cumulant method yields several independent estimates
of the flow. From the point of view of physics, the lowest-order
estimates $v_1\{2\}$, $v_2\{2\}$ are essentially equivalent to the
estimates from the event-plane method, while higher-order estimates
$v_1\{3\}$, $v_2\{4\}$ are hopefully free from nonflow effects. This
is the main motivation of the cumulant method. The two methods are
very different in their practical implementation. The cumulant method
no longer requires one to construct subevents, or to correct for the
event-plane resolution. All flow estimates are derived from a single
generating function of azimuthal correlations. Constructing this
generating function, however, requires more computer time than the
standard flow analysis. Another significant difference between the two
methods is that the cumulant method takes naturally into account
azimuthal asymmetries in the detector acceptance. Hence the flattening
procedures, and the cuts in phase space which are required in the
event-plane method in order to minimize the effects of these
asymmetries, are no longer required. The price to pay for all these
enhancements is increased statistical errors.

We have obtained the first direct, quantitative evidence for
collective motion at these energies: elliptic flow at 158$A$ GeV
has been reconstructed independently from genuine 4, 6 and 8-particle
correlations, and all three results agree within statistical errors
(Fig.~\ref{fig:Vn_int_cum}, top left).  This is confirmed at both
energies by differential analyses of elliptic flow (as a function of
rapidity or transverse momentum) from genuine 4-particle correlations.
In the case of directed flow, nonflow correlations due to momentum
conservation, which are large, have been subtracted.  Furthermore, a
new method of analysis from 3-particle correlations, which is unbiased
by nonflow correlations, has been implemented for the first time at
both energies.

The directed flow of protons reveals a structure which is
characteristic of ultrarelativistic energies, and is not present at
AGS energies. A clear separation appears for the first time between
the central rapidity region, where the proton $v_1$ is essentially
zero, and the target-projectile fragmentation region, where it is
large.  Indeed, at 40$A$ GeV, significant directed flow is
observed only at the most forward rapidities covered by the detector
acceptance (Fig.~\ref{fig:40_V1vsY}, right).  At 158$A$ GeV,
where the acceptance covers smaller values of the scaled rapidity,
$v_1$ values are consistent with zero (Fig.~\ref{fig:158_V1vsY},
right), within statistical errors and possible contributions by
nonflow effects. In the fragmentation region, on the other hand, large
$v_1$ values have been observed by WA98~\cite{WA98:QM99}. At both
energies, the first observation of the ``wiggle'' (i.e., a negative
slope of the proton $v_1$ near mid-rapidity) is reported, but there
are indications that it may be due to nonflow effects.

Surprisingly, the directed flow of pions does not follow the same
behavior as that of protons.  While the proton $v_1$ at central
rapidity is much smaller than at AGS energies, the pion $v_1$ remains
essentially of the same magnitude.  It becomes even larger, in
absolute value, than the proton $v_1$.  This amazing phenomenon, which
has never been observed at lower energies, clearly indicates that the
proton $v_1$ and the pion $v_1$ have different physical origins.  The
directed flow of pions behaves similarly at the two beam energies,
both in magnitude and in shape.  It has a peculiar, essentially flat,
transverse momentum dependence (Figs.~\ref{fig:158_V1vsPt} and
\ref{fig:40_V1vsPt}, left).  Its centrality dependence is also quite
remarkable: it increases in magnitude steadily without saturating up
to the most peripheral collisions.  (Fig.~\ref{fig:v_cen}, top, and
Fig.~\ref{fig:Vn_int_cum}, bottom).

Elliptic flow becomes the dominant azimuthal anisotropy at
ultrarelativistic energies.  While it is smaller than directed flow up
to the top AGS energy, here it becomes larger already at
40$A$ GeV.  This is again an indication that SPS is probing the
truly ultrarelativistic regime.  As a consequence of the larger value,
our estimates of $v_2$ are more accurate than our estimates of $v_1$.
As a function of transverse momentum, $v_2$ increases almost linearly
for pions, and more quadratically for protons, as already seen at
RHIC.  On the other hand, the rapidity dependences are the same for
pions and protons: $v_2$ is approximately constant in the central
rapidity region (see e.g., Fig.~\ref{fig:158_V2vsY}, middle) and drops
in the target-projectile fragmentation regions, roughly where the
proton $v_1$ starts increasing.  At 158$A$ GeV, $v_2$ has a
pronounced centrality dependence (Figs. \ref{fig:v_cen} and
\ref{fig:Vn_int_cum}, left) but, unlike directed flow, it saturates
for very peripheral collisions.  The centrality dependence at 40$A$
GeV (Figs. \ref{fig:v_cen} and \ref{fig:Vn_int_cum}, right) is less
significant.  For protons (shown in the same figures) the tendencies
are the same, but the significance is reduced due to larger
statistical errors.

The energy dependence of $v_2$ looks roughly linear from AGS up to
RHIC energies (Fig. \ref{fig:v2_edep}).  No indication of
non-monotonic behavior is visible, as would be expected from the
softening of the equation of state for a system close to the critical
temperature.  The dependence of elliptic flow (divided by the
eccentricity of the nuclear overlap region) on particle (rapidity)
density also exhibits a smooth increase without significant structure
(Fig. \ref{fig:fscaling}) which would indicate a change in the physics
of rescattering.

\begin{acknowledgments}
This work was supported by the Director, Office of Energy Research,
Division of Nuclear Physics of the Office of High Energy and Nuclear
Physics of the US Department of Energy (DE-ACO3-76SFOOO98 and
DE-FG02-91ER40609), the US National Science Foundation, the
Bundesministerium fur Bildung und Forschung, Germany, the Alexander
von Humboldt Foundation, the UK Engineering and Physical Sciences
Research Council, the Polish State Committee for Scientific Research
(2 P03B 13023, SPB/CERN/P-03/Dz 446/2002-2004, 2 P03B 02418 and 2
P03B 04123), the Hungarian Scientific Research Foundation (T14920,
T32293, and T32648), Hungarian National Science Foundation, OTKA,
(F034707), the EC Marie Curie Foundation, the Polish-German
Foundation, and the ``Actions de Recherche Concert{\'e}es'' of
``Communaut{\'e} Fran{\c c}aise de Belgique'' and IISN-Belgium.
\end{acknowledgments}

\appendix
\section{Interpolation formulas for the cumulants}
\label{s:interpolation}
In this appendix, we give interpolation formulas which were used to
extract the cumulants of multiparticle correlations up to
eight-particle correlations.  Following the procedure proposed in
Ref.~\cite{Borghini:2001vi}, we introduce interpolation points
$z_{p,q}=x_{p,q}+iy_{p,q}$,
\begin{eqnarray}
\label{defxy}
x_{p,q}&\equiv& r_0\sqrt{p}\,\cos\left(\frac{2\,q\,\pi}{9}\right),\cr
y_{p,q}&\equiv& r_0\sqrt{p}\,\sin\left(\frac{2\,q\,\pi}{9}\right),
\end{eqnarray}
for $p=1,\cdots,4$, $q=0,\cdots,8$, and $r_0$ of order unity.  The
generating function of cumulants Eq.\ (\ref{gencum}) is then computed
at these various points: $C_{p,q} \equiv {\cal C}_n( z_{p,q})$, and
averaged
\begin{equation}
C_p\equiv \frac{1}{9}\sum_{q=0}^8 C_{p,q}.
\end{equation}
Finally, the cumulants for correlations of 2, 4, 6, and 8 particles
are given by
\begin{eqnarray}
\label{cum_int}
\displaystyle c_n\{2\} &=& 
\frac{1}{r_0^2\mean{w_n^2}} \left( 4\,C_1- 3\,C_2+\frac{4}{3}C_3 - \frac{1}{4}C_4 \right), 
\cr
\displaystyle c_n\{4\} &=& 
\frac{1}{r_0^4\mean{w_n^2}^2} \left( \frac{-52}{3} C_1+19\,C_2-\frac{28}{3} C_3 + 
\frac{11}{6} C_4 \right), \cr
\displaystyle c_n\{6\} &=& 
\frac{6}{r_0^6\mean{w_n^2}^3} \left(9\, C_1-12\,C_2+7\,C_3-\frac{3}{2}C_4\right), \cr 
\displaystyle c_n\{8\} &=& 
\frac{96}{r_0^8\mean{w_n^2}^4} \left(-C_1+\frac{3}{2}C_2-C_3+\frac{1}{4}C_4\right).
\end{eqnarray}

\section{Statistical errors of the contribution of nonflow effects}
\label{s:nonflow-error}

In this appendix, we derive the statistical uncertainties on the
coefficients $g_n$ in Eqs.\ (\ref{g1-g2}), using the analytical
formulas for the statistical errors of the various estimates
$v_n\{k\}$ given in Refs.~\cite{Borghini:2001vi,Borghini:2002vp}.

Let us start with $g_2$, that measures nonflow correlations in the
second harmonic.  We recall its definition:
\begin{eqnarray}
\label{g2}
g_2 & \equiv & N \cdot \left( c_2\{2\} - \sqrt{-c_2\{4\}} \right), \cr
 & = & N \cdot \left( v_2\{2\}^2 - v_2\{4\}^2 \right),
\end{eqnarray}
where the second line holds provided $c_2\{4\}$ has the right,
negative sign.  The statistical error of $g_2$ can easily be
calculated and reads
\begin{equation}
\label{g2-error}
\left( \delta g_2 \right)^2 =  
4 N^2 \,v_2^2 \left[ (\delta v_2\{2\})^2 + (\delta v_2\{4\})^2 - 
 2\left( \mean{v_2\{2\} v_2\{4\}} - \mean{v_2\{2\}}\mean{v_2\{4\}} \right) 
\right]. 
\end{equation}
where we used $\delta(v_n^2) = 2 v_n \delta v_n$.
Note that since $v_2\{2\}$ and $v_2\{4\}$ are slightly correlated if
they measure an existing $v_2$, the error is smaller than the
mere geometrical mean of the uncertainties on each flow estimate. 
Using formulas given in Ref.~\cite{Borghini:2001vi}, we obtain
\begin{equation}
\label{g2-error2}
\delta g_2 = \frac{N}{M} \, \frac{1}{\sqrt{N_{\rm evts}}} \,
\frac{\sqrt{1 + 4{\chi_2}^2 + 2{\chi_2}^4}}{{\chi_2}^2}\, ,
\end{equation}
where $N$ is the total number of emitted particles, $M$ the
multiplicity used in the analysis, $N_{\rm evts}$ the total number of
events and ${\chi_2}=v_2\sqrt{M}$ the resolution parameter.  The
estimate of $v_2$ from 4-particle cumulants, $v_2\{4\}$, was used to
calculate $\chi_2$.

Eq.~(\ref{g2-error2}) is valid as long as relative errors on both
2-particle and 4-particle cumulants are small.  When relative errors
become large (usually for central collisions where $v_2$ is small, or
for the most peripheral collisions where $M$ is small), it may happen
that the cumulant $c_2\{4\}$ is positive, so that $g_2$ in
Eq.~(\ref{g2}) is undefined.  In this case, we set our 4-particle
estimate $v_2\{4\}$ to zero in the second line of Eq.~(\ref{g2}). We
then estimate the statistical error noting that it is essentially
dominated by that on the 4-particle cumulant $c_2\{4\}$, which is
$\delta c_2\{4\}=2/\left(M^2 \sqrt{N_{\rm
evts}}\right)$\cite{Borghini:2001vi} when the resolution parameter
$\chi_2$ is small. The statistical error of $g_2$ then reads
\begin{equation}
\label{g2-error3}
\delta g_2 = N \sqrt{\delta c_2\{4\}}=
 \frac{N\sqrt{2}}{M\,N_{\rm evts}^{1/4}}.
\end{equation}

Consider now $g_1$, that is, nonflow correlations in the first
harmonic.  We may assume that $c_1\{2\}$ and $c_1\{3\}$ are
uncorrelated since the latter comes from a mixed correlation.
Therefore, the statistical error of $g_1$ can be written as
\begin{equation}
\label{g1-error}
\left( \delta g_1 \right)^2 = 
N^2 \left[ (\delta c_1\{2\})^2 + \frac{(\delta c_1\{3\})^2}{v_2^2} + 
\left(\frac{c_1\{3\}}{{v_2}^2}\,\delta v_2 \right)^2 \right].
\end{equation}
If the estimated $v_1\{3\}$ is reconstructed from $v_2\{4\}$, as is
the case in the 158$A$ GeV analysis, one should use the
statistical error $\delta v_2\{4\}$ in Eq.\ (\ref{g1-error}).
However, at 40$A$ GeV, since we used $v_2\{2\}$ (see
Sec. \ref{s:cumulant}) we have to use $\delta v_2\{2\}$ in Eq.\
(\ref{g1-error}).  The corresponding formulas are given in
\cite{Borghini:2001vi}, while the formula for $\delta c_1\{3\}$ can be
found in Ref.~\cite{Borghini:2002vp}.

\end{document}